\documentclass[journal]{IEEEtran}
\usepackage{cite}
\usepackage{graphicx}
\usepackage{subfigure}
\usepackage{amssymb} 
\usepackage[cmex10]{amsmath} 
\usepackage{multirow}
\usepackage{array}
\usepackage{float}
\usepackage{perpage}
\MakeSorted{table}
\usepackage{flafter}

\begin{document}

\title{Validation of Geant4 simulation of electron energy deposition}

\author{Matej Bati\v{c}, Gabriela Hoff, Maria Grazia Pia, Paolo Saracco, Georg Weidenspointner% <-this % stops a space
\thanks{Manuscript received April 11, 2013.}% <-this % stops a space
\thanks{This work has been partly funded by CNPq BEX6460/10-0 grant, Brazil.}% <-this % stops a space
\thanks{M. Bati\v{c} was with INFN Sezione di Genova, I-16146 Genova, Italy; he is now with  
             Sinergise, 1000 Ljubljana, Slovenia.}
\thanks{G. Hoff is with  
             Pontificia Universidade Catolica do Rio Grande do Sul, Porto Alegre, Brazil.}
\thanks{M. G. Pia and P. Saracco are with INFN Sezione di Genova, Via Dodecaneso 33, I-16146 Genova, Italy 
	(phone: +39 010 3536328, fax: +39 010 313358,
	MariaGrazia.Pia@ge.infn.it,Paolo.Saracco@ge.infn.it).}
\thanks{Georg Weidenspointner is with the Max-Planck-Institut f\"ur
	extraterrestrische Physik, Postfach 1603, 85740 Garching, Germany, and
	with the MPI Halbleiterlabor, Otto-Hahn-Ring 6, 81739 M\"unchen, Germany.}
}

\maketitle

\begin{abstract}
Geant4-based simulations of the energy deposited by electrons in various
materials are quantitatively compared to high precision calorimetric
measurements taken at Sandia Laboratories.
The experimental data concern electron beams of energy between a few tens of keV
and 1 MeV at various incidence angles.
Two experimental scenarios are evaluated: the longitudinal energy deposition
pattern in a finely segmented detector, and the total energy deposited in a larger size
calorimeter.
The simulations are produced with Geant4 versions from 9.1 to 9.6; they involve
models of electron-photon interactions in the standard and low energy
electromagnetic packages, and various implementations of electron multiple scattering.
Significant differences in compatibility with experimental data are observed in
the longitudinal energy deposition patterns produced by the examined Geant4
versions, while the total deposited energy exhibits smaller variations across
the various Geant4 versions, with the exception Geant4 9.4.
The validation analysis, based on statistical methods, shows that the best
compatibility between simulation and experimental energy deposition profiles is
achieved using electromagnetic models based on the EEDL and EPDL evaluated data
libraries with Geant4 9.1.
The results document the accuracy achievable in the simulation of the energy
deposited by low energy electrons with Geant4; they provide guidance for 
application in similar experimental scenarios and for improving Geant4.
\end{abstract}

% -----------------------------------------------------------------------------------------

\section{Introduction}
\label{sec_intro}
\PARstart{T}{he} simulation of the interactions with matter of electrons and of
their secondary particles is one of the main tasks of any Monte Carlo codes for particle
transport.
The resulting energy deposition is relevant to a wide variety of experimental
applications, where electrons contribute to determine experimental observables either as
primary or secondary particles.

High precision experimental measurements
\cite{sandia79,lockwood73,sandia80,lockwood75,lockwood76} were performed at the
Sandia National Laboratories specifically for the validation of the ITS
(Integrated Tiger Series) \cite{tiger} simulation code: they concern electrons with energies
ranging from a few tens of keV to 1~MeV, and involve various target materials
and electron incidence angles.
These experimental data are still regarded as the most comprehensive reference
for benchmarking the simulation of energy deposition by low energy electrons:
they have been exploited in the validation
\cite{kawrakow,chibani,penelopebench,carrier,ivanchenko,kim,kadri,jun, tns_sandia} of numerous general
purpose Monte Carlo codes other than the ITS system, for which the measurements
were originally intended, such as EGS \cite{egs4}, EGSnrc \cite{egsnrc}, Geant4
\cite{g4nim,g4tns}, MCNP \cite{mcnp}, MCNPX \cite{mcnpx} and
Penelope \cite{penelope}.
%Most of these comparisons are limited to a qualitative, visual appraisal of the
%compatibility of simulated and experimental data, while the validation study documented in
%\cite{tns_sandia} is characterized by a quantitative estimate, based on rigorous
%statistical methods.

The validation of simulated electron energy deposition in \cite{tns_sandia}
concerns two versions of Geant4, 8.1p02 and 9.1: the latter was the latest
version available at the time when the article was written.
Some differences in compatibility with experiment were observed between the two
Geant4 versions, which were ascribed to evolutions in Geant4 multiple
scattering implementation.

Statements of improvements to Geant4 simulation of electromagnetic interactions
have been reported in the literature
\cite{em_mc2010,em_chep2010,em_radecs2011,em_chep2012} since the publication of
\cite{tns_sandia}, and a multiple scattering model specifically addressing the
transport of electrons \cite{kadri_goudsmit} has been introduced in the Geant4
toolkit.
This paper documents quantitatively how these evolutions in Geant4
electromagnetic physics domain affect the accuracy of the simulation of the
energy deposited by low energy electrons: it reports comparisons between
experimental data in \cite{sandia79,sandia80,lockwood73,lockwood75} and
simulations based on Geant4 versions from 9.1 to 9.6, which span five years'
Geant4 development.

In this respect, it is worthwhile to note that several versions of Geant4 are
actively used in the experimental community at any given time, not limited to the
latest release: in fact, despite the fast release rate of Geant4 of one or two new
versions per year, often complemented by correction patches, experimental
projects usually require a stable simulation production environment for large
portions of their life-cycle and retain a Geant4 version in their
simulation productions for an extended period, even though new versions may become available in
the meantime.

Besides the effects due to the evolution of Geant4 electromagnetic physics, this
paper evaluates quantitatively another issue of experimental relevance: the
sensitivity to physics modeling features in relation to the geometrical
granularity of the detector.

The results of this validation analysis provide guidance to experimental users
in optimizing the configuration of Geant4-based applications in scenarios
concerned by the simulation of the energy deposited by low energy electrons.
This investigation may be relevant also to high energy experiments, since low
energy electrons contribute to the determination of the characteristics of electromagnetic
showers initiated by high energy particles, or the signal produced in
detectors in general.
The validation tests reported here contribute to improve Geant4 by objectively
identifying areas where its capability of reproducing experimental measurements
could profit from more refined physics modeling or software engineering methods.

% -----------------------------------------------------------------------------------------

\section{Strategy of this study}
\label{sec_strategy}

This validation study covers two experimental scenarios: the
longitudinal pattern of the energy deposited by electrons in a segmented
calorimeter, and the total energy deposited in a bulk calorimeter.
%They correspond to the measurements respectively 
%reported in \cite{sandia79,lockwood73} and \cite{sandia80,lockwood75}.
Test cases, characterized by electron energy, beam incidence angle and target
material, are reproduced in the simulation for both scenarios according to the
respective experimental references \cite{sandia79,sandia80}.
The test cases involving uranium as target material are not considered in the
validation process, since concerns were expressed about the presence of 
systematic errors in the calorimeter data for this material \cite{sandia80}.

The physics configuration activated in the simulation (selection of Geant4
processes, models and secondary production thresholds) is the same in both
scenarios; only the geometrical configuration and the scored observable differ,
as they reflect the respective experimental set-up.
This application design feature allows the evaluation of the sensitivity of different
observables to the physics modeling options available in Geant4 and to the evolution of Geant4 kernel. 

In both scenarios three sets of electron-photon interaction models are
evaluated: 
\begin{itemize}
\item the models based on the EEDL (Evaluated Electron Data Library)
\cite{eedl} and EPDL (Evaluated Photon Data Library) \cite{epdl97}, also known
as ``Livermore models'', included in Geant4 ``low energy'' electromagnetic package \cite{lowe_e,lowe_chep,lowe_nss},
\item the models reengineered from the Penelope \cite{penelope} Monte Carlo code,
also belonging to the `'low energy" package,
\item the models
included in Geant4 ``standard'' electromagnetic package \cite{standard}.
\end{itemize}
In addition, the effects of different multiple scattering models on the energy 
deposition patterns corresponding to the two scenarios are estimated.

Simulations are produced with six Geant4 versions released between late 2007 and
late 2012; correction patches to these versions released by the end of February
2013 are applied on top of the original versions.
%Due to the fast pace at which new Geant4 versions are produced, several versions are 
%still actively used in experimental applications, which often require a stable 
%simulation configuration over an extended portion of the experimental life-cycle.
For convenience, the Geant4 versions evaluated in this study are identified
through their original version number; the corresponding patched versions used
to produce the results reported in this paper are listed in Table
\ref{tab_versions}.

\begin{table}
\begin{center}
\caption{Geant4 versions subject to test}
\begin{tabular}{cc}
%\hline
{\bf Version Identifier}	& {\bf Patched Geant4 Version}	\\
\hline
9.1 		& 9.1p03	\\
9.2		& 9.2p04	\\
9.3 		& 9.3p02	\\
9.4 		& 9.4p04	\\
9.5 		& 9.5p01	\\	 
9.6 		& 9.6p01	\\
\hline
\end{tabular}
\label{tab_versions}
\end{center}
\end{table}

%The Geant4-based application used to produce the simulated longitudinal energy
%deposition pattern is the same as in \cite{tns_sandia} for all Geant4 versions,
%apart from modifications to the instantiation of Geant4 objects required by
%changes of class interfaces in Geant4 kernel in versions later than 9.1.
%Only the geometry configuration, and the associated scoring of the energy
%deposited in the sensistive volume, is modified in the simulation for the
%validation of total deposited energy with respect to the experimental data in
%\cite{sandia80}; the physics component of the simulation application is
%identical for both scenarios.

%- total deposited energy and longitudinal energy deposition pattern- 

The compatibility between simulated and experimental data is assessed by means
of statistical methods.
The statistical analysis is articulated over two levels: first the compatibility
between simulated and experimental data is evaluated for each test case on the
basis of goodness-of-fit tests, independently for each Geant4 physics
configuration and version; next differences in compatibility with experiment
are evaluated for different categories of data.
The categories subject to analysis are different electromagnetic physics
settings within a given Geant4 version, and different Geant4 versions with the
same physics settings in the simulation application.
The analysis of categorical data exploits contingency tables based on the
outcome of goodness-of-fit tests over different categories.
%and different multiple scattering settings for the same Geant4 version and
%electron-photon settings.

% -----------------------------------------------------------------------------------------

\section{Reference experimental data}
\label{sec_exp}

The reference data exploited in the validation process derive from high
precision measurements of
longitudinal energy deposition and total deposited energy reported in \cite{sandia79, sandia80} respectively.
The transverse energy deposition was not measured.
The two sets of measurements use the same experimental technique, although
they concern two distinct observables.
The experimental apparatus and measurement techniques are described in detail in
\cite{sandia79,lockwood73,sandia80,lockwood75}; only a brief overview is
provided here to facilitate the comprehension of the results reported in this
paper.

The experimental set-up involved an electron beam impinging on a
target equipped with a calorimeter.  
The energy and incidence angle of the beam varied in the range
from 25~keV to 1.033~MeV, and from 0$^\circ$ to 83.5$^\circ$ respectively. 
The target was configured as a semi-infinite geometry; its thickness was
larger than the range of the most energetic electrons, and its diameter 
was adequate to contain the resulting electromagnetic shower even for non-orthogonal 
beam incidence (apart from possible leakage of Bremsstrahlung photons).
The experimental configuration for the measurement of the longitudinal energy
deposition consisted of a front slab of passive material, a calorimeter and a
so-called ``infinite'' plate, all of the same material; the thickness of the
front slab was varied to measure the deposited energy as a function of
depth.
The measurement depth was determined by the sum of the thickness of the front slab and one-half the calorimeter thickness. 
The calorimeter coincided with the whole target for the measurements 
of the total deposited energy.
For the reader's convenience, the thicknesses of the calorimeters utilized in the
two experimental set-ups are summarized in Table~\ref{tab_calothick}.

% Table generated by Excel2LaTeX from sheet 'caloThick'
\begin{table}[htbp]
  \centering
  \caption{Calorimeter thickness in the two experimental configurations}
    \begin{tabular}{lcc}
    \hline
    \textbf{Target} & \textbf{Longitudinal profile} & \textbf{Total deposited energy} \\
                          & \textbf{Reference \cite{sandia79}} & \textbf{Reference \cite{sandia80}} \\
          & (mm)  & (mm) \\
    \hline
    \textbf{Be} & 0.024 & 3.81 \\
    \textbf{C} & 0.092 & 4.78 \\
    \textbf{Al} & 0.019 & 2.64 \\
    \textbf{Ti} &       & 1.63 \\
    \textbf{Fe} & 0.025 &  \\
    \textbf{Cu} & 0.024 &  \\
    \textbf{Mo} & 0.005 & 0.91 \\
    \textbf{Ta} & 0.010 & 0.53 \\
    \textbf{U} & 0.008 & 0.50 \\
    \hline
    \end{tabular}%
  \label{tab_calothick}%
\end{table}%

% Table generated by Excel2LaTeX from sheet 'TNS_conf79'
\begin{table}
  \centering
  \caption{Test cases for longitudinal energy deposition}
    \begin{tabular}{cccc}
    \hline
    {\bf Target} & \textbf{Z}     & \textbf{Energy} (keV)    & \textbf{Angle} (degrees) \\
\hline
    {\multirow{5}[0]{*}{\bf Be}} & \multirow{5}[0]{*}{4} & 58    & 0 \\
     &       & 109   & 0 \\
     &       & 314   & 0 \\
    &       & 521   & 0 \\
     &       & 1033  & 0 \\
\hline
    {\bf C} & 6     & 1000  & 0 \\
\hline
    {\multirow{6}[0]{*}{\bf Al}} & \multirow{6}[0]{*}{13} & 314   & 0 \\
     &       & 521   & 0 \\
    &       & 1033  & 0 \\
     &       & 314   & 60 \\
     &       & 521   & 60 \\
     &       & 1033  & 60 \\
\hline
    {\multirow{3}[0]{*}{\bf Fe}} & \multirow{3}[0]{*}{26} & 300   & 0 \\
    &       & 500   & 0 \\
    &       & 1000  & 0 \\
\hline
    {\multirow{2}[0]{*}{\bf Cu}} & \multirow{2}[0]{*}{29} & 300   & 0 \\
    &       & 500   & 0 \\
\hline
    {\multirow{7}[0]{*}{\bf Mo}} & \multirow{7}[0]{*}{42} & 100   & 0 \\
   &       & 300   & 0 \\
    &       & 500   & 0 \\
    &       & 1000  & 0 \\
     &       & 300   & 60 \\
     &       & 500   & 60 \\
     &       & 1000  & 60 \\
\hline
    {\multirow{6}[0]{*}{\bf Ta}} & \multirow{6}[0]{*}{73} & 300   & 0 \\
    &       & 500   & 0 \\
    &       & 1000  & 0 \\
    &       & 500   & 30 \\
    &       & 500   & 60 \\
    &       & 1000  & 60 \\
\hline
    {\multirow{4}[0]{*}{\bf U}} & \multirow{4}[0]{*}{92} & 300   & 0 \\
     &       & 500   & 0 \\
     &       & 1000  & 0 \\
     &       & 1000  & 60 \\
    \hline
    \end{tabular}%
  \label{tab_config79}%
\end{table}%

% Table generated by Excel2LaTeX from sheet 'TNS_conf79'
\begin{table}
  \centering
  \caption{Test cases for total energy deposition}
    \begin{tabular}{cccccccccc}
        \hline
    \multicolumn{2}{c}{\textbf{Target}} &  \textbf{E} (keV) & \multicolumn{7}{c}{\textbf{Angle} (degrees)} \\
           Element       & Z   &    & \textbf{0} & \textbf{16} & \textbf{31} & \textbf{46} & \textbf{61} & \textbf{76} & \textbf{83.5} \\
        \hline
    \multirow{4}[0]{*}{\textbf{Be}} & \multirow{4}[0]{*}{4} & 1033  & x      & x     & x     & x     & x     & x     & x \\
          &       & 521   & x     & x     & x     & x     & x     & x     & x \\
          &       & 314   &  x    & x     & x     & x     & x     & x     &  \\
          &       & 109   &      &       & x     & x     &       &       &  \\
    \hline
    &  &  & \multicolumn{7}{c}{{\textbf{Angle} (degrees)}} \\
    \textbf{} &       & \textbf{} & \textbf{0} & \textbf{15} & \textbf{30} & \textbf{45} & \textbf{60} & \textbf{75} & \textbf{82.5} \\
\hline
    \multirow{7}[0]{*}{\textbf{C}} & \multirow{7}[0]{*}{6} & 1000  & x     & x     & x     & x     & x     & x     & x \\
          &       & 500   & x     & x     & x     & x     & x     & x     & x \\
          &       & 300   & x     & x     & x     & x     & x     & x     & x \\
          &       & 100   & x     & x     & x     & x     & x     & x     &  \\
          &       & 75    & x     & x     & x     & x     & x     & x     &  \\
          &       & 50    & x     & x     & x     & x     & x     & x     &  \\
          &       & 25    & x     & x     & x     & x     & x     & x     &  \\
    \hline
%          &       & \textbf{} & \textbf{0} & \textbf{15} & \textbf{30} & \textbf{45} & \textbf{60} & \textbf{75} &  \\
    \multirow{7}[0]{*}{\textbf{Al}} & \multirow{7}[0]{*}{13} & 1033  & x     & x     & x     & x     & x     & x     &  \\
          &       & 521   & x     & x     & x     & x     & x     & x     &  \\
          &       & 314   & x     & x     & x     & x     & x     & x     &  \\
          &       & 109   & x     & x     & x     & x     & x     & x     &  \\
          &       & 84    & x     & x     & x     & x     & x     & x     &  \\
          &       & 58    & x     & x     & x     & x     & x     & x     &  \\
          &       & 32    & x     & x     & x     & x     & x     & x     &  \\
    \hline
%          &       & \textbf{} & \textbf{0} & \textbf{15} & \textbf{30} & \textbf{45} & \textbf{60} & \textbf{75} &  \\
    \multirow{4}[0]{*}{\textbf{Ti}} & \multirow{4}[0]{*}{22} & 1033  & x     & x     & x     & x     & x     & x     &  \\
          &       & 521   & x     & x     & x     & x     & x     & x     &  \\
          &       & 314   & x     & x     & x     & x     & x     & x     &  \\
          &       & 109   & x     & x     & x     & x     & x     & x     &  \\
    \hline
%          &       &       & \textbf{0} & \textbf{15} & \textbf{30} & \textbf{45} & \textbf{60} & \textbf{75} &  \\
    \multirow{4}[0]{*}{\textbf{Mo}} & \multirow{4}[0]{*}{42} & 1033  & x     & x     & x     & x     & x     & x     &  \\
          &       & 521   & x     & x     & x     & x     & x     & x     &  \\
          &       & 314   & x     & x     & x     & x     & x     & x     &  \\
          &       & 109   & x     & x     & x     & x     & x     &       &  \\
    \hline
%          &       &       & \textbf{0} &  & \textbf{30} & & \textbf{60} &       &  \\
    \multirow{3}[0]{*}{\textbf{Ta}} & \multirow{3}[0]{*}{73} & 1033  & x     &       &       &       & x     &       &  \\
          &       & 521   & x     &       & x     &       & x     &       &  \\
          &       & 314   & x     &       &       &       &       &       &  \\
    \hline
%          &       &       & \textbf{0} & \textbf{15} & \textbf{30} & \textbf{45} & \textbf{60} & \textbf{75} &  \\
    \multirow{7}[0]{*}{\textbf{U}} & \multirow{7}[0]{*}{92} & 1033  & x     & x     & x     & x     & x     & x     &  \\
          &       & 521   & x     & x     & x     & x     & x     & x     &  \\
          &       & 314   & x     & x     & x     & x     & x     & x     &  \\
          &       & 109   & x     & x     & x     & x     & x     & x     &  \\
          &       & 84    & x     & x     & x     & x     & x     & x     &  \\
          &       & 58    & x     & x     & x     & x     & x     &       &  \\
          &       & 32    & x     & x     & x     & x     & x     &       &  \\
    \hline
    \end{tabular}%
  \label{tab_config80}%
\end{table}%

The depth at which the deposited energy was measured was expressed as a fraction
of the continuous slowing-down approximation (CSDA) \cite{csda} range.
Some values of the CSDA range of electrons reported in \cite{sandia79} 
differ from those reported by the ESTAR database \cite{estar} of NIST
(National Institute of Standards and Technology), which is considered an
authoritative reference; nevertheless, since the CSDA range is only used as a
scaling factor in the expression of the measurement depth, it does not play any
physical role in the features of the data, therefore any discrepancies with
respect to other references, or to its true value, do not affect the results of
goodness-of-fit tests.

The experimental configurations of the test cases for longitudinal energy
deposition are summarized in Table~\ref{tab_config79}, and those for total
energy deposition in Table~\ref{tab_config80}, where available combinations of
beam energy and incident angle in the experiment are identified by a "x".

The uncertainty in the measurements of the deposited energy
is reported in the experimental references as varying between 1.2\% and 2.2\%
in the different configurations.  
An extensive discussion of the analysis of experimental uncertainties and their
dependence on penetration depth is reported in section V of \cite{tns_sandia};
they are handled in this study accordingly.
%The details of the discussion are not duplicated here due to its length.
The correction of nominal experimental uncertainties as a function of
penetration depth is applied to the data as discussed in \cite{tns_sandia};
nevertheless for the present study it profited from a larger
simulated data sample than the one used in \cite{tns_sandia}, which allowed a
more precise estimate of the error behaviour as a function of depth.
Nevertheless, although the numerical values of the $\chi^{2}$ test statistic
slightly differ when using the original scaling corrections of \cite{tns_sandia}
or the refined ones, the outcome of
the goodness-of-fit tests is the same as reported in \cite{tns_sandia}
in terms of rejection of the null hypothesis (i.e. of
compatibility of experimental and simulated distributions).

%The nominal experimental uncertainties were rescaled as a function of
%penetration depth prior to submitting the data to the goodness-of-fit $\chi^{2}$
%test; the correction procedure profited from a larger simulated data sample than
%the one used in \cite{tns_sandia}, which allowed a more precise estimate of the error
%behaviour as a function of depth.
%It was verified that the update to the error dependence on depth does not
%introduce any systematic effect in the test results: although the numerical
%values of the $\chi^{2}$ test statistic calculated over the Geant4 9.1 data of
%\cite{tns_sandia} slightly differ when using the original scaling corrections
%or the refined ones, the outcome of the goodness-of-fit test in terms
%of rejection of the null hypothesis (i.e. of compatibility of experimental and
%simulated distributions) is not affected.

% -----------------------------------------------------------------------------------------
\section{Simulation configuration}
\label{sec_simu}

The Geant4-based application used for the validation tests reported in this
paper is the same as for the tests reported in \cite{tns_sandia}, apart from the
changes needed to instantiate Geant4 kernel objects whose class interfaces were
modified in the evolution from version 9.1 to 9.6.
A brief overview is provided here; more extensive details can be found in \cite{tns_sandia}.

The energy spectrum of the primary electrons is modeled according to a gaussian
distribution, with width defined by the uncertainty on the beam energy reported
in the experimental references \cite{sandia79, sandia80}.
The beam direction is set in the simulations according to the incident angle
corresponding to each test case.

\begin{figure} 
\centerline{\includegraphics[angle=0,width=7cm]{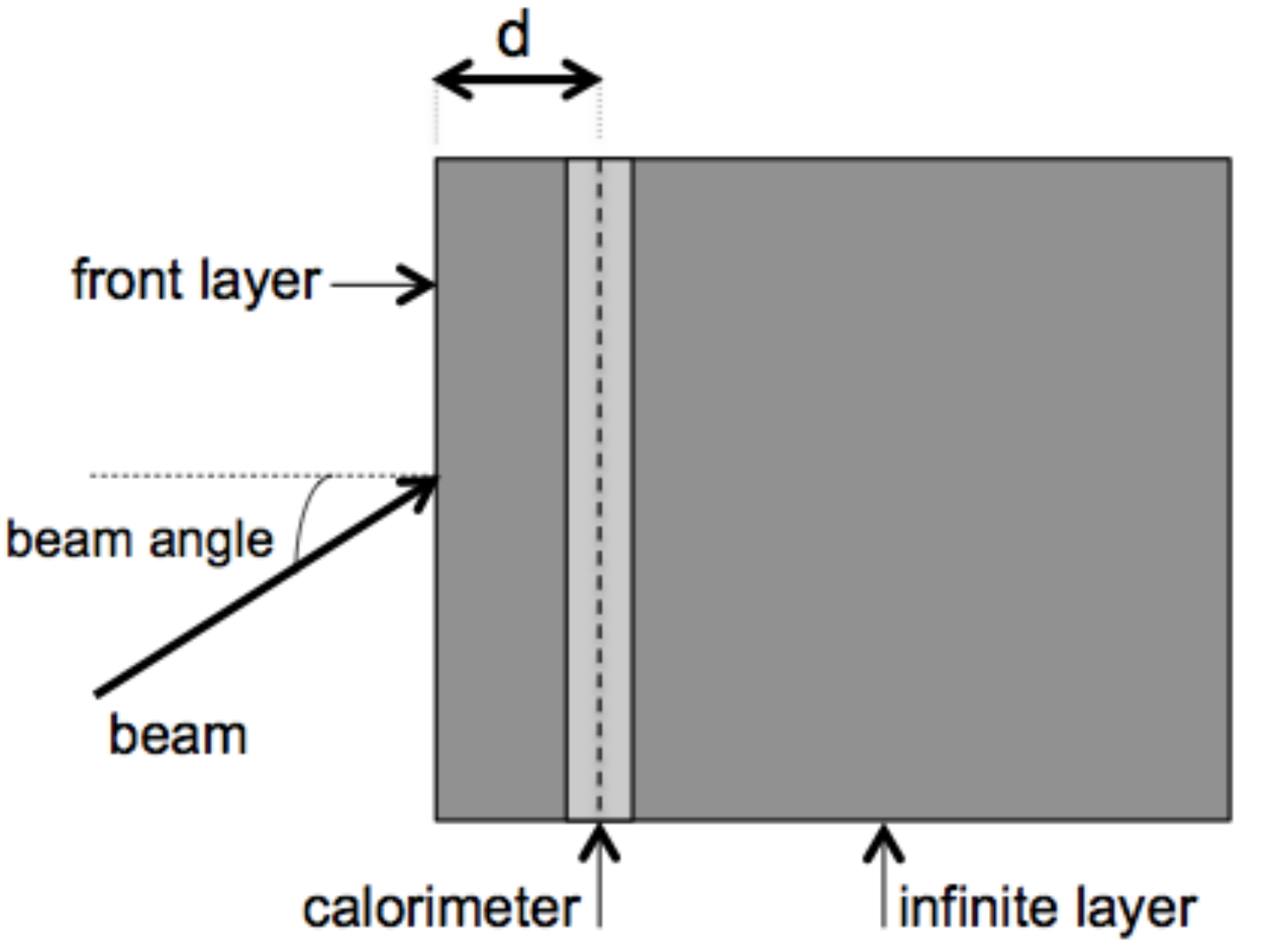}}
\caption{Sketch of the geometrical configuration corresponding to the
experimental set-up of \cite{sandia79} for the test of longitudinal energy
deposition. The lentgth indicated as ``d'' in the figure represents the depth at
which the energy deposition is scored.}
\label{fig_sketch79}
\end{figure}

\begin{figure} 
\centerline{\includegraphics[angle=0,width=7cm]{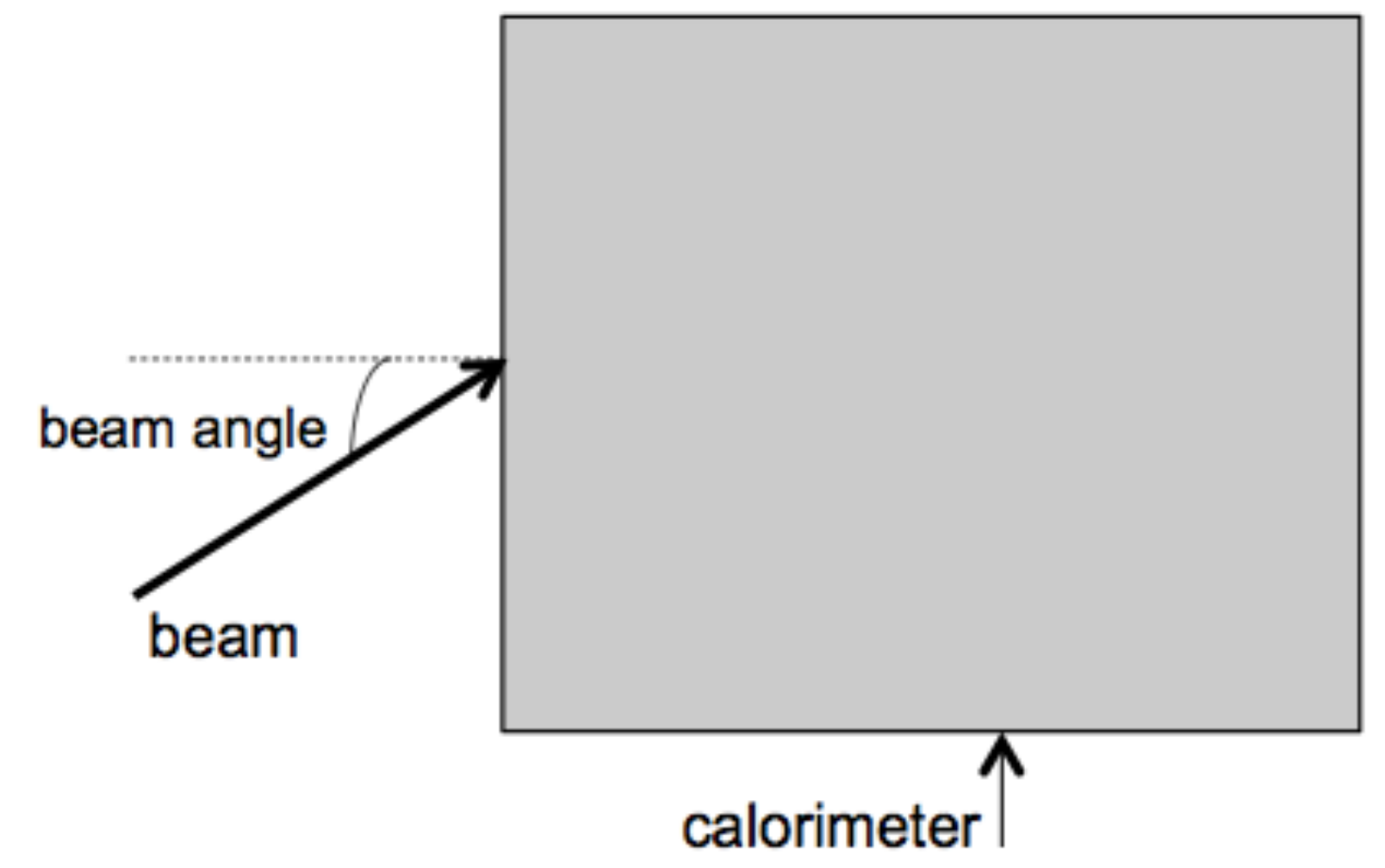}}
\caption{Sketch of the geometrical configuration corresponding to the
experimental set-up of \cite{sandia80} for the test of total energy deposition.}
\label{fig_sketch80}
\end{figure}

The geometry configuration reproducing the experimental set-up for the
determination of the longitudinal energy deposition is the same as described in
\cite{tns_sandia}.
For the simulation of the total deposited energy, the geometry configuration
consists of a single sensitive volume, corresponding to the calorimeter in the
experimental set-up.
The two configurations are sketched in Fig.~\ref{fig_sketch79} and \ref{fig_sketch80} 
respectively.

The physics interactions that can be configured in the simulation are:
ionization, Bremsstrahlung and multiple scattering for electrons, Rayleigh and
Compton scattering, photoionization and conversion for photons.
With the exception of multiple scattering, the physics configuration of the
simulation is identified in the following as ``electron-photon settings''.

The simulations are configured with three alternative electron-photon settings:
those based on the EEDL and EPDL evaluated data libraries (also
known as the ``Livermore'' library), those reengineered from the Penelope Monte Carlo
code and those implemented in the Geant4 ``standard'' electromagnetic package.
%The electron-photon processes activated in the simulation are listed in 
%Table \ref{tab_ephoton}; 
$\gamma$ conversion is not relevant at the energies considered in this study.
Functionality for the simulation of Rayleigh scattering is available in Geant4,
based on EPDL and reengineered from Penelope, for all versions considered in
this paper, and is activated in the simulations produced for this paper.
Functionality for the simulation of Rayleigh scattering, implemented in the
\textit{G4XrayRayleighModel} class, was first introduced in Geant4
``standard'' electromagnetic package in the 9.5 version; nevertheless, due to 
its questionable physical behavior observed in \cite{tns_rayleigh}, it is not
included in the simulation configurations examined in this paper.
Atomic deexcitation \cite{tns_relax} following the creation of a vacancy in atomic shell occupation 
is activated in the physics processes and models to which it is pertinent.
Since it was verified that the compatibility of the simulated energy deposition 
with experimental data is insensitive to the generation of Auger electrons in all
the test cases of \cite{sandia79,sandia80}, this time-consuming
atomic deexcitation component was not enabled
to reduce the computational resources required for the production of simulated data.

%% Table generated by Excel2LaTeX from sheet 'p-value R'
%\begin{table}
%  \centering
%  \caption{Processes (excluding multiple scattering) simulated in the three electron-photon settings subject to evaluation}
%    \begin{tabular}{lcc}
% %   \hline
%          & \multicolumn{1}{c}{\textbf{EEDL-EPDL, Penelope}} & \textbf{Standard} \\
%%          & \multicolumn{1}{c}{\textbf{Penelope}} &  \\
%\hline
%    \multirow{4}[0]{*}{Photons} & Rayleigh scattering & - \\
%          & \multicolumn{2}{c}{Compton scattering} \\
%          & \multicolumn{2}{c}{Photoelectric effect } \\
%          & \multicolumn{2}{c}{($\gamma$ conversion)} \\
%\hline
%    \multicolumn{1}{l}{\multirow{2}[0]{*}{Electrons}} & \multicolumn{2}{c}{Ionization} \\
%    \multicolumn{1}{l}{} & \multicolumn{2}{c}{Bremsstrahlung} \\
%    \hline
%    \end{tabular}%
%  \label{tab_ephoton}%
%\end{table}%

The Penelope-like models reengineered in Geant4 versions from 9.1 to 9.4
reproduce those implemented in Penelope 2001 version \cite{penelope2001}, while
in Geant4 9.5 they have been updated to those implemented in Penelope 2008
\cite{penelope2008}: the results reported in section~\ref{sec_results79} as
``Penelope'' correspond accordingly to the reengineered version of the original code.
Models based on both Penelope 2008 and 2001 are available in Geant4 9.5: the
validation results reported in section~\ref{sec_results79} as ``Penelope''
correspond to the activation of models reengineered from Penelope 2008, while
results based on reengineered Penelope 2001 models are explicitly indicated as
``Penelope 2001''.

Unless otherwise specified, the simulation results reported in this paper
are produced with the default multiple scattering configuration corresponding to
each Geant4 version. 
The Urban multiple scattering model \cite{urban2002,urban,urban2006}, based on
Lewis' theory \cite{lewis}, is instantiated by default in the electron multiple
scattering process; several variants of this model have been implemented
in Geant4.
The main features of the default multiple scattering configuration 
associated with the Geant4 versions examined in this paper are
summarized in Table~\ref{tab_mscattdef}.
Information about the parameters listed in Table~\ref{tab_mscattdef}
is available in \cite{elles}.

%: the
%algorithm was originally implemented in the \textit{G4MultipleScattering} class
%and later evolutions \textit{G4MultipleScattering52},
%\textit{G4MultipleScattering71}, then in \textit{G4MscModel},
%\textit{G4MscModel71}, \textit{G4UrbanMscModel}, \textit{G4UrbanMscModel90},
%\textit{G4UrbanMscModel2}, \textit{G4UrbanMscModel92},
%\textit{G4UrbanMscModel93}, \textit{G4UrbanMscModel95} and
%\textit{G4UrbanMscModel96}.

Validation results are also reported with an implementation of the
Goudsmit-Saunderson multiple scattering algorithm \cite{kadri_goudsmit}, which
was first released in Geant4 version 9.3.

%The simulation are produced with default values of the empirical parameters of
%Geant4 multiple scattering algorithm listed in Table~\ref{tab_mscattdef}, unless
%differently specified.

% Table generated by Excel2LaTeX from sheet 'TNS_conf79'
\begin{table}
  \centering
  \caption{Default multiple scattering settings}
    \begin{tabular}{clccc}
   \hline
    \textbf{Geant4} & \textbf{Multiple Scattering} & \textbf{Range} &  & \textbf{Geom} \\
   \textbf{Version} &\textbf{Model Class} & \textbf{Factor} & \textbf{skin} & \textbf{Factor} \\
    \hline
    \textbf{9.1} & G4UrbanMscModel 		& 0.02   & 0     	& 2.5 \\
    \textbf{9.2} & G4UrbanMscModel2 	& 0.02  & 3     	& 2.5 \\
    \textbf{9.3} & G4UrbanMscModel92 	& 0.04  & 3     	& 2.5 \\
    \textbf{9.4} & G4UrbanMscModel93 	& 0.04   & 1     	& 2.5 \\
    \textbf{9.5} & G4UrbanMscModel95	& 0.04   & 1     	& 2.5 \\
    \textbf{9.6} & G4UrbanMscModel95 	& 0.04   & 1     	& 2.5 \\
    \hline
    \end{tabular}%
  \label{tab_mscattdef}%
\end{table}%

%% Table generated by Excel2LaTeX from sheet 'TNS_conf79'
%\begin{table}
%  \centering
%  \caption{Default values of multiple scattering parameters}
%    \begin{tabular}{cccccc}
%    \hline
%    \textbf{Geant4} & \textbf{Range} & \textbf{Step} & \textbf{Lateral} & \textbf{skin} & \textbf{geom } \\
%   \textbf{Version} & \textbf{Factor} & \textbf{Limitation} & \textbf{ Displacement} & & \textbf{Factor} \\
%    \hline
%%    \textbf{8.1.p02} & 0.02  &       &       &       & 3.5 \\
%    \textbf{9.1.p03} & 0.02  & fUseSafety & true & 0     & 2.5 \\
%    \textbf{9.2.p04} & 0.02  & fUseSafety & true & 3     & 2.5 \\
%    \textbf{9.3.p02} & 0.04  & fUseSafety & true & 3     & 2.5 \\
%    \textbf{9.4.p04} & 0.04  & fUseSafety & true & 1     & 2.5 \\
%    \textbf{9.5.p01} & 0.04  & fUseSafety & true & 1     & 2.5 \\
%    \textbf{9.6} & 0.04  & fUseSafety & true & 1     & 2.5 \\
%    \hline
%    \end{tabular}%
%  \label{tab_mscattpar}%
%\end{table}%

Production thresholds are defined for each target material to enable the
production of secondary particles with energy above 250~eV in simulations
involving models based on the EEDL and EPDL data libraries and originating from
Penelope, and above 1~keV in simulations activating models in Geant4 standard
electromagnetic package.

The maximum step limit is set to 1~$\mu$m based on the optimization
described in \cite{kadri}.
This setting has been further verified in the context of the validation process.

The energy deposited by primary and secondary particles is scored in the
sensitive volume corresponding to the calorimeter in the experimental set-up.

The size of the simulated event samples is subject to the requirement that 
the statistical uncertainties of the simulated data are negligible with respect
to the experimental uncertainties.
Unless differently specified, the simulated data sample is based on one million
primary electrons generated in each test configuration.
The statistical uncertainties of the simulated data are taken into account in the
calculation of the test statistic of goodness of fit tests.

% -----------------------------------------------------------------------------------------
\section{Data analysis methods}
\label{sec_analysis}

The data analysis addresses various issues related to the validation of
Geant4-based simulation of the energy deposited by electrons: the evaluation of the 
capability of Geant4 to produce results consistent with measurements in the various
experimental configurations, the comparison of the simulation accuracy
achievable with different Geant4 physics configurations in the user application,
and the evolution of compatibility with experiment when the same test case is
simulated with the same nominal physics configuration, but with different Geant4
versions.

The statistical analysis takes into account the relationship between the samples of simulated data
 that are subject to comparison, as determined by the simulation
configuration.
Simulations using different sets of electron-photon models produce unrelated
samples, as the three electromagnetic options available in Geant4
implement distinct conceptual alternatives in the treatment of particle interactions with
matter, while samples deriving from simulations that differ only for a secondary
option (e.g. a feature in the calculation of multiple scattering), or that share
identical physics configurations, are to some extent related.

%The outcome of simulations resulting from the use of different full sets of
%electron-photon models can be considered unrelated data samples, as indeed the
%three electromagnetic options available in the Geant4 toolkit are intended to
%implement conceptual alternatives in the treatment of the interactions of
%electrons and photons with matter,

The approach adopted in the statistical analysis takes into account the physics
configuration that characterizes how the data samples were produced; appropriate
statistical methods are applied to each analysis scenario to quantify the
compatibility of simulation with experimental data.

%Categorical statistical tests determine whether the differences in compatibility
%with experiment observed across the various categories can be explained only by
%chance, or should be interpreted as deriving from intrinsic behavioral characteristics.

The significance level for the rejection of the null hypothesis is set at 0.01
for all tests, unless otherwise specified.

The statistical data analysis reported in the following sections exploits the 
Statistical Toolkit \cite{gof1,gof2}  and R \cite{R}.

%Geant4 physics settings, e.g. of simulations based on two different sets of
%Geant4 electron-photon models, 

Categorical statistical tests determine whether the differences in compatibility
with experiment observed across the various categories can be explained only by
chance, or should be interpreted as deriving from intrinsic behavioral characteristics.
The approach adopted in the categorical analysis is related to the physics
configuration that characterizes how the data samples were produced: 
statistical tests pertinent to independent or to related data samples
are applied as appropriate.

%simulations
%resulting from the use of different full sets of electron-photon models are
%considered independent data samples, as indeed the three electromagnetic options
%available in the Geant4 toolkit are intended to implement conceptual
%alternatives in the treatment of the interactions of electrons and photons with
%matter, while simulations differing only for a secondary feature (e.g. a model
%or a parameter in the multiple scattering process), or resulting from identical
%physics configurations in different Geant4 versions, are treated as matched
%pairs, and subject to statistical tests pertinent to related data samples.

% -----------------------------------------------------------------------------------------
\subsection{Evaluation of individual test cases}
\label{sec_chi2}
% Table generated by Excel2LaTeX from sheet 'TNS_conf79'
\begin{table}
  \centering
  \caption{Configuration of contingency tables for the evaluation of different Geant4 physics settings}
    \begin{tabular}{l|ll}
    \hline
          & \textbf{Model A} & \textbf{Model B} \\
    \hline
    \textbf{Pass} & N$_{pass A}$ & N$_{pass B}$ \\
    \textbf{Fail} & N$_{fail A}$ & N$_{fail B}$ \\
    \hline
    \end{tabular}%
  \label{tab_exconting}%
\end{table}%

% Table generated by Excel2LaTeX from sheet 'TNS_conf79'
\begin{table}
  \centering
  \caption{Configuration of contingency tables for the evaluation of different Geant4 versions}
    \begin{tabular}{l|ll}
    \hline
          & \textbf{Version A Pass} & \textbf{Version A Fail} \\
    \hline
    \textbf{Version B Pass} & N$_{pass\,A, \, pass\,B}$ & N$_{fail\,A, \, pass\,B}$ \\
    \textbf{Version B Fail} & N$_{pass\,A, \, fail\,B}$ & N$_{fail\,A, \, fail\,B}$ \\
    \hline
    \end{tabular}%
  \label{tab_exmcnemar}%
\end{table}%

The evaluation of the simulation accuracy in each experimental configuration is
based on the $\chi^2$ test \cite{bock}.
This goodness-of-fit test takes experimental uncertainties into account explicitly in the
calculation of the test statistic.
The null hypothesis for this test is defined as 
the equivalence of  the simulated and experimental distributions.
%to derive from the same parent distribution.
The outcome of this test is classified as ``fail'', if the null hypothesis is rejected, as
``pass'' otherwise.

This stage of the validation analysis closely follows the procedure described in
\cite{tns_sandia} and reuses previous assessments, such as the treatment of
outliers in the experimental data and of the tails of the longitudinal energy
deposition distribution in the calculation of the $\chi^2$ test statistic.

% -----------------------------------------------------------------------------------------
\subsection{Evaluation of unrelated data samples}
\label{sec_conting}

This component of the analysis evaluates the difference in compatibility with
experiment across independent samples.
This is the case, for instance, with simulations based on entirely different modeling 
approaches to describe the interactions of electrons and photons with matter
(e.g. based on the interpolation of data libraries or implementing analytical models).

The differences in the behavior of the two categories are quantified by means of
contingency tables derived from the results of the $\chi^2$ test described in
section \ref{sec_chi2}.
Contingency tables for this evaluation are built by counting the number of test
cases where the null hypothesis is rejected or not rejected by the $\chi^2$
test at the defined significance level for each category of data.
An example of their configuration is shown in Table~\ref{tab_exconting}.

%The second stage of the statistical analysis quantifies the differences of the
%simulation models in compatibility with experiment.
%It consists of a categorical analysis based on contingency tables, which derive
%from the results of the $\chi^2$ test: the outcome of this test is classified as
%``fail'' or ``pass'', according respectively to whether the hypothesis of
%compatibility of experimental and calculated data is rejected or not.
%The categorical analysis takes as a reference the simulation model exhibiting
%the largest efficiency at reproducing experimental data; the other models are
%compared to it, to determine whether they exhibit statistically significant
%differences of compatibility with measurements.

In the analysis of contingency tables the null hypothesis is that there is no
relationship between the two categories; in physical terms it means that the two
categories under examination (e.g. two sets of Geant4 electron-photon models)
are equivalent regarding the compatibility with experiment of their respective simulation outcome.
%The rejection of the null hypothesis would correspond to observing a
%statistically significant difference between the two categories in 
%consistency with experimental measurements: for instance, a significantly
%different behaviour of simulations based on two sets of Geant4 electromagnetic
%processes in ``passing'' or ``failing'' the $\chi^2$ test over the whole set of
%experimental configurations.

A variety of tests is applied to determine the statistical significance of the
difference between the two categories of data subject to evaluation: Pearson's
$\chi^2$ test \cite{pearson} (when the number of entries in each cell of the
table is greater than 5), Fisher's exact test \cite{fisher}, Barnard's test
\cite{barnard}, Boschloo's test \cite{boschloo} and Suissa and Schuster's 
\cite{suissa} calculation of a Z-pooled statistic.
The use of several tests mitigates the risk of introducing systematic effects
in the validation results due to peculiarities in the mathematical formulation
of the test statistic.

Fisher's test is widely used in the analysis of contingency tables.
It is based on a model in which both the row and column sums are
fixed in advance, which seldom occurs in experimental practice; it remains valid
for the cases in which the row or column totals, or both, are not fixed,
but in these cases it tends to be conservative, yielding a larger p-value than
the true significance of the test \cite{agresti}.
Barnard's test is deemed more powerful than Fisher's exact test in some
configurations of 2$\times$2 contingency tables \cite{andres_1994,andres_2004},
although it is computationally more intensive.
Boschloo's test and the Z-pooled statistic are also considered more powerful
than Fisher's exact test \cite{boschloo, suissa}.

% is independent of the two categories .

%In a two-sided test the alternative hypothesis is formulated as the two 
%categories subject to test being different regarding their compatibility between 
%simulated and measured data.
%The rejection of the null hypothesis in favour of the alternative hypothesis
%would mean that the compatibility with experiment is significantly different for
%the two simulation configurations subject to evaluation, without
%distinguishing whether either configuration would be responsible for a greater
%proportion of test cases compatible with experiment.
%
%In a one-sided test the alternative hypothesis is formulated as one of the
%categories subject to test exhibiting better compatibility with experimental data.
%A one-tailed test evaluates the significance of differences in a
%specific direction: the rejection of the null hypothesis in favour of the
%alternative hypothesis would distinguish differences as associated with 
%either physics configuration producing more accurate simulation results.

% -----------------------------------------------------------------------------------------

\begin{figure*} 
\centerline{\includegraphics[angle=0,width=18cm]{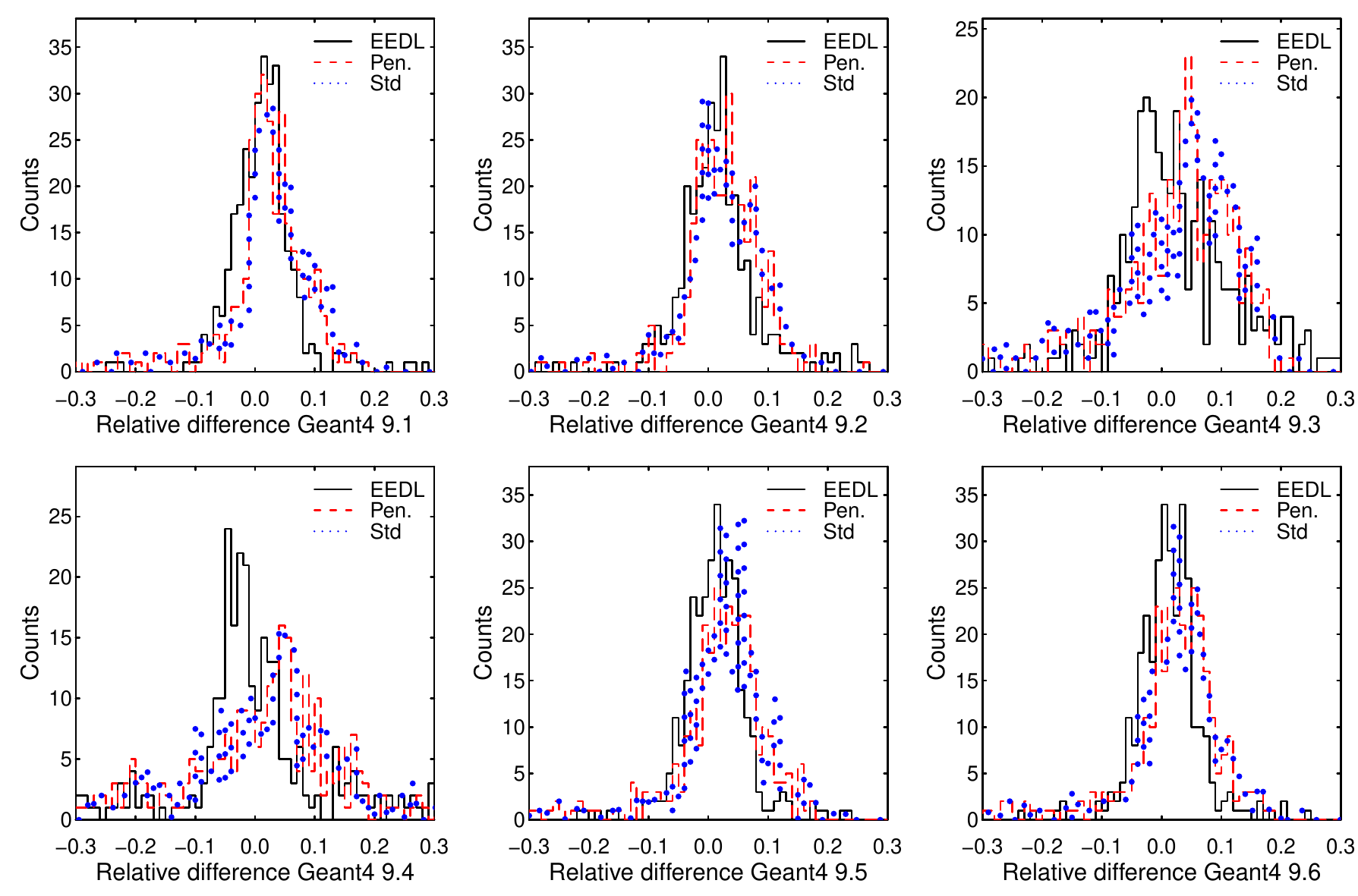}}
\caption{Relative difference between simulated and experimental longitudinal
profiles of deposited energy produced with Geant4 versions 9.1 to 9.6; the
simulations use one of three sets of electron-photon models: based on EEDL and
EPDL evaluated data libraries (solid black line), reengineered from Penelope(dashed red histogram identified as ``Pen.'') or included in the
``standard'' electromagnetic package (dotted blue histogram identified as ``Std'').
The relative difference at a given depth is defined as $(E_{simulated}-E_{experimental})/E_{experimental}$, where $E$ is
the deposited energy at the considered depth. 
Color codes are reported in the caption of this figure and the following to
facilitate the appraisal for readers having access to a color version of the
paper; different line types allow distinguishing multiple histograms in 
a black and white version of the paper.}
\label{fig_diff_exp79}
\end{figure*}

% -----------------------------------------------------------------------------------------
\subsection{Evaluation of related data samples}
\label{sec_mcnemar}

This analysis evaluates the difference in compatibility with
experiment of dependent data samples across data categories.
In this type of analysis each subject (e.g. a physics configuration in the user
application) serves in both situations being evaluated (e.g. two Geant4
versions): if the result of the test is significant (i.e. the null hypothesis is
rejected), one can conclude that there is a high likelihood that the two
situations represent populations with different behaviors.
This type of analysis also applies when one examines two closely related subjects:
for instance, when one wants to estimate the effect of a secondary feature (e.g.
a multiple scattering parameter), while the main electron-photon physics
settings that characterize the simulation are common to both configurations
being evaluated.

%subject is matched to another
%closely related subject across two situations being evaluated: for instance, when one
%wants to estimate the effect of a secondary feature (e.g. a multiple scattering
%parameter), without changing the main electron-photon physics settings that
%characterize the simulation.

%It concerns the analysis of the evolution of the compatibility with experiment 
%over different Geant4 version, where the outcome of each test case simulated with 
%the same user settings is matched across two versions subject to evaluation.
%It is also applied to comparisons concerning the same Geant4 version and main
%simulation settings, where a secondary feature (e.g. a multiple scattering parameter)
%is modified, without changing the characterizing electron-photon physics configuration.

Appropriate 2$\times$2 contingency tables are built for this purpose, based on the
results of the $\chi^2$ test of section \ref{sec_chi2}:
%These tables are different from those described in section \ref{sec_conting}:
they report on one diagonal the number of test cases where both categories (e.g.
Geant4 simulation configurations) subject to evaluation ``pass'' or ``fail'' the
$\chi^2$ test of section \ref{sec_chi2}, and on the other diagonal the number of
test cases where one category ``passes'' the $\chi^2$ test, while the other
one``fails''.
An example of their configuration is shown in Table~\ref{tab_exmcnemar}.

McNemar's test \cite{mcnemar} is applied to the analysis of related data samples.
This test focuses on the significance of the discordant results, i.e. the number of test cases where one category ``passes'' 
the $\chi^2$ test and the other one ``fails''.
The null hypothesis for McNemar's test is that the proportions of 
discordant results is the same in the two cells corresponding to 
``pass-fail'' or ``fail-pass'' associated with the two categories subject to test.

%test cases
%concerning simulated and measured energy deposition are the same 
%across the two categories subject to test, e.g.
%before and after the evolution of Geant4 electromagnetic physics over two versions.

The calculation of McNemar's test is performed using either the $\chi^2$
asymptotic distribution or the binomial distribution \cite{bennett}:
the former calculation method is usually identified simply as ``McNemar test'',
while the latter is known as ``McNemar exact test''.
Yates' \cite{yates} continuity correction may be applied to the calculation of the
$\chi^2$ statistic to account for cells with a small number of entries.
According to \cite{lui_2001}, McNemar test uncorrected for continuity is more
powerful than the exact test, and performs well even when the number of
discordant pairs is as low as 6, while both the exact test and the corrected
McNemar test are conservative.

\tabcolsep=6pt
\begin{table*}
  \centering
  \caption{P-values of the $\chi^2$ tests for longitudinal energy deposition: simulations with electron-photon models based on EEDL-EPDL}
    \begin{tabular}{lrcc|cccccc}
    \hline
    \textbf{Target} & \textbf{Z} & \textbf{E } & \textbf{angle}   & \multicolumn{6}{c}{\textbf{Geant4 version}} \\
    \textbf{} &       & (kev) & (degrees) & \textbf{9.1} & \textbf{9.2} & \textbf{9.3} & \textbf{9.4} & \textbf{9.5} & \textbf{9.6} \\
    \hline
    Be    & 4     & 58    & 0     & 0.071 & 0.014 & 0.124 & 0.311 & 0.149 & 0.156 \\
    Be    & 4     & 109   & 0     & 0.021 & $<0.001$ & $<0.001$ & $<0.001$ & 0.015 & 0.013 \\
    Be    & 4     & 314   & 0     & 0.015 & 0.764 & $<0.001$ & $<0.001$ & 0.013 & 0.014 \\
    Be    & 4     & 521   & 0     & 0.092 & 0.967 & $<0.001$ & $<0.001$ & 0.832 & 0.793 \\
    Be    & 4     & 1033  & 0     & $<0.001$ & $<0.001$ & $<0.001$ & $<0.001$ & $<0.001$ & $<0.001$ \\
    C     & 6     & 1000  & 0     & 0.917 & 0.994 & $<0.001$ & $<0.001$ & 0.290 & 0.346 \\
    Al    & 13    & 314   & 0     & 0.182 & $<0.001$ & $<0.001$ & $<0.001$ & 0.004 & 0.007 \\
    Al    & 13    & 521   & 0     & 0.574 & $<0.001$ & $<0.001$ & $<0.001$ & 0.091 & 0.089 \\
    Al    & 13    & 1033  & 0     & 0.484 & 0.123 & $<0.001$ & $<0.001$ & $<0.001$ & $<0.001$ \\
    Al    & 13    & 314   & 60    & 0.396 & 0.596 & $<0.001$ & $<0.001$ & 0.001 & 0.002 \\
    Al    & 13    & 521   & 60    & 0.137 & 0.011 & 0.001 & $<0.001$ & 0.056 & 0.086 \\
    Al    & 13    & 1033  & 60    & $<0.001$ & $<0.001$ & $<0.001$ & $<0.001$ & $<0.001$ & $<0.001$ \\
    Fe    & 26    & 300   & 0     & 0.832 & $<0.001$ & 0.351 & 0.741 & 0.787 & 0.742 \\
    Fe    & 26    & 500   & 0     & 0.055 & $<0.001$ & 0.314 & 0.003 & 0.814 & 0.808 \\
    Fe    & 26    & 1000  & 0     & $<0.001$ & $<0.001$ & 0.169 & 0.003 & $<0.001$ & $<0.001$ \\
    Cu    & 29    & 300   & 0     & $<0.001$ & $<0.001$ & $<0.001$ & $<0.001$ & $<0.001$ & $<0.001$ \\
    Cu    & 29    & 500   & 0     & $<0.001$ & $<0.001$ & $<0.001$ & $<0.001$ & $<0.001$ & $<0.001$ \\
    Mo    & 42    & 100   & 0     & $<0.001$ & $<0.001$ & $<0.001$ & $<0.001$ & $<0.001$ & $<0.001$ \\
    Mo    & 42    & 300   & 0     & 0.062 & $<0.001$ & 0.001 & $<0.001$ & 0.008 & 0.002 \\
    Mo    & 42    & 500   & 0     & 0.020 & $<0.001$ & $<0.001$ & 0.001 & 0.128 & 0.115 \\
    Mo    & 42    & 1000  & 0     & $<0.001$ & $<0.001$ & $<0.001$ & $<0.001$ & $<0.001$ & $<0.001$ \\
    Mo    & 42    & 300   & 60    & 0.023 & 0.002 & 0.049 & 0.043 & 0.029 & 0.022 \\
    Mo    & 42    & 500   & 60    & 0.022 & $<0.001$ & 0.011 & 0.006 & 0.003 & 0.007 \\
    Mo    & 42    & 1000  & 60    & 0.037 & $<0.001$ & 0.010 & 0.028 & 0.001 & 0.002 \\
    Ta    & 73    & 300   & 0     & 0.043 & 0.511 & 0.242 & 0.272 & 0.364 & 0.294 \\
    Ta    & 73    & 500   & 0     & 0.025 & 0.003 & $<0.001$ & $<0.001$ & 0.012 & 0.019 \\
    Ta    & 73    & 1000  & 0     & 0.030 & $<0.001$ & $<0.001$ & $<0.001$ & 0.002 & 0.001 \\
    Ta    & 73    & 500   & 60    & 0.011 & 0.003 & 0.040 & 0.042 & 0.010 & 0.007 \\
    Ta    & 73    & 1000  & 60    & $<0.001$ & $<0.001$ & $<0.001$ & $<0.001$ & $<0.001$ & $<0.001$ \\
    Ta    & 73    & 500   & 30    & 0.034 & 0.005 & 0.004 & 0.006 & 0.020 & 0.017 \\
%    U     & 92    & 300   & 0     & $<0.001$ & $<0.001$ & $<0.001$ & $<0.001$ & $<0.001$ & $<0.001$ \\
%    U     & 92    & 500   & 0     & $<0.001$ & $<0.001$ & $<0.001$ & $<0.001$ & $<0.001$ & $<0.001$ \\
%    U     & 92    & 1000  & 0     & $<0.001$ & $<0.001$ & $<0.001$ & $<0.001$ & $<0.001$ & $<0.001$ \\
%    U     & 92    & 1000  & 60    & $<0.001$ & $<0.001$ & $<0.001$ & $<0.001$ & $<0.001$ & $<0.001$ \\
    \hline
    \end{tabular}%
  \label{tab_pvalue79_liv}%
\end{table*}%
\tabcolsep=6pt

% Table generated by Excel2LaTeX from sheet 'p-value R'
\begin{table*}
  \centering
  \caption{P-values of the $\chi^2$ tests for longitudinal energy deposition: simulations with Penelope-like electron-photon models }
    \begin{tabular}{lrcc|cccccc}
    \hline
    \textbf{Target} & \textbf{Z} & \textbf{E } & \textbf{angle}   & \multicolumn{6}{c}{\textbf{Geant4 version}} \\
    \textbf{} &       & (kev) & (degrees) & \textbf{9.1} & \textbf{9.2} & \textbf{9.3} & \textbf{9.4} & \textbf{9.5} & \textbf{9.6} \\
    \hline
    Be    & 4     & 58    & 0     & 0.018 & 0.002 & 0.245 & 0.123 & 0.060 & 0.069 \\
    Be    & 4     & 109   & 0     & $<0.001$ & $<0.001$ & $<0.001$ & $<0.001$ & $<0.001$ & $<0.001$ \\
    Be    & 4     & 314   & 0     & $<0.001$ & $<0.001$ & $<0.001$ & $<0.001$ & $<0.001$ & $<0.001$ \\
    Be    & 4     & 521   & 0     & 0.270 & 0.002 & $<0.001$ & $<0.001$ & 0.002 & 0.003 \\
    Be    & 4     & 1033  & 0     & 0.885 & 0.192 & $<0.001$ & $<0.001$ & 0.195 & 0.206 \\
    C     & 6     & 1000  & 0     & 0.997 & 0.448 & $<0.001$ & $<0.001$ & 0.346 & 0.407 \\
    Al    & 13    & 314   & 0     & 0.145 & $<0.001$ & $<0.001$ & $<0.001$ & 0.001 & 0.001 \\
    Al    & 13    & 521   & 0     & 0.350 & $<0.001$ & $<0.001$ & $<0.001$ & 0.004 & 0.001 \\
    Al    & 13    & 1033  & 0     & 0.174 & $<0.001$ & $<0.001$ & $<0.001$ & $<0.001$ & $<0.001$ \\
    Al    & 13    & 314   & 60    & 0.275 & 0.702 & $<0.001$ & $<0.001$ & 0.005 & 0.007 \\
    Al    & 13    & 521   & 60    & 0.006 & 0.001 & $<0.001$ & $<0.001$ & 0.001 & 0.003 \\
    Al    & 13    & 1033  & 60    & $<0.001$ & $<0.001$ & $<0.001$ & $<0.001$ & $<0.001$ & $<0.001$ \\
    Fe    & 26    & 300   & 0     & $<0.001$ & $<0.001$ & $<0.001$ & $<0.001$ & $<0.001$ & $<0.001$ \\
    Fe    & 26    & 500   & 0     & $<0.001$ & $<0.001$ & $<0.001$ & $<0.001$ & $<0.001$ & $<0.001$ \\
    Fe    & 26    & 1000  & 0     & $<0.001$ & $<0.001$ & $<0.001$ & $<0.001$ & $<0.001$ & $<0.001$ \\
    Cu    & 29    & 300   & 0     & $<0.001$ & 0.001 & $<0.001$ & $<0.001$ & $<0.001$ & $<0.001$ \\
    Cu    & 29    & 500   & 0     & $<0.001$ & $<0.001$ & $<0.001$ & $<0.001$ & $<0.001$ & $<0.001$ \\
    Mo    & 42    & 100   & 0     & $<0.001$ & $<0.001$ & $<0.001$ & $<0.001$ & $<0.001$ & $<0.001$ \\
    Mo    & 42    & 300   & 0     & $<0.001$ & $<0.001$ & $<0.001$ & $<0.001$ & $<0.001$ & $<0.001$ \\
    Mo    & 42    & 500   & 0     & $<0.001$ & $<0.001$ & $<0.001$ & $<0.001$ & $<0.001$ & $<0.001$ \\
    Mo    & 42    & 1000  & 0     & 0.001 & 0.769 & 0.044 & $<0.001$ & $<0.001$ & $<0.001$ \\
    Mo    & 42    & 300   & 60    & $<0.001$ & $<0.001$ & $<0.001$ & $<0.001$ & $<0.001$ & $<0.001$ \\
    Mo    & 42    & 500   & 60    & $<0.001$ & $<0.001$ & $<0.001$ & $<0.001$ & $<0.001$ & $<0.001$ \\
    Mo    & 42    & 1000  & 60    & $<0.001$ & $<0.001$ & $<0.001$ & $<0.001$ & $<0.001$ & $<0.001$ \\
    Ta    & 73    & 300   & 0     & $<0.001$ & $<0.001$ & $<0.001$ & $<0.001$ & $<0.001$ & $<0.001$ \\
    Ta    & 73    & 500   & 0     & $<0.001$ & $<0.001$ & $<0.001$ & $<0.001$ & $<0.001$ & $<0.001$ \\
    Ta    & 73    & 1000  & 0     & $<0.001$ & 0.001 & 0.119 & 0.137 & $<0.001$ & $<0.001$ \\
    Ta    & 73    & 500   & 60    & $<0.001$ & $<0.001$ & $<0.001$ & $<0.001$ & $<0.001$ & $<0.001$ \\
    Ta    & 73    & 1000  & 60    & 0.327 & 0.489 & 0.176 & 0.177 & 0.136 & 0.257 \\
    Ta    & 73    & 500   & 30    & $<0.001$ & 0.001 & $<0.001$ & $<0.001$ & $<0.001$ & $<0.001$ \\
%    U     & 92    & 300   & 0     & $<0.001$ & $<0.001$ & $<0.001$ & $<0.001$ & $<0.001$ & $<0.001$ \\
%    U     & 92    & 500   & 0     & $<0.001$ & $<0.001$ & $<0.001$ & $<0.001$ & $<0.001$ & $<0.001$ \\
%    U     & 92    & 1000  & 0     & $<0.001$ & $<0.001$ & $<0.001$ & $<0.001$ & $<0.001$ & $<0.001$ \\
%    U     & 92    & 1000  & 60    & $<0.001$ & $<0.001$ & $<0.001$ & $<0.001$ & $<0.001$ & $<0.001$ \\
   \hline
    \end{tabular}%
  \label{tab_pvalue79_pen}%
\end{table*}%

% Table generated by Excel2LaTeX from sheet 'p-value R'

\begin{table*}
  \centering
  \caption{P-values of the $\chi^2$ tests for longitudinal energy deposition: simulations with Standard electron-photon models }
   \begin{tabular}{lrcc|cccccc}
    \hline
    \textbf{Target} & \textbf{Z} & \textbf{E } & \textbf{angle}   & \multicolumn{6}{c}{\textbf{Geant4 version}} \\
    \textbf{} &       & (kev) & (degrees) & \textbf{9.1} & \textbf{9.2} & \textbf{9.3} & \textbf{9.4} & \textbf{9.5} & \textbf{9.6} \\
    \hline
    Be    & 4     & 58    & 0     & 0.009 & $<0.001$ & 0.374 & 0.031 & 0.009 & 0.009 \\
    Be    & 4     & 109   & 0     & $<0.001$ & $<0.001$ & $<0.001$ & $<0.001$ & $<0.001$ & $<0.001$ \\
    Be    & 4     & 314   & 0     & $<0.001$ & $<0.001$ & $<0.001$ & $<0.001$ & $<0.001$ & $<0.001$ \\
    Be    & 4     & 521   & 0     & 0.464 & $<0.001$ & $<0.001$ & $<0.001$ & $<0.001$ & 0.001 \\
    Be    & 4     & 1033  & 0     & 0.805 & 0.180 & $<0.001$ & $<0.001$ & 0.080 & 0.090 \\
    C     & 6     & 1000  & 0     & 0.014 & 0.850 & $<0.001$ & $<0.001$ & 0.869 & 0.859 \\
    Al    & 13    & 314   & 0     & 0.176 & $<0.001$ & $<0.001$ & $<0.001$ & 0.007 & 0.007 \\
    Al    & 13    & 521   & 0     & 0.124 & $<0.001$ & $<0.001$ & $<0.001$ & 0.062 & 0.030 \\
    Al    & 13    & 1033  & 0     & 0.087 & $<0.001$ & $<0.001$ & $<0.001$ & $<0.001$ & $<0.001$ \\
    Al    & 13    & 314   & 60    & 0.286 & 0.921 & $<0.001$ & $<0.001$ & 0.015 & 0.016 \\
    Al    & 13    & 521   & 60    & $<0.001$ & $<0.001$ & $<0.001$ & $<0.001$ & 0.001 & 0.003 \\
    Al    & 13    & 1033  & 60    & $<0.001$ & $<0.001$ & $<0.001$ & $<0.001$ & $<0.001$ & $<0.001$ \\
    Fe    & 26    & 300   & 0     & $<0.001$ & $<0.001$ & $<0.001$ & $<0.001$ & $<0.001$ & $<0.001$ \\
    Fe    & 26    & 500   & 0     & $<0.001$ & $<0.001$ & $<0.001$ & $<0.001$ & $<0.001$ & $<0.001$ \\
    Fe    & 26    & 1000  & 0     & $<0.001$ & $<0.001$ & $<0.001$ & $<0.001$ & $<0.001$ & $<0.001$ \\
    Cu    & 29    & 300   & 0     & $<0.001$ & 0.010 & $<0.001$ & $<0.001$ & $<0.001$ & $<0.001$ \\
    Cu    & 29    & 500   & 0     & $<0.001$ & $<0.001$ & $<0.001$ & $<0.001$ & $<0.001$ & $<0.001$ \\
    Mo    & 42    & 100   & 0     & $<0.001$ & $<0.001$ & $<0.001$ & $<0.001$ & $<0.001$ & $<0.001$ \\
    Mo    & 42    & 300   & 0     & $<0.001$ & $<0.001$ & $<0.001$ & $<0.001$ & $<0.001$ & $<0.001$ \\
    Mo    & 42    & 500   & 0     & $<0.001$ & $<0.001$ & $<0.001$ & $<0.001$ & $<0.001$ & $<0.001$ \\
    Mo    & 42    & 1000  & 0     & $<0.001$ & 0.181 & $<0.001$ & $<0.001$ & $<0.001$ & $<0.001$ \\
    Mo    & 42    & 300   & 60    & $<0.001$ & $<0.001$ & $<0.001$ & $<0.001$ & $<0.001$ & $<0.001$ \\
    Mo    & 42    & 500   & 60    & $<0.001$ & $<0.001$ & $<0.001$ & $<0.001$ & $<0.001$ & $<0.001$ \\
    Mo    & 42    & 1000  & 60    & $<0.001$ & $<0.001$ & $<0.001$ & $<0.001$ & $<0.001$ & $<0.001$ \\
    Ta    & 73    & 300   & 0     & $<0.001$ & $<0.001$ & $<0.001$ & $<0.001$ & $<0.001$ & $<0.001$ \\
    Ta    & 73    & 500   & 0     & $<0.001$ & $<0.001$ & $<0.001$ & $<0.001$ & $<0.001$ & $<0.001$ \\
    Ta    & 73    & 1000  & 0     & $<0.001$ & $<0.001$ & $<0.001$ & 0.006 & $<0.001$ & $<0.001$ \\
    Ta    & 73    & 500   & 60    & $<0.001$ & $<0.001$ & $<0.001$ & $<0.001$ & $<0.001$ & $<0.001$ \\
    Ta    & 73    & 1000  & 60    & 0.001 & 0.016 & $<0.001$ & 0.007 & 0.151 & 0.139 \\
    Ta    & 73    & 500   & 30    & $<0.001$ & $<0.001$ & $<0.001$ & $<0.001$ & $<0.001$ & $<0.001$ \\
%    U     & 92    & 300   & 0     & $<0.001$ & $<0.001$ & $<0.001$ & $<0.001$ & $<0.001$ & $<0.001$ \\
%    U     & 92    & 500   & 0     & $<0.001$ & $<0.001$ & $<0.001$ & $<0.001$ & $<0.001$ & $<0.001$ \\
%    U     & 92    & 1000  & 0     & $<0.001$ & $<0.001$ & $<0.001$ & $<0.001$ & $<0.001$ & $<0.001$ \\
%    U     & 92    & 1000  & 60    & $<0.001$ & $<0.001$ & $<0.001$ & $<0.001$ & $<0.001$ & $<0.001$ \\
    \hline
    \end{tabular}%
  \label{tab_pvalue79_std}%
\end{table*}%
\tabcolsep=6pt

% -----------------------------------------------------------------------------------------
\section{Results: energy deposition profile}
\label{sec_results79}

This part of the validation process concerns the comparisons of Geant4-based
simulations with the measurements of deposited energy as a function of
penetration depth reported in \cite{sandia79}.
Various issues are investigated:
\begin{itemize}
\item the effect of different Geant4 electron-photon models on simulation accuracy, 
\item the effect of the improvements to Geant4 electromagnetic physics described
in \cite{em_mc2010,em_chep2010,em_radecs2011,em_chep2012} on the accuracy of the
simulation of the energy deposition profile,
\item the effect of different modeling options and empirical parameters in
the implementation of multiple scattering on simulation accuracy.
\end{itemize}

%The results of the analysis are reported in the following sections for each of
%the topics addressed by the validation process.

% -----------------------------------------------------------------------------------------
\subsection{General features}
\label{sec_sandia79_gen}

The relative difference between simulated and experimental data of \cite{sandia79} for 
each electron-photon model and Geant4 version is shown in Fig.~\ref{fig_diff_exp79}.
The relative difference at a given depth is defined as $(E_{simulated}-E_{experimental})/E_{experimental}$, where $E$ is
the deposited energy at the considered depth.
These distributions encompass all experimental test cases except those involving
uranium targets.

\begin{figure}
\centerline{\includegraphics[angle=0,width=9cm]{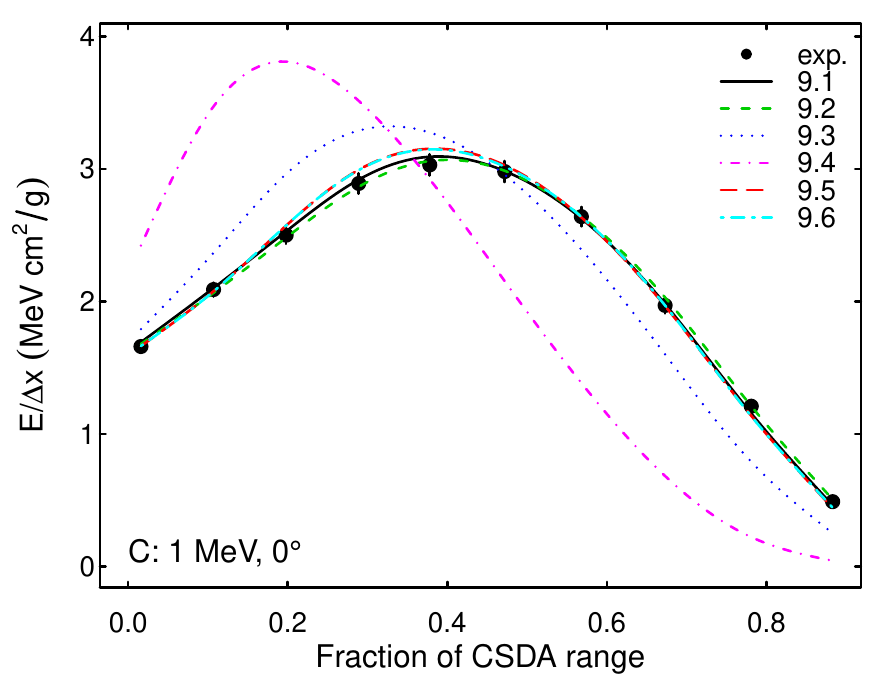}}
\caption{Longitudinal energy deposition in carbon: experimental data from \cite{sandia79} (black dots) and
simulations with Geant4 versions 9.1 to 9.6, using models based on EEDL and EPDL
evaluated data libraries.
The error bars of the experimental data points are not visible, when they are
smaller than the symbol size.}
\label{fig_C79}
\end{figure}

\begin{figure*}
\centerline{\includegraphics[angle=0,width=18cm]{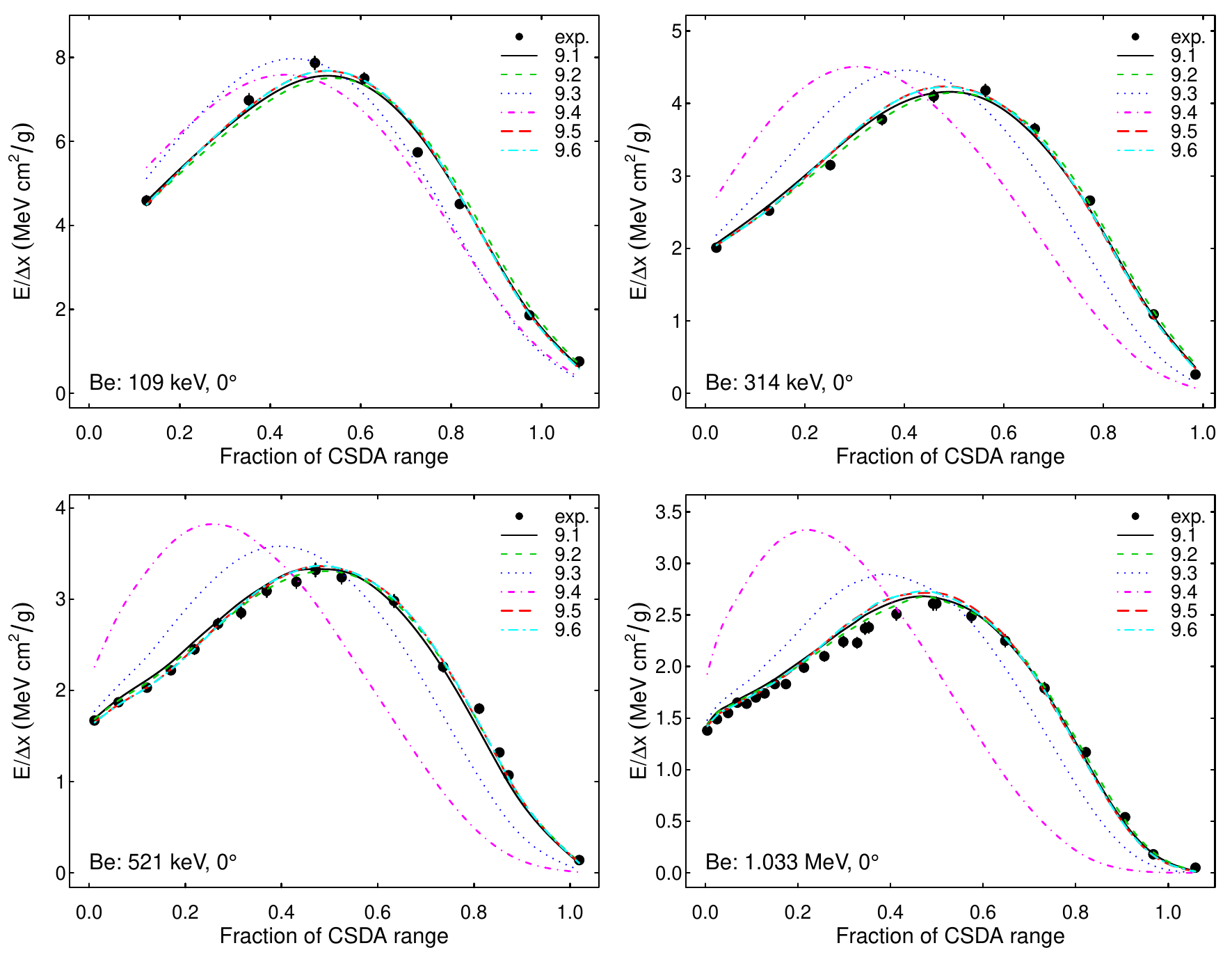}}
\caption{Longitudinal energy deposition in beryllium: experimental
data from \cite{sandia79} (black dots) and simulations with Geant4 versions 9.1
to 9.6, using models based on EEDL and EPDL evaluated data libraries. 
The error bars of the experimental data points are not visible, when they are
smaller than the symbol size.
The data at 58 keV are not shown due to their limited visual interest, since 
they encompass only two points.}
\label{fig_Be79}
\end{figure*}

\begin{figure*}
\centerline{\includegraphics[angle=0,width=18cm]{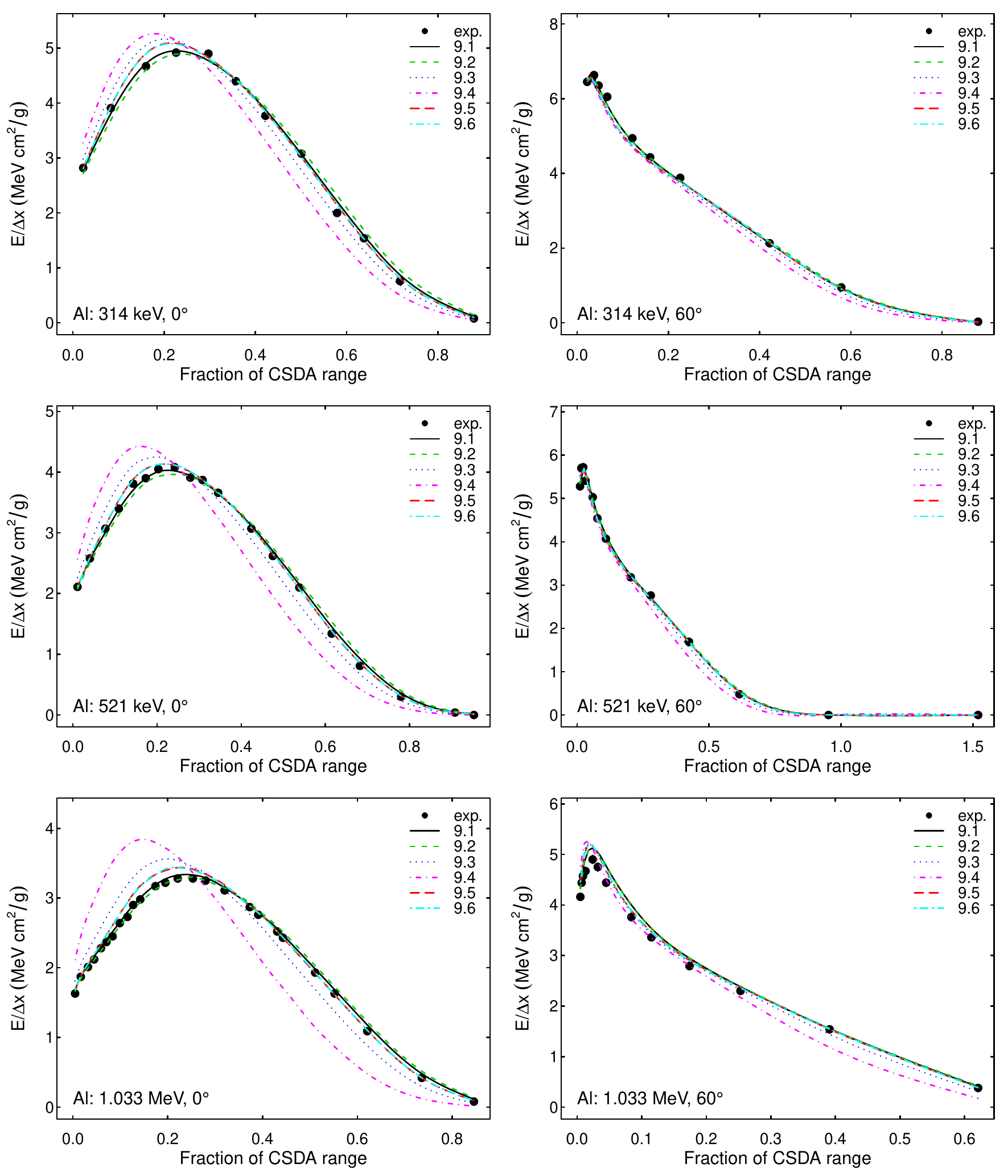}}
\caption{Longitudinal energy deposition in aluminium: experimental data from
\cite{sandia79} (black dots) and simulations with Geant4 versions 9.1 to 9.6,
using models based on EEDL and EPDL evaluated data libraries.
The error bars of the experimental data points are not visible, when they are
smaller than the symbol size.}
\label{fig_Al79}
\end{figure*}

\begin{figure}[htbp]
\centerline{\includegraphics[angle=0,width=9cm]{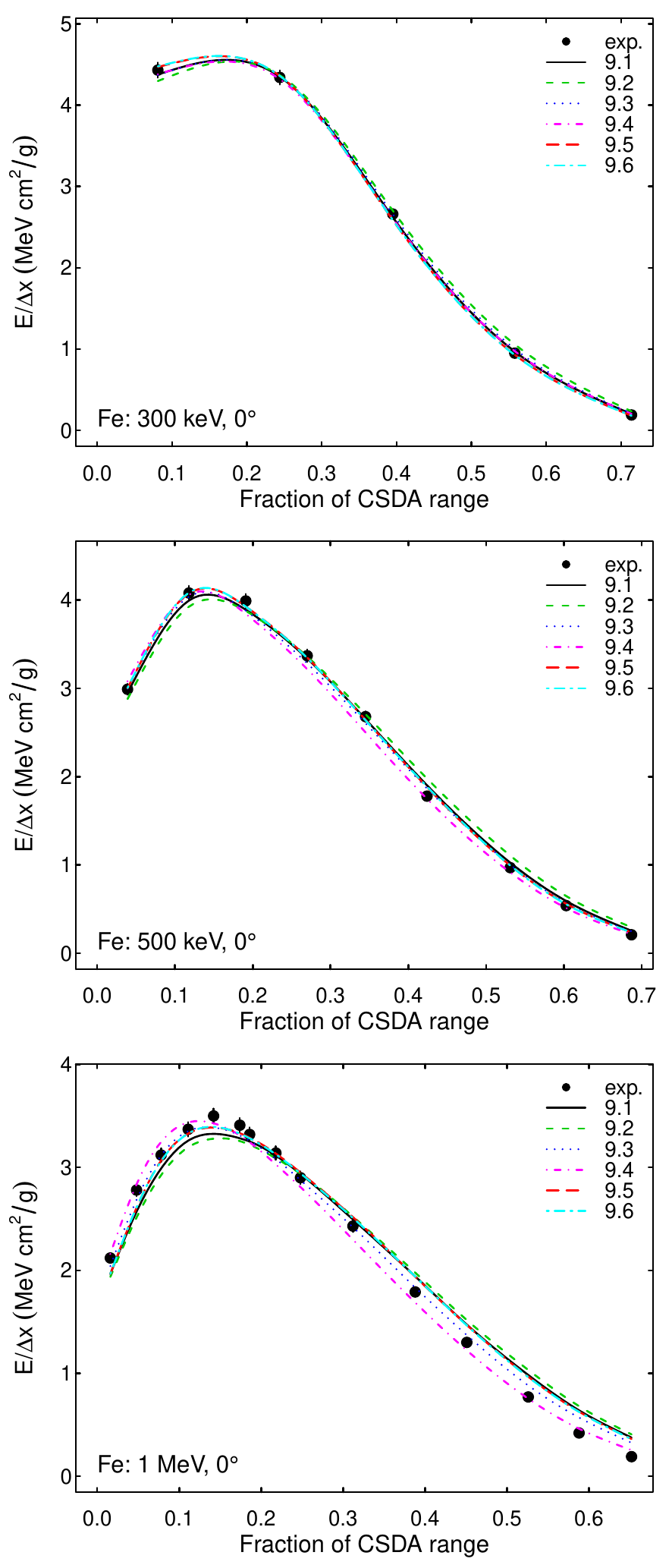}}
%\caption{Longitudinal energy deposition in iron (for orthogonal incidence) 
%and in molybdenum (with 60$^\circ$ incidence of the electorn beam): experimental data from
%\cite{sandia79} (black dots) and simulations with Geant4 versions 9.1 to 9.6,
%using models based on EEDL and EPDL evaluated data libraries.} 
\caption{Longitudinal energy deposition in iron produced by an orthogonally
incident electron beam: experimental data from \cite{sandia79} (black dots) and
simulations with Geant4 versions 9.1 to 9.6, using models based on EEDL and EPDL
evaluated data libraries.
The error bars of the experimental data points are not visible, when they are
smaller than the symbol size.}
\label{fig_Fe79}
\end{figure}

\begin{figure}[htbp]
\centerline{\includegraphics[angle=0,width=9cm]{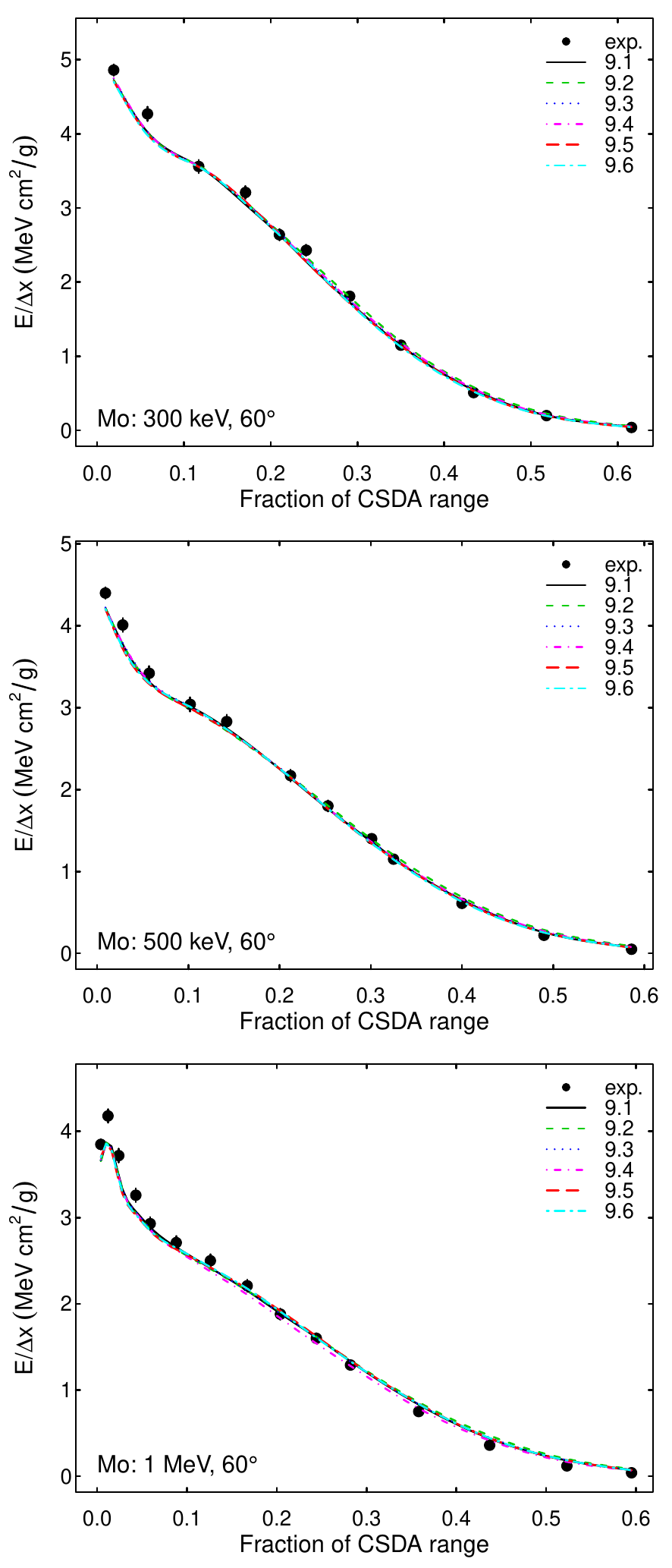}}
\caption{Longitudinal energy deposition in molybdenum produced by an electron beam
with 60$^{\circ}$ angle of incidence: experimental data from \cite{sandia79} (black dots) and
simulations with Geant4 versions 9.1 to 9.6, using models based on EEDL and EPDL
evaluated data libraries.
The error bars of the experimental data points are not visible, when they are
smaller than the symbol size.}
\label{fig_Mo6079}
\end{figure}

\begin{figure*}
\centerline{\includegraphics[angle=0,width=18cm]{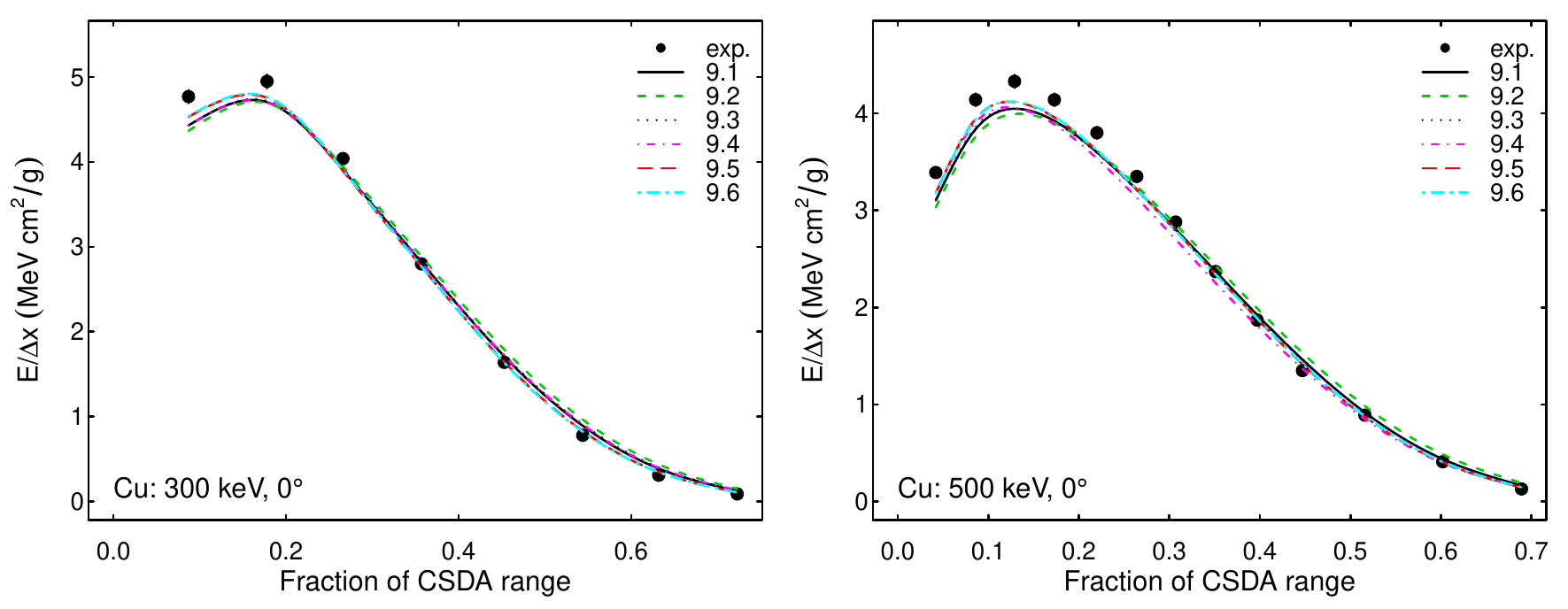}}
\caption{Longitudinal energy deposition in copper: experimental data from
\cite{sandia79} (black dots) and simulations with Geant4 versions 9.1 to 9.6,
using models based on EEDL and EPDL evaluated data libraries.
The error bars of the experimental data points are not visible, when they are
smaller than the symbol size.} 
\label{fig_Cu79}
\end{figure*}

\begin{figure*} 
\centerline{\includegraphics[angle=0,width=18cm]{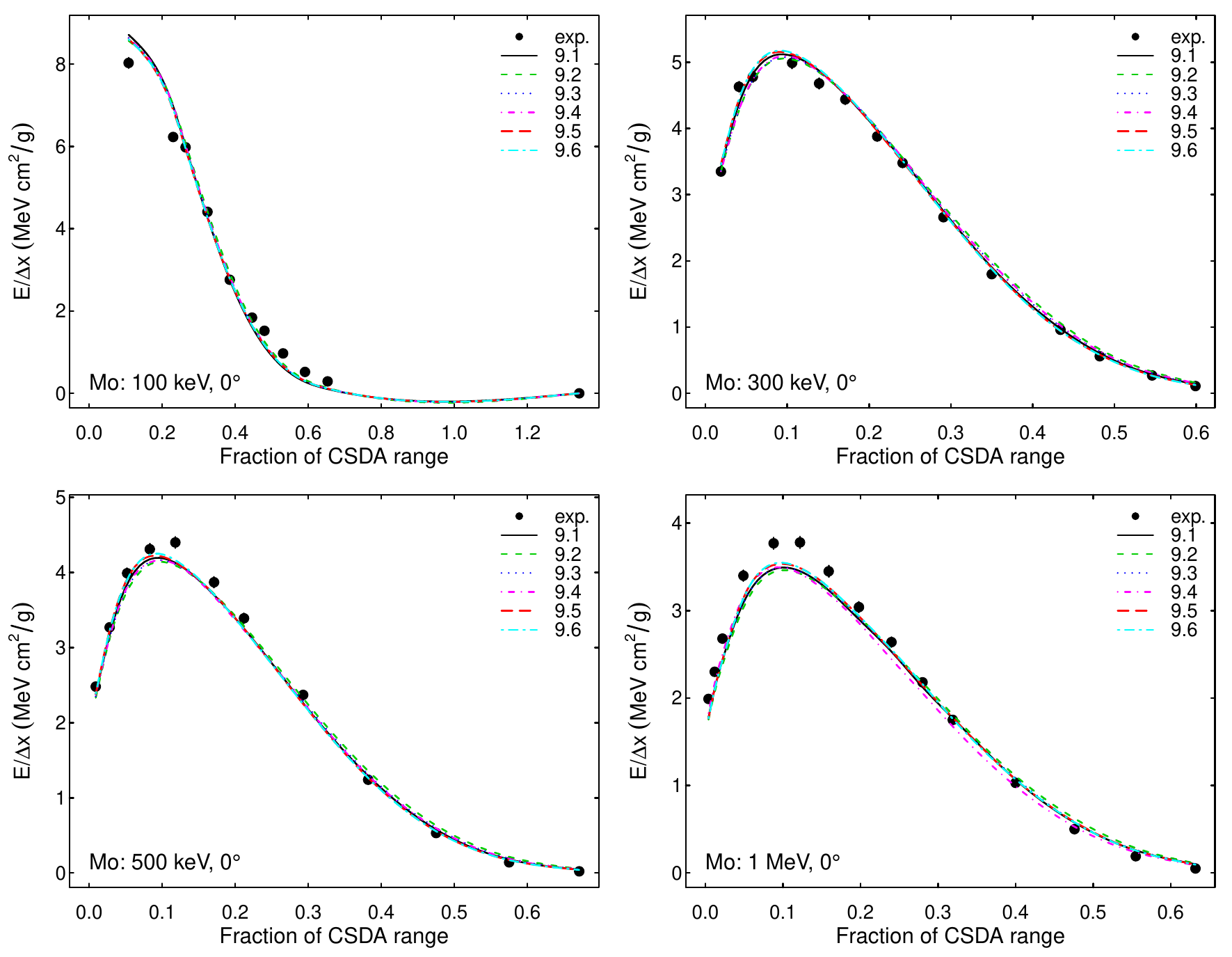}}
\caption{Longitudinal energy deposition in molybdenum, orthogonally incident
beam: experimental data from \cite{sandia79} (black dots) and simulations with
Geant4 versions 9.1 to 9.6, using models based on EEDL and EPDL evaluated data
libraries.
The error bars of the experimental data points are not visible, when they are
smaller than the symbol size.}
\label{fig_Mo079} 
\end{figure*}

\begin{figure*} 
\centerline{\includegraphics[angle=0,width=18cm]{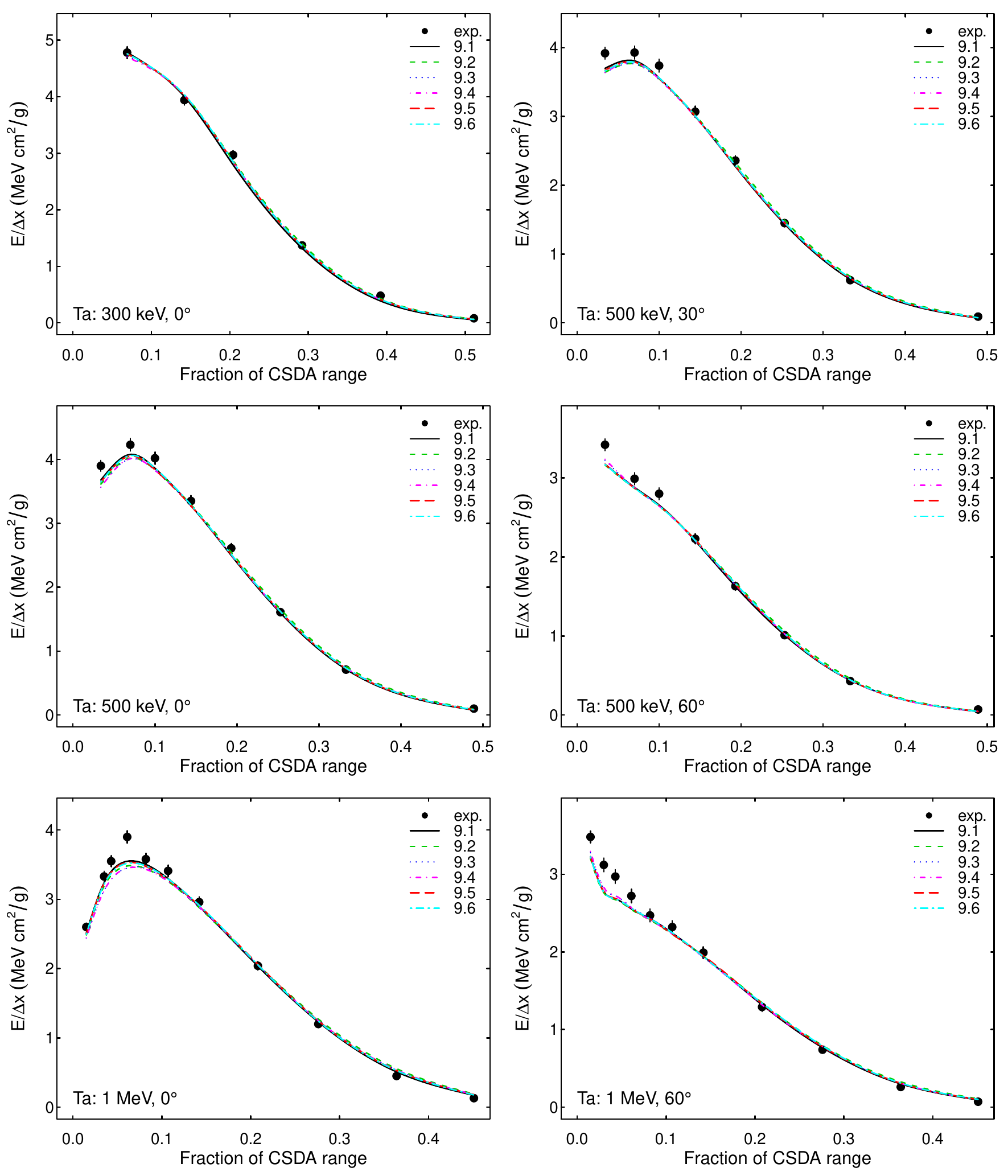}}
\caption{Longitudinal energy deposition in tantalum: experimental data from
\cite{sandia79} (black dots) and simulations with Geant4 versions 9.1 to 9.6,
using models based on EEDL and EPDL evaluated data libraries.
The error bars of the experimental data points are not visible, when they are
smaller than the symbol size.}
\label{fig_Ta79}
\end{figure*}

\begin{figure*} 
\centerline{\includegraphics[angle=0,width=18cm]{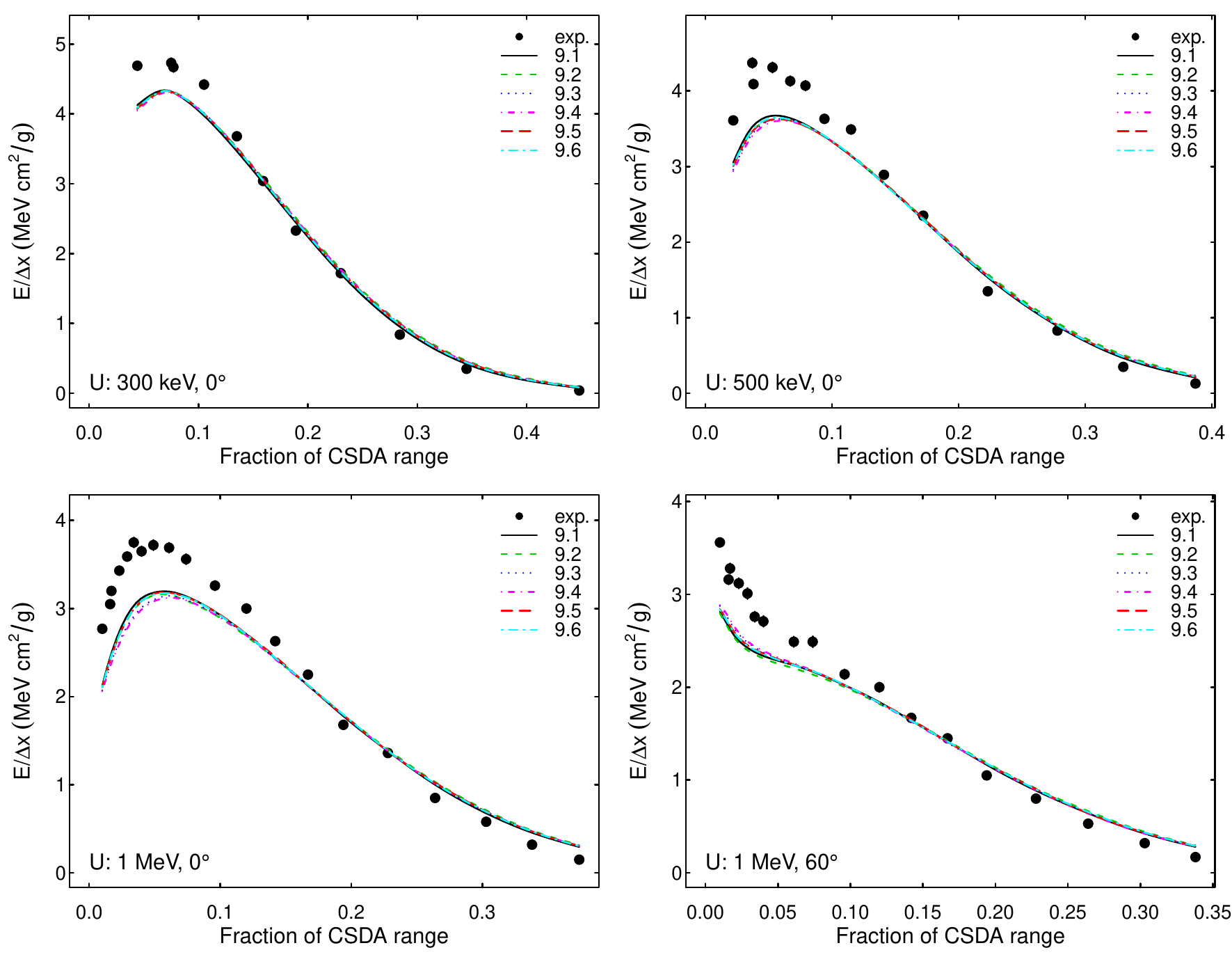}}
\caption{Longitudinal energy deposition in uranium: experimental data from
\cite{sandia79} (black dots) and simulations with Geant4 versions 9.1 to 9.6,
using models based on EEDL and EPDL evaluated data libraries.
The error bars of the experimental data points are not visible, when they are
smaller than the symbol size.}
\label{fig_U79}
\end{figure*}

The longitudinal energy profiles produced by simulations with the electron-photon
models based on EEDL-EPDL evaluated data libraries and default multiple
scattering settings are shown for all test cases and Geant4 versions in
Fig.~\ref{fig_C79}-\ref{fig_U79}.
The plots also report the experimental data of \cite{sandia79}.
The quantity reported in the plots is the energy deposited in the
calorimeter divided by the thickness of the calorimeter.
The associated depth is determined by the sum of the thickness of the front slab
and one-half the calorimeter thickness, and is expressed as a fraction of the
CSDA range.

The p-values resulting from the $\chi^2$ test over all experimental configurations
are listed in Tables \ref{tab_pvalue79_liv}, \ref{tab_pvalue79_pen} and
\ref{tab_pvalue79_std}, for Geant4 electron-photon interaction
models based on EEDL-EPDL, originating from Penelope and implemented in the
``standard'' electromagnetic package respectively.

The p-values obtained for Geant4 9.1 are numerically different from those listed
in \cite{tns_sandia}.
The simulations were produced with two different releases of Geant4 code:
the original 9.1 version for \cite{tns_sandia} and patch \textit{p03} on top of
the original version for the data examined in this paper.
Moreover, the simulation productions were executed with different seeds of the
random number generators, and the experimental error scale factors were slightly
different in the two analyses, as explained in section \ref{sec_exp}.
Nevertheless, the outcome of the $\chi^2$ test in terms of rejection of the null
hypothesis with 0.01 significance is the same for the data evaluated in this
paper and in \cite{tns_sandia}.

The number of test cases that pass the $\chi^2$ test, i.e. having p-value greater
than the defined 0.01 significance level, is reported in Table~\ref{tab_pass79}:
the largest number of successes of the $\chi^2$ test is observed with Geant4
9.1, using electron-photon models based on the EEDL-EPDL evaluated data
libraries.

For convenience, the ``efficiency'' of a Geant4 simulation configuration is
defined as the fraction of test cases in which the $\chi^2$ test does not reject
the null hypothesis at 0.01 level of significance: this quantifies the capability
of that simulation configuration to produce results statistically consistent
with experiment over the whole set of experimental conditions.
The test cases with uranium targets are not considered 
in the calculation of the efficiencies, as discussed in section \ref{sec_strategy}.

The efficiency of Geant4 configurations is plotted in Fig.~\ref{fig_eff79} as a
function of Geant4 version for the three sets of electron-photon models
examined in this paper. 
%The results reported in Fig.~\ref{fig_eff79} were produced with the same
%settings in each Geant4 version, apart from the electron-photon processes 
%listed in Table~\ref{tab_processes; 
%within the context of each Geant4 version, one can
%assume that other effects affect the outcome of the simulations in a similar
%manner.
In all experimental configurations the best efficiency is obtained with Geant4
electron-photon models based on the EEDL and EPDL evaluated data libraries.
According to Table~\ref{tab_pass79} and Fig.~\ref{fig_eff79}, the best
compatibility with experimental data is achieved with Geant4 version 9.1.

% -----------------------------------------------------------------------------

% -----------------------------------------------------------------------------------------
\subsection{Evaluation of electron-photon modeling alternatives}
\label{sec_sandia79_mod}

This analysis compares the relative capability of Geant4 electron-photon models
of producing simulations consistent with experimental data.

Electron-photon interactions determine the main characteristics of the simulated
energy deposition.
It is worthwhile to note that electron-photon models are not solely
responsible for the overall compatibility with experimental data: other Geant4
physics components, such as the simulation of multiple scattering and of energy
loss fluctuations, contribute to the shape of the energy deposition as well.
Therefore, the appraisal of the relative merits of different electron-photon
models is performed within the context of a given Geant4 version, where other
factors which may affect the simulation can be considered constant and common to
all test cases.

The input to the comparison of the accuracy of the three modeling alternatives
consists of the outcome of the $\chi^2$ test for compatibility with experimental
data.
Table \ref{tab_pass79} summarizes the number of test cases where the $\chi^2$
test rejects or does not reject the hypothesis of equivalence between
simulated and experimental energy deposition profiles.
Contingency tables are
built based on these results, as described in Table~\ref{tab_exconting}, and
the modeling alternatives they encompass are compared by means of the tests
discussed in section~\ref{sec_conting}.
The  p-values resulting from the tests on these contingency tables 
are reported in Tables~\ref{tab_livpen}, \ref{tab_livstd}
and \ref{tab_penstd}.

% Table generated by Excel2LaTeX from sheet 'p-value R'
\begin{table}[htbp]
  \centering
  \caption{Number of test cases of longitudinal energy deposition that pass the $\chi^2$ test }
    \begin{tabular}{llcccccc}
    \hline
    &  & \multicolumn{6}{c}{\textbf{Geant4 version}} \\
    & \textbf{Geant4 models} & \textbf{9.1} & \textbf{9.2} & \textbf{9.3} & \textbf{9.4} & \textbf{9.5} & \textbf{9.6} \\
    \hline
    \multicolumn{1}{c}{\multirow{3}[0]{*}{\bf Pass}} & EEDL-EPDL & 22    & 8     & 8     & 6     & 14    & 14 \\
    \multicolumn{1}{c}{} & Penelope & 9     & 5     & 4     & 3     & 4     & 4 \\
    \multicolumn{1}{c}{} & Standard & 7     & 5     & 1     & 1     & 5     & 5 \\
\hline
    \multicolumn{1}{c}{\multirow{3}[0]{*}{\bf Fail}} & EEDL-EPDL & 8     & 22    & 22    & 24    & 16    & 16 \\
    \multicolumn{1}{c}{} & Penelope & 21    & 25    & 26    & 27    & 26    & 26 \\
    \multicolumn{1}{c}{} & Standard & 23    & 25    & 29    & 29    & 25    & 25 \\
    \hline
   \end{tabular}%
  \label{tab_pass79}%
\end{table}%

% Table generated by Excel2LaTeX from sheet 'R Fisher'
\begin{table}[tbp]
  \centering
\caption{P-values of tests comparing the compatibility of simulations using 
Geant4 electron-photon models based on EEDL-EPDL and models originating from Penelope with experimental data}
    \begin{tabular}{lcccccc}
    \hline
& \multicolumn{6}{c}{\textbf{Geant4 version}} \\
    \textbf{Test} & \textbf{9.1} & \textbf{9.2} & \textbf{9.3} & \textbf{9.4} & \textbf{9.5} & \textbf{9.6} \\
    \hline
    \textbf{Fisher} & 0.002 & 0.532 & 0.333 & 0.472 & 0.010 & 0.010 \\
    \textbf{Pearson $\chi^2$} & 0.001 & 0.347 & -  & - & - & - \\
    \textbf{Barnard} & 0.001 & 0.527 & 0.245 & 0.308 & 0.006 & 0.005 \\
    \textbf{Z-pooled} & 0.001 & 0.527 & 0.245 & 0.308 & 0.006 & 0.006 \\
    \textbf{Boschloo} & 0.001 & 0.378 & 0.245 & 0.366 & 0.006 & 0.006 \\
    \hline
    \end{tabular}%
  \label{tab_livpen}%
\end{table}%

% Table generated by Excel2LaTeX from sheet 'R Fisher'
\begin{table}[tbp]
  \centering
\caption{P-values of tests comparing the compatibility of simulations using 
Geant4 electron-photon models based on EEDL-EPDL and models in the standard package  with experimental data}
    \begin{tabular}{lcccccc}
    \hline
& \multicolumn{6}{c}{\textbf{Geant4 version}} \\
    \textbf{Test} & \textbf{9.1} & \textbf{9.2} & \textbf{9.3} & \textbf{9.4} & \textbf{9.5} & \textbf{9.6} \\
    \hline
    \textbf{Fisher} & $<0.001$ & 0.532 & 0.026 & 0.103 & 0.025 & 0.025 \\
    \textbf{Pearson $\chi^2$} & $<0.001$ & 0.347 & - & - & 0.012 & 0.012 \\
    \textbf{Barnard} & $<0.001$ & 0.527 & 0.014 & 0.055 & 0.014 & 0.014 \\
    \textbf{Z-pooled} & $<0.001$ & 0.527 & 0.014 & 0.055 & 0.014 & 0.014 \\
    \textbf{Boschloo} & $<0.001$ & 0.378 & 0.014 & 0.056 & 0.014 & 0.014 \\
    \hline
    \end{tabular}%
  \label{tab_livstd}%
\end{table}%

% Table generated by Excel2LaTeX from sheet 'R Fisher'
\begin{table}[tbp]
  \centering
\caption{P-values of tests comparing the compatibility of simulations using 
Geant4 electron-photon models originating from Penelope and models in the standard package with experimental data}
    \begin{tabular}{lcccccc}
    \hline
& \multicolumn{6}{c}{\textbf{Geant4 version}} \\
    \textbf{Test} & \textbf{9.1} & \textbf{9.2} & \textbf{9.3} & \textbf{9.4} & \textbf{9.5} & \textbf{9.6} \\
    \hline
    \textbf{Fisher} & 0.771 & 1.000 & 0.353 & 0.612 & 1     & 1 \\
    \textbf{Pearson $\chi^2$} & 0.559 & 1.000 & 0.161 & 0.301 & 0.718 & 0.718 \\
    \textbf{Barnard} & 0.680 & 1.000 & 0.236 & 0.362 & 0.815 & 0.815 \\
    \textbf{Z-pooled} & 0.680 & 1.000 & 0.236 & 0.362 & 0.815 & 0.815 \\
    \textbf{Boschloo} & 0.617 & 1.000 & 0.245 & 0.519 & 1     & 1 \\
    \hline
    \end{tabular}%
  \label{tab_penstd}%
\end{table}%

Table~\ref{tab_livpen} shows that Geant4 electron-photon models
based on EEDL-EPDL and the Penelope-like models result in significantly
different accuracies of the longitudinal energy deposition pattern of electrons
simulated by Geant4 versions 9.1, 9.5 and 9.6. 
The significance level of these tests is 0.01.
Regarding the marginal compatibility between the two categories in Geant4
versions 9.5 and 9.6 according to Fisher's exact test, one should take into
account that this test is conservative, when applied to experimental designs
that do not constrain both row and column totals in the associated contingency
tables.
The two sets of models produce statistically equivalent results in Geant4 
versions 9.2, 9.3 and 9.4.
It is worthwhile to note that the ``efficiency'', reported in
Fig.~\ref{fig_eff79}, is quite low for all electron-photon models in these
versions, not exceeding 0.3 for any model: this observation suggests that the
analysis of contingency tables for these Geant4 versions would have scarce
discriminating power to appreciate differences in accuracy associated with any
model.
%since in most test cases the $\chi^2$ test rejects the hypothesis of
%compatibility with measurements irrespective of which set of electron-photon
%models is activated in the simulation.

According to Table~\ref{tab_livstd}, the hypothesis of equivalent compatibility
with experiment for the electron-photon models based on EEDL-EPDL
and the models in the ``standard'' package is rejected in Geant4 version 9.1, while
it is not rejected in the later versions.
The efficiency of both sets of models is lower in Geant4 versions later than
9.1, therefore the result of the statistical tests cannot be explained by an
improvement of the accuracy of the models in Geant4 standard electromagnetic
package.
This result could be due to a worse degradation of the accuracy of the models
based on EEDL-EPDL than of the standard ones in versions later than 9.1, or
could be explained by the loss of discriminating power of the tests over
contingency tables, when only a small number of test cases ``pass'' the
$\chi^2$ test of compatibility between simulation and experiment.

Based on the results in Tables~\ref{tab_livpen} and \ref{tab_livstd}, one can
conclude that the three sets of electron-photon models contribute to
significantly different simulation accuracy when using Geant4 9.1, while, along with
a general degradation of simulation accuracy in later Geant4 versions, the
selection of electron-photon models is less critical when using other, more recent
versions.

The simulations based on Penelope-like and standard electron-photon models exhibit 
statistically equivalent compatibility with experimental data: the null hypothesis of
equivalence of the two categories is never rejected in any of the test cases
summarized in Table \ref{tab_penstd}.

% -----------------------------------------------------------------------------------------

\subsection{Evolution of Penelope-like electron-photon modeling}

%No significant difference is observed regarding the compatibility with
%experimental data using electron-photon models reengineered from Penelope 2001
%and Penelope 2008.
The general trend of the results reported in section \ref{sec_sandia79_mod} does
not suggest any major variations in the compatibility of the Penelope-like
models with experimental data, where differences could have arisen due to the
update from Penelope 2001 to Penelope 2008.
% enforced from Geant4 9.5 version.
Since classes reengineered from both Penelope versions cohexist in Geant4 9.5, it is
possible to compare their accuracy directly in that context.

Longitudinal energy deposition profiles were produced with Geant4 9.5 using
both implementations of Penelope-like electron-photon models, keeping all other
simulation features unchanged, and were subject to the same analysis procedure.
The results of the $\chi^2$ test appear similar for the two implementations: the
number of test cases that ``pass'' the $\chi^2$ test is 4 and 5, using code
reengineered from Penelope 2008 and 2001 respectively, while the hypothesis
of compatibility with experimental measurements is rejected in 26 and 25 test cases respectively.

The analysis of the resulting 2$\times$2 table does not reject the
hypothesis of equivalent compatibility with experiment using Penelope 2008 or 2001: the p-value of
McNemar's test is 0.317, while it is 1 for McNemar's exact test.
These results suggest that the reengineered Penelope 2008 models
do not represent a significant improvement with respect to the 2001 ones
in the experimental scenario subject to evaluation.

%This result can be explained by the loss of discriminant power of the
%categorical tests, when the compatibility with measurements is significantly
%degraded for any physics settings in Geant4 versions later than 9.1.

%observe that the hypothesis of equivalent simulation accuracy produced by Geant4
%EEDL-EPDL and other electron-photon models is rejected only in Geant4 version
%9.1 with 0.01 significance.

% -----------------------------------------------------------------------------------------
\subsection{Evolution over Geant4 versions}

\begin{figure}
\centerline{\includegraphics[angle=0,width=9cm]{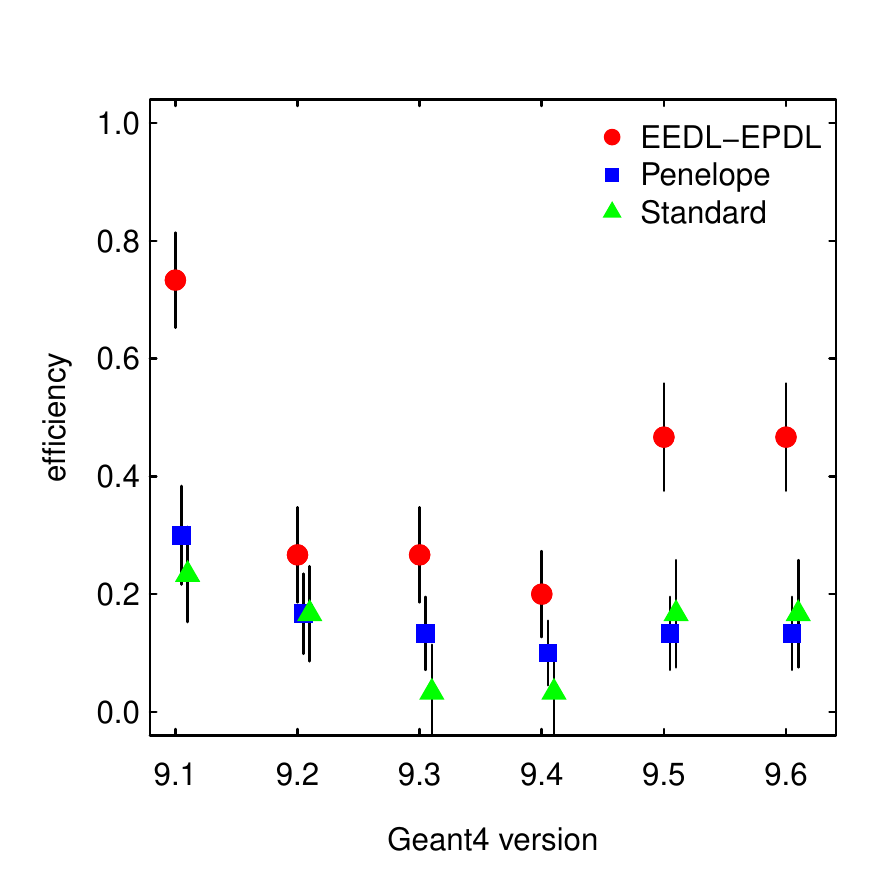}}
\caption{Efficiency of Geant4 simulation configurations for
reproducing experimental longitudinal energy deposition, as a function of Geant4
version. The efficiency is shown for three sets of Geant4 electron-photon models: based on EEDL-EPDL
evaluated data libraries (red circles), originating from Penelope (blue squares)
and in the "standard" package (green triangles).
%The efficiency is defined as the fraction of test cases in which the $\chi^2$
%test does not reject the null hypothesis at 0.01 level of significance.
The symbols representing the data points are slightly shifted along the x-axis
to improve the clarity of the plot.}
\label{fig_eff79}
\end{figure}

One can observe in Fig.~\ref{fig_eff79} a similar trend associated with all
Geant4 electron-photon models: for each modeling option the highest efficiency
is achieved with Geant4 9.1, the lowest with Geant4 9.4, while some improvement
with respect to the results of Geant4 9.4 is visible with Geant4 9.5 and 9.6,
although the efficiency associated with these versions remains lower than with
Geant4 9.1.
This trend suggests that the accuracy of the simulation is also influenced by
other Geant4 features, which appear independent from the electron-photon
settings selected in the user application.
% and subject to evolution over the versions evaluated in this paper.
This issue was already discussed in \cite{tns_sandia}, although in that case the
observation was limited to two Geant4 versions, while a larger simulation data
sample is analyzed here.

The equivalence in compatibility with experiment of different Geant4 versions is
estimated by means of an analysis of matched pairs, to which McNemar's
\cite{mcnemar} test is applied.
Matched pairs consist of simulations with identical physics settings in the user
application, whose compatibility with experimental data is evaluated before and
after a Geant4 kernel evolution, i.e. based on a given Geant4 version and on a
later one.

%This statistical analysis quantifies the effect of the improvements in Geant4
%electromagnetic physics stated in
%\cite{em_mc2010,em_chep2010,em_radecs2011,em_chep2012} on the accuracy of the
%simulated energy deposition, with respect to the results previously reported in
%\cite{tns_sandia}.
%In this respect it is worthwhile to note that in the evolution from Geant4 9.1
%to 9.6 versions no major modifications occurred in Geant4 kernel other than
%electromagnetic physics, which would be relevant to the simulation scenarios
%considered in this study.

%The relative capability of Geant4 versions of producing longitudinal energy
%deposition patterns compatible with measured data is evaluated by means of 
%McNemar's test.

The test is applied to 2$\times$2 tables built as specified in
Table~\ref{tab_exmcnemar}; matched pairs consist of test cases involving Geant4
9.1 and a later version.
The p-values for matched pairs concerning Geant4 versions 9.2 to 9.6 are
reported in Table \ref{tab_mcnemar} for a calculation based on the $\chi^2$
asymptotic distribution and for the "exact" test.
The null hypothesis of equivalent compatibility with experimental data 
with Geant4 9.1 and with any later versions is rejected with 0.01 significance.

% Table generated by Excel2LaTeX from sheet 'R Fisher'
\begin{table}
  \centering
\caption{P-value of McNemar's test comparing the compatibility with experimental
longitudinal energy deposition profiles simulated by Geant4 9.1 and those of later
versions}
    \begin{tabular}{lccccc}
    \hline
          & \multicolumn{5}{c}{\textbf{Geant4 version compared to 9.1}} \\
    \textbf{McNemar test} & \textbf{9.2} & \textbf{9.3} & \textbf{9.4} & \textbf{9.5} & \textbf{9.6} \\
  \hline    
\textbf{$\chi^2$} & $<0.001$ & $<0.001$ & $<0.001$ & 0.005 & 0.005 \\
    \textbf{Exact} & $<0.001$ & 0.001 & $<0.001$ & 0.008 & 0.008 \\
%    \textbf{Exact 1-side} & $<0.001$ & $<0.001$ & $<0.001$ & 0.004 & 0.004 \\
    \hline
    \end{tabular}%
  \label{tab_mcnemar}%
\end{table}%

%The rejection of the null hypothesis in favour of the alternative hypothesis
%in the one-sided McNemar test concerns differences in a specific direction;
%the outcome of the test excludes that the energy deposition profiles are
%more accurately simulated in Geant4 versions later than 9.1.
%
%as the accuracy of the simulation being significantly better with Geant4 9.1
%with respect to Geant4 9.2 to 9.6 versions.
%The null hypothesis 

A thorough investigation of the causes of the deterioration of the compatibility
between simulated and experimental energy deposition profiles is outside the
scope of this paper; presumably, changes to multiple scattering code and
possibly to other electromagnetic physics features contribute to the results of
Table~\ref{tab_exmcnemar}.
A limited assessment is documented in the following section.

% -----------------------------------------------------------------------------------------
\subsection{Multiple scattering settings}
\label{sec_ms_param}

The behavior of Geant4 multiple scattering process can be configured
by selecting a model to be instantiated and setting the
values of some empirical parameters embedded in the software implementation.
While a  thorough assessment of the optimization of all such options
is outside the scope of this paper, some limited investigation was
performed to assess to which extent the evolution of default multiple scattering
settings, summarized in Table \ref{tab_mscattdef}, could contribute to the
accuracy of the simulation.

%The Geant4 versions examined in this paper encompass one or more of variants
%of the Urban multiple scattering model.
%classes.

The accuracy of energy deposition profiles was estimated by performing
simulations with a given Geant4 version in different conditions: the default configuration
of electron multiple scattering and modified configurations, where the current default
settings associated with that version were replaced by settings implemented as
default values in previous Geant4 versions.
These simulations were performed with Geant4 version 9.5p01, which at the time these studies were made was the latest released version;
%The research project documented in this paper spanned several months' work;
since all the validation tests reported in this paper show 
largely similar, or even identical results with Geant4 9.5 and 9.6,
this investigation was not repeated with Geant4 9.6.

The \textit{skin} parameter was modified from the
default value of 1 in Geant4 9.5 to 3, which is the default value in Geant4 9.2 and 9.3; the
\textit{range factor} parameter, which is set to 0.04 by default in Geant4 9.5,
was changed to 0.02, which is the default value in Geant4 9.1 and 9.2.
The default multiple scattering model was modified from \textit{G4UrbanMscModel95}
to \textit{G4UrbanMscModel93}, which is the default model in Geant4 9.4.
Only one default setting was modified in each simulation production to best
appraise its effect.
The electron-photon models based on EEDL-EPDL were used in all these simulations; all
the other settings of the simulations were identical.
%apart from those concerning multiple scattering.

The number of test cases that pass, or fail the $\chi^2$ test of compatibility with
the experimental energy deposition profiles is listed in Table
\ref{tab_mscattparam} for the various configurations subject to evaluation, 
along with the resulting ``efficiency''.

% Table generated by Excel2LaTeX from sheet 'p-value R'
\tabcolsep=4pt 
\begin{table}[htbp]
  \centering
  \caption{Effect of multiple scattering configuration on the compatibility with experimental energy deposition profiles in simulations produced with Geant4 9.5}
    \begin{tabular}{lcccc}
    \hline
          & \textbf{Default} & \textbf{Model} & \textbf{skin} & \textbf{Range Factor} \\
          & \textbf{configuration}  & Urban93 & 3 & 0.02 \\
    \hline
    \textbf{Pass} & 14    & 7     & 14    & 15 \\
    \textbf{Fail} & 16    & 23    & 16    & 15 \\
    \textbf{Efficiency} & 0.47$\pm$0.09  & 0.23$\pm$0.08  & 0.47$\pm$0.09  & 0.50$\pm$0.09 \\
    \hline
    \end{tabular}%
  \label{tab_mscattparam}%
\end{table}%
\tabcolsep=6pt

The effect of the changes to the \textit{skin} and \textit{range factor}
parameters on the compatibility with experimental data is negligible.
Regarding the test cases that ``pass'' the $\chi^2$ test of compatibility with
experimental data,
the result with modified \textit{skin} is identical to that with the default
multiple scattering configuration (apart some numerical differences of the 
individual p-values, which anyway do not alter the outcome of the test at the defined level of significance).
A small improvement in compatibility with experiment is obtained with the
modified \textit{range factor}; nevertheless, this increase in ``efficiency'' is
statistically insignificant.

More significant effects are observed by changing the multiple scattering model.
The compatibility with the experimental data from \cite{sandia79} appears to be
severely affected by the selection of the multiple scattering modeling
option. 
Table \ref{tab_mscattparam} shows that the ``efficiency'' of the
simulation drops from 0.47$\pm$0.09 with the default configuration to
0.23$\pm$0.08 with \textit{G4UrbanMscModel93}.
The efficiency obtained with Geant4 9.5 and \textit{G4UrbanMscModel93} is
comparable to that obtained with Geant4 9.4, where the same model is
instantiated in the default multiple scattering configuration.
This result hints that \textit{G4UrbanMscModel93} could be responsible
for the lower compatibility with experimental data of simulations produced with
Geant4 9.4 generally observed in the validation tests documented in this paper.
Scarce documentation about the physics features, software implementation and
evolution of Urban multiple scattering calculation is available in the
literature and in Geant4 software manuals; therefore the identification of the
features in the two examined variants of the Urban model that are responsible
for significantly different behavior is not practically feasible within the
scope of this paper.
\subsection{Multiple scattering with Goudsmit-Saunderson and Urban  models}
\label{sec_goudsmit}

% Table generated by Excel2LaTeX from sheet 'p-value GS'
\tabcolsep=4pt
\begin{table}[tbp]
  \centering
  \caption{P-values of the $\chi^2$ tests for longitudinal energy deposition: simulations with electron-photon models based on EEDL-EPDL and Goudsmit-Saunderson multiple scattering model}
    \begin{tabular}{lrcc|cccc}
    \hline
     \multicolumn{2}{c}{\textbf{Target}}  & \textbf{E } & \textbf{Angle}   & \multicolumn{4}{c}{\textbf{Geant4 version}} \\
    & Z       & (kev) & (degrees) & \textbf{9.3} & \textbf{9.4} & \textbf{9.5} & \textbf{9.6} \\
      \hline
    Be    & 4     & 58    & 0     & 0.124 & 0.352 & 0.435 & 0.446 \\
    Be    & 4     & 109   & 0     & $<0.001$ & 0.264 & 0.089 & 0.084 \\
    Be    & 4     & 314   & 0     & $<0.001$ & $<0.001$ & $<0.001$ & $<0.001$ \\
    Be    & 4     & 521   & 0     & $<0.001$ & $<0.001$ & $<0.001$ & $<0.001$ \\
    Be    & 4     & 1033  & 0     & $<0.001$ & $<0.001$ & $<0.001$ & $<0.001$ \\
    C     & 6     & 1000  & 0     & $<0.001$ & $<0.001$ & $<0.001$ & $<0.001$ \\
    Al    & 13    & 314   & 0     & $<0.001$ & $<0.001$ & $<0.001$ & $<0.001$ \\
    Al    & 13    & 521   & 0     & $<0.001$ & $<0.001$ & $<0.001$ & $<0.001$ \\
    Al    & 13    & 1033  & 0     & $<0.001$ & $<0.001$ & $<0.001$ & $<0.001$ \\
    Al    & 13    & 314   & 60    & $<0.001$ & 0.003 & 0.005 & 0.003 \\
    Al    & 13    & 521   & 60    & 0.214 & $<0.001$ & $<0.001$ & $<0.001$ \\
    Al    & 13    & 1033  & 60    & $<0.001$ & $<0.001$ & $<0.001$ & $<0.001$ \\
    Fe    & 26    & 300   & 0     & 0.024 & 0.320 & 0.468 & 0.400 \\
    Fe    & 26    & 500   & 0     & $<0.001$ & $<0.001$ & $<0.001$ & $<0.001$ \\
    Fe    & 26    & 1000  & 0     & $<0.001$ & $<0.001$ & $<0.001$ & $<0.001$ \\
    Cu    & 29    & 300   & 0     & $<0.001$ & $<0.001$ & $<0.001$ & $<0.001$ \\
    Cu    & 29    & 500   & 0     & $<0.001$ & $<0.001$ & $<0.001$ & $<0.001$ \\
    Mo    & 42    & 100   & 0     & $<0.001$ & $<0.001$ & $<0.001$ & $<0.001$ \\
    Mo    & 42    & 300   & 0     & $<0.001$ & $<0.001$ & $<0.001$ & $<0.001$ \\
    Mo    & 42    & 500   & 0     & $<0.001$ & 0.014 & 0.005 & 0.011 \\
    Mo    & 42    & 1000  & 0     & $<0.001$ & $<0.001$ & $<0.001$ & $<0.001$ \\
    Mo    & 42    & 300   & 60    & $<0.001$ & $<0.001$ & $<0.001$ & $<0.001$ \\
    Mo    & 42    & 500   & 60    & $<0.001$ & $<0.001$ & $<0.001$ & $<0.001$ \\
    Mo    & 42    & 1000  & 60    & $<0.001$ & $<0.001$ & $<0.001$ & $<0.001$ \\
    Ta    & 73    & 300   & 0     & 0.076 & 0.010 & 0.010 & 0.009 \\
    Ta    & 73    & 500   & 0     & 0.042 & 0.020 & 0.019 & 0.032 \\
    Ta    & 73    & 1000  & 0     & $<0.001$ & 0.003 & 0.002 & 0.007 \\
    Ta    & 73    & 500   & 60    & $<0.001$ & $<0.001$ & $<0.001$ & $<0.001$ \\
    Ta    & 73    & 1000  & 60    & $<0.001$ & $<0.001$ & $<0.001$ & $<0.001$ \\
    Ta    & 73    & 500   & 30    & 0.006 & 0.004 & 0.003 & 0.004 \\
%    U     & 92    & 300   & 0     & $<0.001$ & $<0.001$ & $<0.001$ & $<0.001$ \\
%    U     & 92    & 500   & 0     & $<0.001$ & $<0.001$ & $<0.001$ & $<0.001$ \\
%    U     & 92    & 1000  & 0     & $<0.001$ & $<0.001$ & $<0.001$ & $<0.001$ \\
%    U     & 92    & 1000  & 60    & $<0.001$ & $<0.001$ & $<0.001$ & $<0.001$ \\
    \hline
    \end{tabular}%
  \label{tab_pvgoudsmit}%
\end{table}%
\tabcolsep=6pt

\begin{figure}
\centerline{\includegraphics[angle=0,width=9cm]{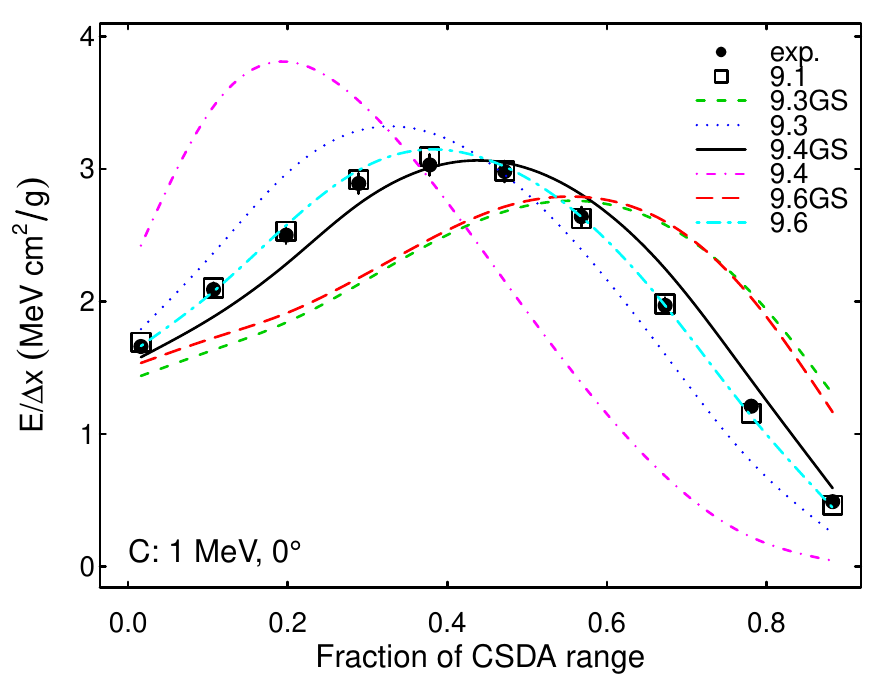}}
\caption{Longitudinal energy deposition in carbon produced with two multiple scattering models implemented in Geant4: Goudsmit-Saunderson and variants of the Urban model. The plots shows experimental data from \cite{sandia79} (black dots) and
simulations with Geant4 versions 9.1 to 9.6, using models based on EEDL and EPDL
evaluated data libraries.
The Goudsmit-Saunderson multiple scattering model was first released in Geant4 version 9.3.
Simulation results for Geant4 9.5 are omitted for better clarity of the plot,
since they are very similar to those of Geant4 9.6.
The error bars of the experimental data points are not visible, when they are
smaller than the symbol size.}
\label{fig_C79GS}
\end{figure}

An implementation based on Goudsmit-Saunderson calculations
\cite{goudsmit1,goudsmit2}, specialized for the transport of electrons,
has been available in Geant4 for the simulation of
multiple scattering since  version 9.3 \cite{kadri_goudsmit}.
An example of energy deposition profiles obtained activating this
multiple scattering model in the simulation application is illustrated in 
Fig. \ref{fig_C79GS}, where  profiles produced with the default variant 
of the Urban model are also shown.

%This model 
%represent an alternative with respect to the modeling approach of the Urban model \cite{urban},
%which is has been available in Geant4 since its first version, and has
%evolved through several modeling variants.
The contribution of the Goudsmit-Saunderson multiple scattering implementation 
to simulation accuracy has been estimated by evaluating the 
longitudinal energy deposition patterns produced in a physics configuration,
where this model is activated in place of the default ``Urban'' algorithm.
These simulations were produced with 
the electron-photon models based on EEDL-EPDL data libraries: thanks to the
higher ``efficiency'' associated with this set of electron-photon models with
respect to the two other modeling options, statistically significant variations of the
simulation accuracy are more easily visible in the distributions produced with
different multiple scattering options along with EEDL-EPDL electron-photon models.

The p-values resulting from the $\chi^2$ tests in the experimental configurations 
of \cite{sandia79} are listed in Table~\ref{tab_pvgoudsmit}.
The number of test cases that pass or fail the $\chi^2$ test with 0.01
significance is reported in Table~\ref{tab_passfailGS}.
The efficiency, defined as in section\ref{sec_sandia79_gen}, is listed in Table~\ref{tab_effGS}.
The results produced with the Goudsmit-Saunderson multiple scattering model
appear similar in all Geant4 versions.

The Goudsmit-Saunderson multiple scattering model produces similar results 
to the Urban92 model of  Geant4 9.3 and the Urban93 model of Geant4
9.4, while one can observe qualitatively that in Geant4 9.5 and 9.6 the
simulations produced with the Urban95 multiple scattering model are compatible
with experimental data in a larger fraction of test cases than those produced
with the Goudsmit-Saunderson model.
This qualitative observation is confirmed by the statistical comparison of the
compatibility with experimental data associated with either of the two multiple scattering models.

The relative contribution of the two multiple scattering models to the
compatibility of simulation with experimental data is analyzed 
%considering the data sample as constituted by matched pairs;
%the 2$\times$2 tables built for each Geant4 version as illustrated in
%Table~\ref{tab_exmcnemar} are analyzed 
by means of McNemar's test;
% which are characterized by the same
%beam and target parameters and electron-photon physics configuration.
the resulting p-values
are listed in Table~\ref{tab_gsmcnemar}, both for the plain McNemar test and for
the ``exact'' test.

The results in Table~\ref{tab_gsmcnemar} reject the hypothesis of equivalent
compatibility with experimental data using Urban95 or Goudsmit-Saunderson
multiple scattering models.
The null hypothesis is not rejected for the Urban92 and Urban93 models;
nevertheless, one should take into account the generally low compatibility
with experimental data achieved with Geant4 9.3 and 9.4.
% irrespective of thephysics options activated in the simulation.

The outcome of this quantitative analysis suggests that, regarding the accuracy
of the resulting energy deposition, the Goudsmit-Saunderson multiple scattering
model is not preferable to the Urban model in the simulation of the longitudinal
profile of the energy deposited by low energy electrons.

% Table generated by Excel2LaTeX from sheet 'p-value GS'
\begin{table}
  \centering
  \caption{Number of test cases that pass or fail the $\chi^2$ test with Goudsmit-Saunderson and Urban multiple scattering models}
    \begin{tabular}{lccccc}
    \hline
     &     & \multicolumn{4}{c}{\textbf{Geant4 version}} \\
    \textbf{Model}& \textbf{$\chi^2 $ test}  & \textbf{9.3} & \textbf{9.4} & \textbf{9.5} & \textbf{9.6} \\
    \hline
    Goudsmit & Pass  & 5     & 6     & 5     & 5 \\
    Saunderson & Fail  & 25    & 24    & 25    & 25 \\
\hline
    \multicolumn{1}{l}{\multirow{2}[0]{*}{Urban}} & Pass  & 8     & 6     & 14    & 14 \\
    \multicolumn{1}{l}{} & Fail  & 22    & 24    & 16    & 16 \\
    \hline
    \end{tabular}%
  \label{tab_passfailGS}%
\end{table}%

% Table generated by Excel2LaTeX from sheet 'p-value GS'
%\begin{table}
%  \centering
%  \caption{Add caption}
%    \begin{tabular}{lcccc}
%    \hline
%    \textbf{Multiple} & \multicolumn{4}{c}{\textbf{Geant4 version}} \\
%    \textbf{scattering} & \textbf{9.3} & \textbf{9.4} & \textbf{9.5} & \textbf{9.6} \\
%\hline
%    Goudsmit & 0.17 $\pm$ 0.07  & 0.20 $\pm$ 0.07 & 0.17  $\pm$ 0.07& 0.17 \\
%    Urban & 0.27 $\pm$ 0.08  & 0.20  $\pm$ 0.07 & 0.47  $\pm$ 0.09 & 0.47 $\pm$ 0.09\\
%    \hline
%    \end{tabular}%
%  \label{tab:addlabel}%
%\end{table}%

% Table generated by Excel2LaTeX from sheet 'p-value GS'
\tabcolsep=4pt
\begin{table}
  \centering
  \caption{Efficiency for  energy deposition profiles with Goudsmit-Saunderson and Urban multiple scattering model}
    \begin{tabular}{rcccc}
    \hline
     & \multicolumn{4}{c}{\textbf{Geant4 version}} \\
    \textbf{Model} & \textbf{9.3} & \textbf{9.4} & \textbf{9.5} & \textbf{9.6} \\
 \hline
    Goudsmit & \multirow{2}[0]{*}{0.17 $\pm$ 0.07} & \multirow{2}[0]{*}{0.20 $\pm$ 0.07} & \multirow{2}[0]{*}{0.17 $\pm$ 0.07} & \multirow{2}[0]{*}{0.17 $\pm$ 0.07} \\
    Saunderson &       &       &       &  \\
    Urban & 0.27 $\pm$ 0.08  & 0.20 $\pm$ 0.07  & 0.47 $\pm$ 0.09 & 0.47 $\pm$ 0.09 \\
    \hline
    \end{tabular}%
  \label{tab_effGS}%
\end{table}%
\tabcolsep=6pt

% Table generated by Excel2LaTeX from sheet 'p-value GS'
\begin{table}
  \centering
  \caption{P-values of McNemar's test comparing the compatibility with experimental energy deposition profiles of simulations using Urban or Goudsmit-Saunderson multiple scattering models}
    \begin{tabular}{lcccc}
   \hline
     & \multicolumn{4}{c}{\textbf{Geant4 version compared to 9.1}} \\
    \textbf{Test} & \multicolumn{1}{c}{\textbf{9.3}} & \multicolumn{1}{c}{\textbf{9.4}} & \multicolumn{1}{c}{\textbf{9.5}} & \multicolumn{1}{c}{\textbf{9.6}} \\
   \hline
    {\bf McNemar exact} & 0.453 & 1.000 & 0.004 & 0.004 \\
    {\bf McNemar} & 0.257 & 1.000 & 0.003 & 0.003 \\
    \hline
    \end{tabular}%
  \label{tab_gsmcnemar}%
\end{table}%

\begin{figure*} 
\centerline{\includegraphics[angle=0,width=18cm]{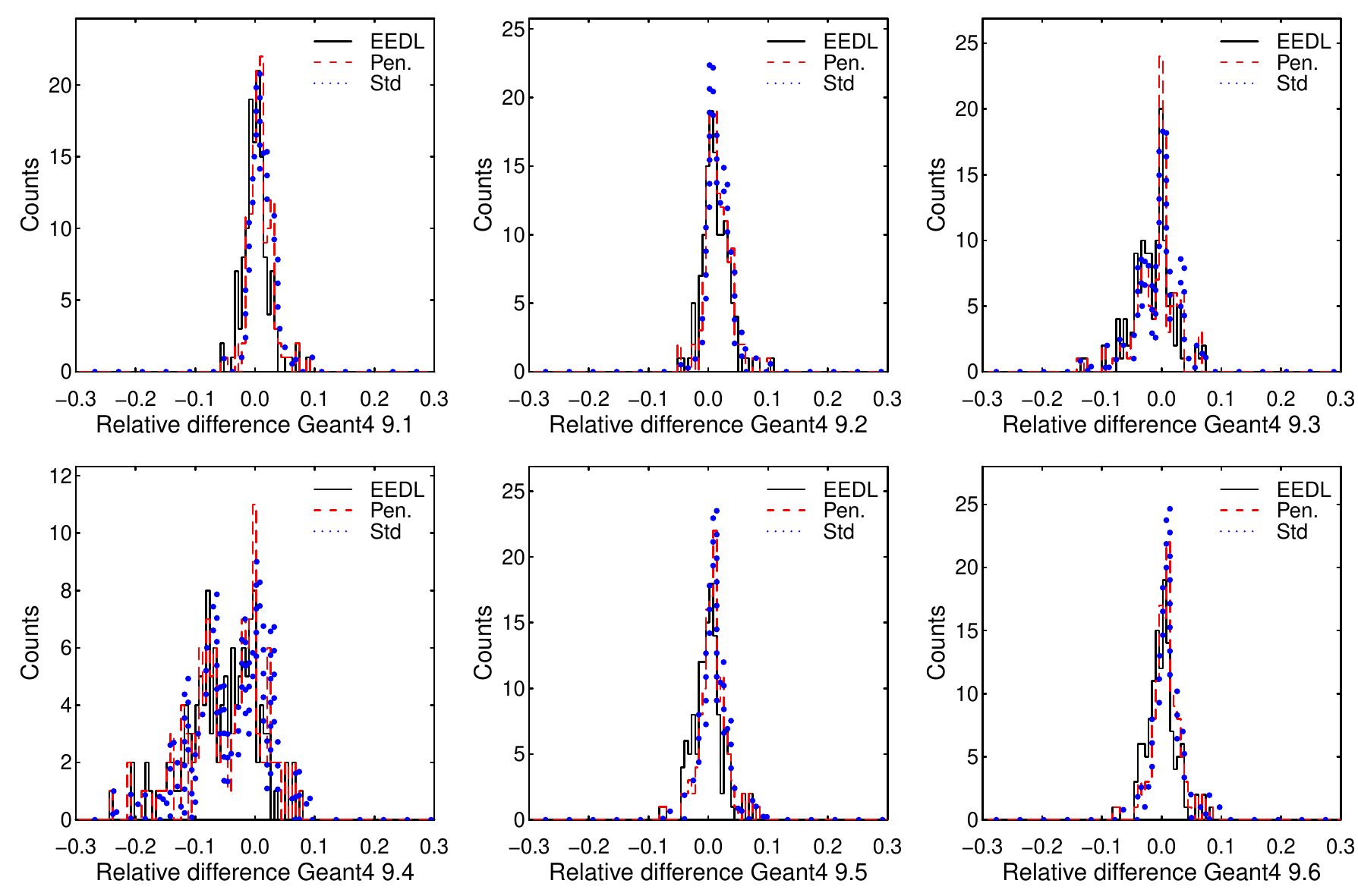}}
\caption{Relative difference between simulated and experimental total 
deposited energy produced with Geant4 versions 9.1 to 9.6; the
simulations use one of three sets of electron-photon models: based on EEDL and
EPDL evaluated data libraries (solid black line), reengineered from Penelope(dashed red histogram identified as ``Pen.'') use included in the
``standard'' electromagnetic package (dotted blue histogram identified as ``Std'').
The relative difference is defined as $(E_{simulated}-E_{experimental})/E_{experimental}$, where $E$ is
the total energy deposited in the calorimeter.}
\label{fig_diff_exp80}
\end{figure*}

% -----------------------------------------------------------------------------------------

\section{Results: total energy deposition}
\label{sec_results80}

This part of the validation process concerns the comparison of simulations with 
measurements of total deposited energy reported in \cite{sandia80}.

The validation analysis reported in the following is limited to the test cases
concerning beryllium, carbon, aluminium, titanium and tantalum targets: as
discussed in section \ref{sec_strategy}, the uranium data appear to be affected by
systematic effects \cite{sandia80}, while the measurements with molybdenum
targets performed with two experimental techniques reported in \cite{sandia80}
exhibit in some cases differences exceeding three standard deviations (up to 8
standard deviations), which also hint to the presence of some systematic effects
in the total energy measurements concerning this target material.
It should be noted that in other cases for which measurements with two 
experimental techniques are reported in \cite{sandia80} the two sets
of experimental values differ in average by 1.2 standard deviation.
The apparent systematic effects in the measurements with molybdenum
are indeed prone to bias the validation results: depending on which set of 
experimental data is taken as a reference for the validation of the simulation,
one would draw different conclusions regarding the compatibility of
simulations involving molybdenum targets resulting from different Geant4 
electron-photon models.
Since the experimental references do not provide sufficient information
to discriminate the reliability of the two sets of experimental data, neither 
experimental references for molybdenum were included in the validation process.

Two further measurements were not considered in the validation analysis:
they concern the energy deposition
in carbon resulting from 25~keV electrons, irrespective of the angle of
incidence of the beam, and from 75~keV electrons incident at 75 degrees.
These test cases appear to generate largely inconsistent results with 
respect to all the others; due to insufficient details in the experimental references,
it is impossible to ascertain whether these measurements are affected by 
systematic effects, or the simulation models are inadequate to reproduce 
them correctly.
Since other test cases at similar energy and incidence angle appear to be reproduced 
adequately by all simulation models, one may suspect indeed the presence of
systematic effects.

The relative difference between simulated and experimentally measured total energy
deposition is shown in Fig.~\ref{fig_diff_exp80} for the electron-photon models
and Geant4 versions evaluated in this paper.
The relative difference is defined as $(E_{simulated}-E_{experimental})/E_{experimental}$, where $E$ is
the total energy deposited in the calorimeter.
%these distributions encompass the test cases concerning beryllium, carbon, 
%aluminium, titanium and tantalum targets.
The horizontal scale in these plots spans the same range as in
Fig.~\ref{fig_diff_exp79}, which concerns relative difference between simulated
and experimental longitudinal energy deposition profiles: it is qualitatively
evident that the distributions in Fig.~\ref{fig_diff_exp80} are in general narrower than those in 
Fig.~\ref{fig_diff_exp79}.

\begin{figure}
\centerline{\includegraphics[angle=0,width=9cm]{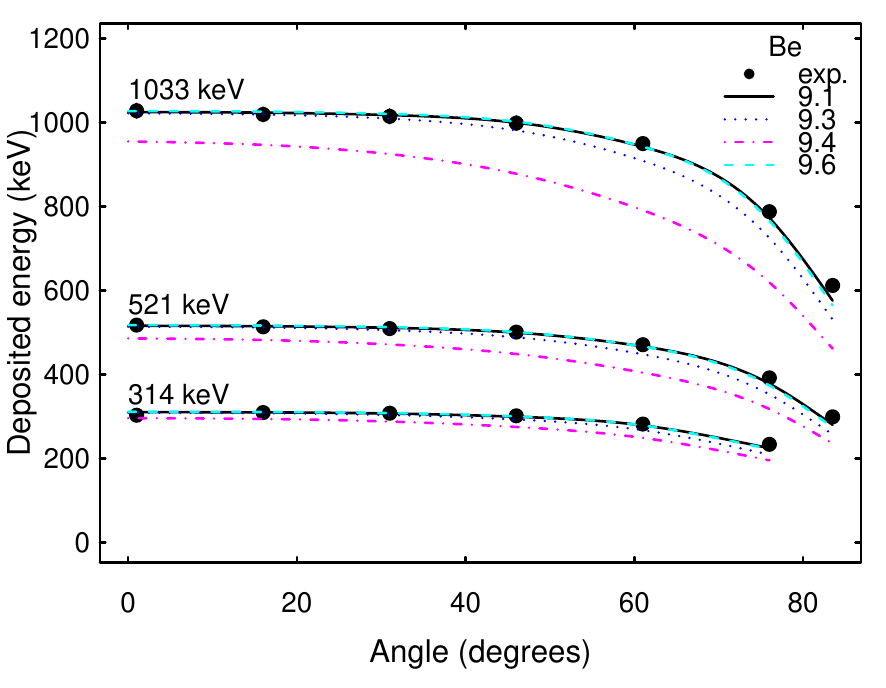}}
\caption{Total energy deposited in beryllium: experimental data from
\cite{sandia80} (black dots) and simulations with Geant4 versions 9.1, 9.3, 9.4 and 9.6,
using models based on EEDL and EPDL evaluated data libraries.
The error bars of the experimental data points are not visible, when they are
smaller than the symbol size.
The results of simulations with Geant4 versions 9.2 and 9.5 are not shown for better clarity of the plot, since they 
are very close to those of the 9.6 version.}
\label{fig_Be80}
\end{figure}

\begin{figure} 
\centerline{\includegraphics[angle=0,width=9cm]{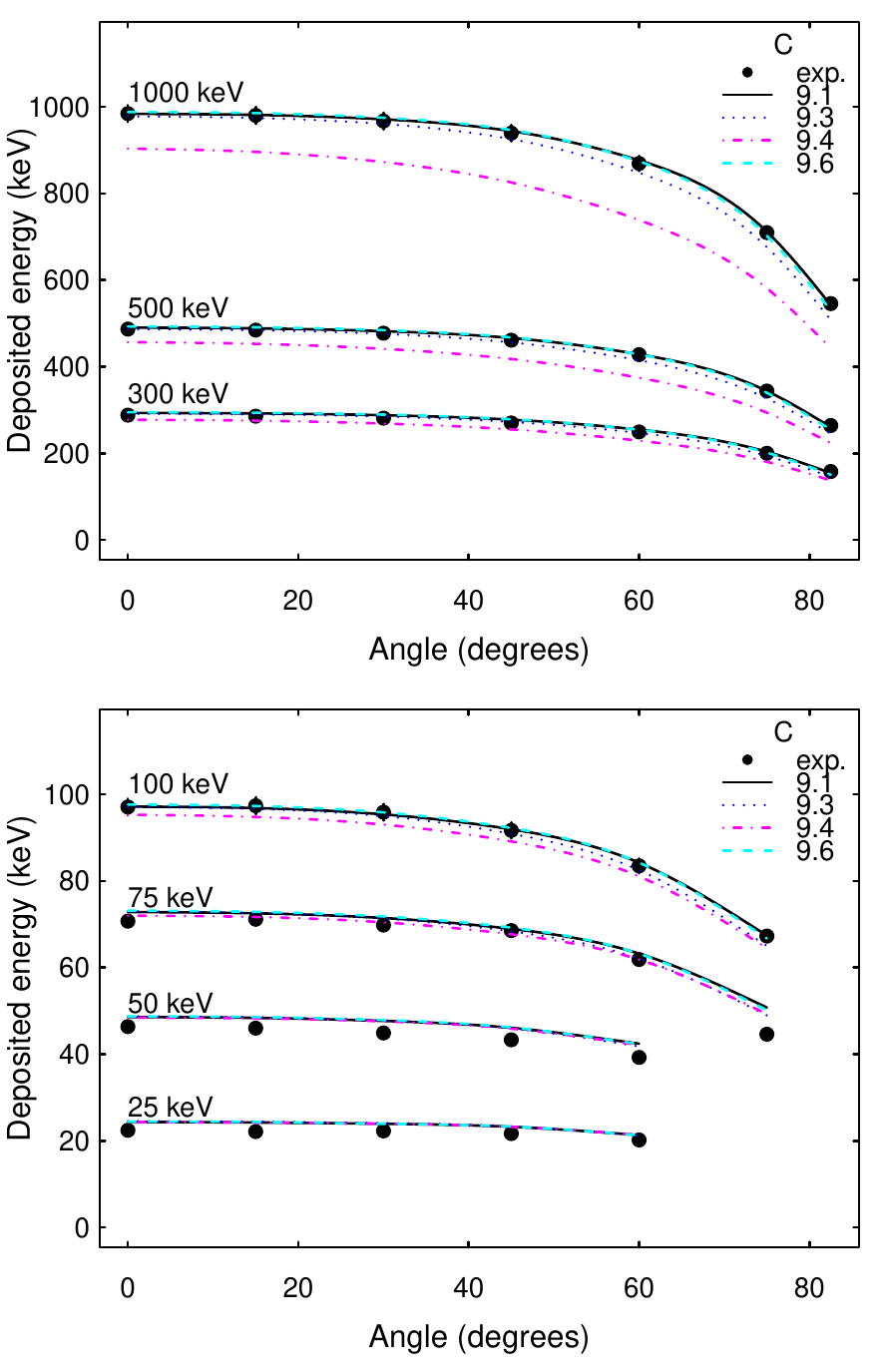}}
\caption{Total energy deposited in carbon: experimental data from
\cite{sandia80} (black dots) and simulations with Geant4 versions 9.1, 9.3, 9.4 and 9.6,
using models based on EEDL and EPDL evaluated data libraries.
The error bars of the experimental data points are not visible, when they are
smaller than the symbol size.
The results of simulations with Geant4 versions 9.2 and 9.5 are not shown for better clarity of the plots, since they 
are very close to those of the 9.6 version.}
\label{fig_C80}
\end{figure}

\begin{figure}
\centerline{\includegraphics[angle=0,width=9cm]{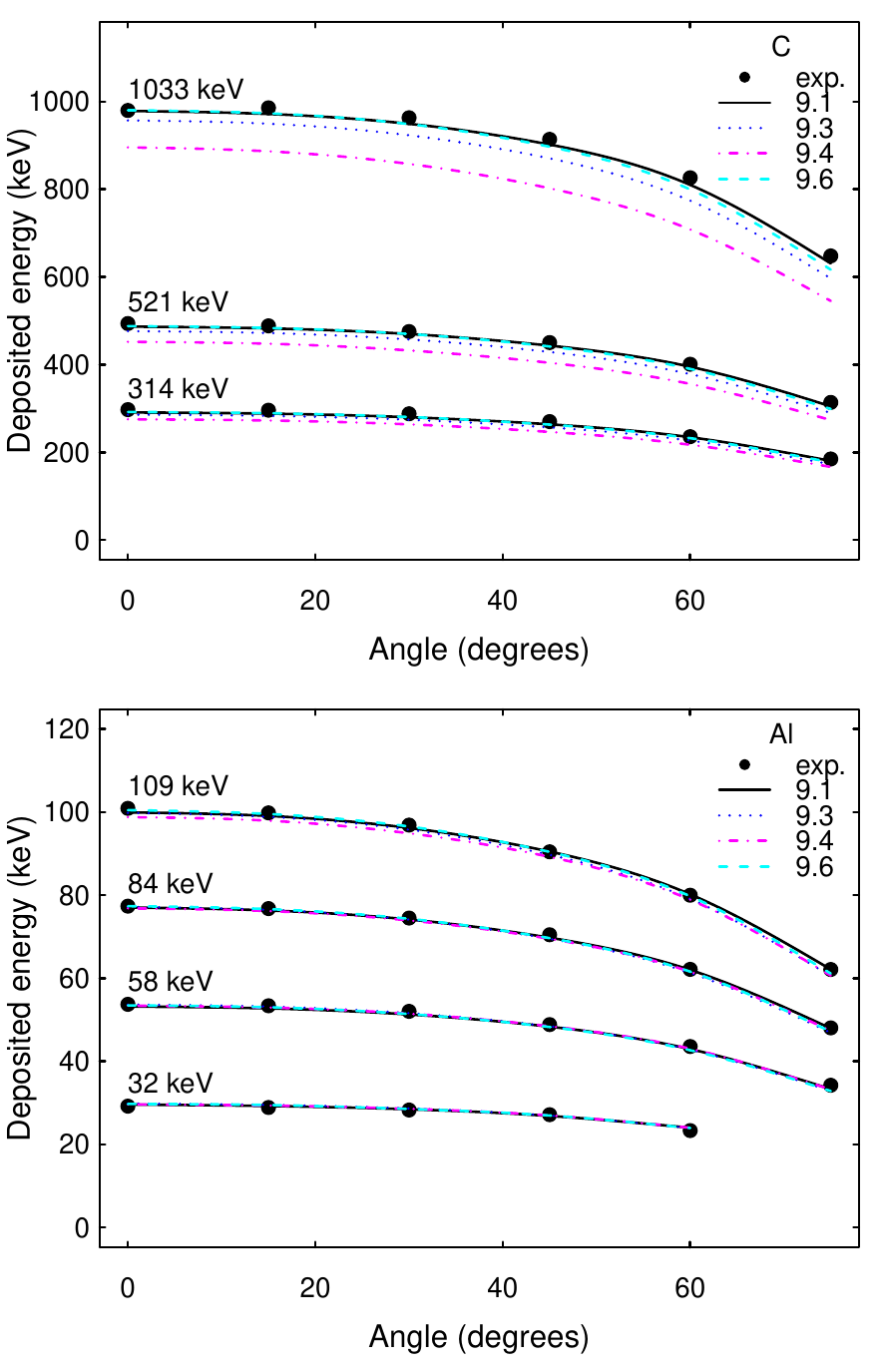}}
\caption{Total energy deposited in aluminium: experimental data from
\cite{sandia80} (black dots) and simulations with Geant4 versions 9.1, 9.3, 9.4 and 9.6,
using models based on EEDL and EPDL evaluated data libraries.
The error bars of the experimental data points are not visible, when they are
smaller than the symbol size.
The results of simulations with Geant4 versions 9.2 and 9.5 are not shown for better clarity of the plots, since they 
are very close to those of the 9.6 version.}
\label{fig_Al80}
\end{figure}

\begin{figure}
\centerline{\includegraphics[angle=0,width=9cm]{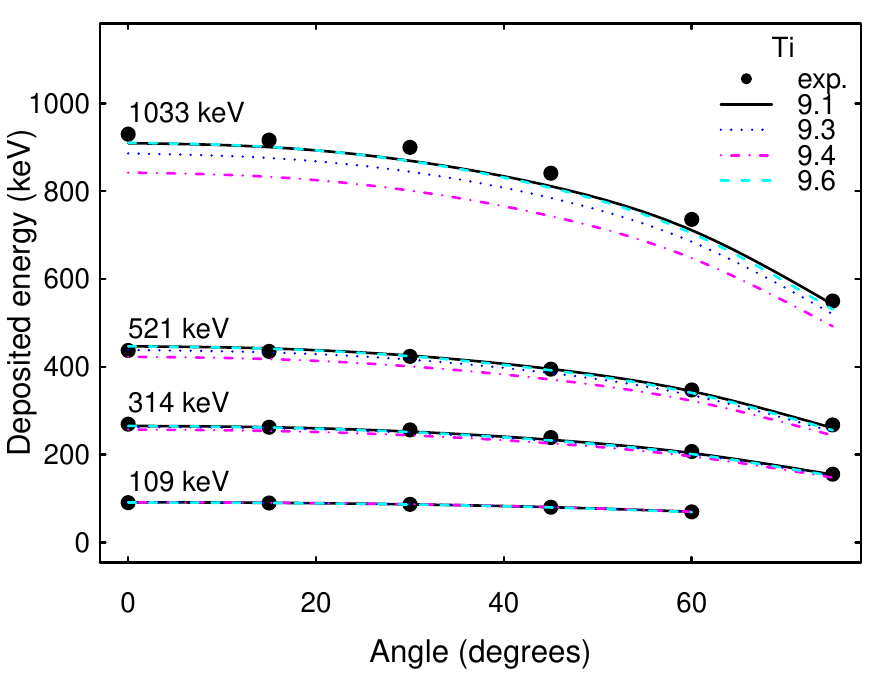}}
\caption{Total energy deposited in titanium: experimental data from
\cite{sandia80} (black dots) and simulations with Geant4 versions 9.1, 9.3, 9.4 and 9.6,
using models based on EEDL and EPDL evaluated data libraries.
The error bars of the experimental data points are not visible, when they are
smaller than the symbol size.
The results of simulations with Geant4 versions 9.2 and 9.5 are not shown for better clarity of the plot, since they 
are very close to those of the 9.6 version.}
\label{fig_Ti80}
\end{figure}

\begin{figure}[p] 
\centerline{\includegraphics[angle=0,width=9cm]{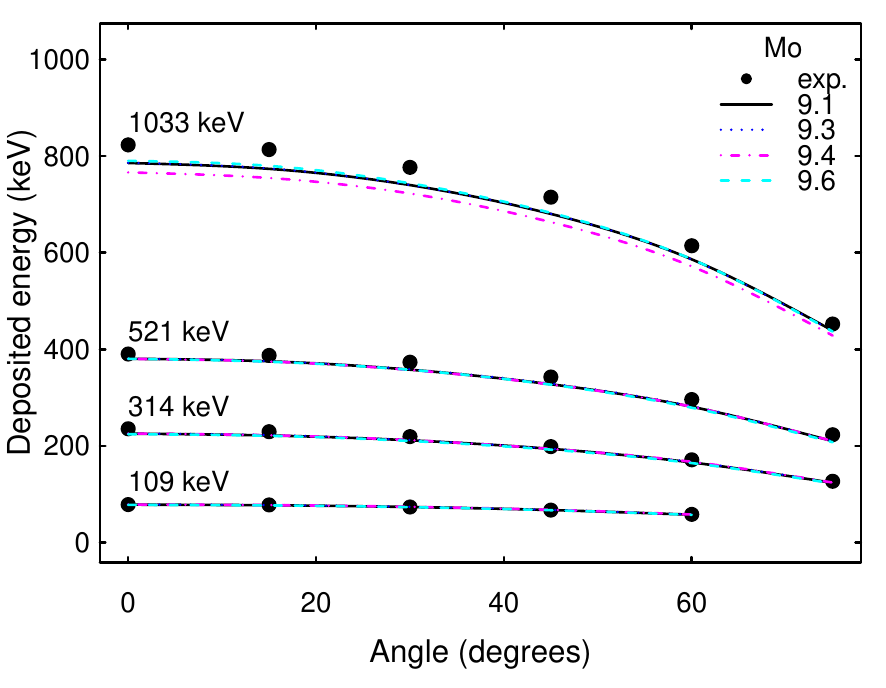}}
\caption{Total energy deposited in molybdenum: experimental data from
\cite{sandia80} (black dots) and simulations with Geant4 versions 9.1, 9.3, 9.4 and 9.6,
using models based on EEDL and EPDL evaluated data libraries.
The error bars of the experimental data points are not visible, when they are
smaller than the symbol size.
The results of simulations with Geant4 versions 9.2 and 9.5 are not shown for better clarity of the plot, since they 
are very close to those of the 9.6 version.}
\label{fig_mo80}
\end{figure}

\begin{figure}
\centerline{\includegraphics[angle=0,width=9cm]{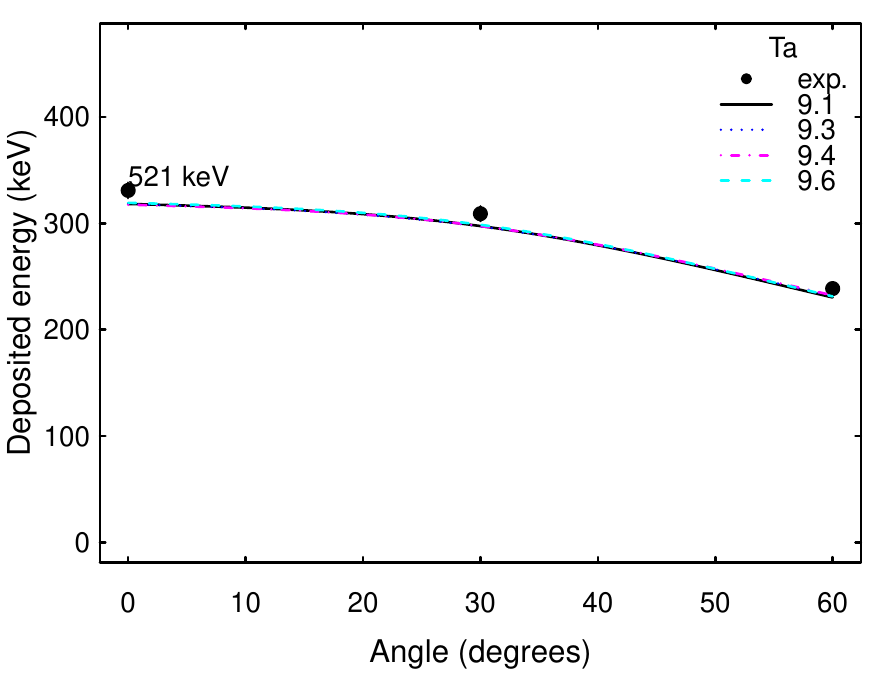}}
\caption{Total energy deposited in tantalum: experimental data from
\cite{sandia80} (black dots) and simulations with Geant4 versions 9.1, 9.3, 9.4 and 9.6,
using models based on EEDL and EPDL evaluated data libraries.
The error bars of the experimental data points are not visible, when they are
smaller than the symbol size.
The results of simulations with Geant4 versions 9.2 and 9.5 are not shown for better clarity of the plot, since they 
are very close to those of the 9.6 version.}
\label{fig_Ta80}
\end{figure}

\begin{figure}
\centerline{\includegraphics[angle=0,width=9cm]{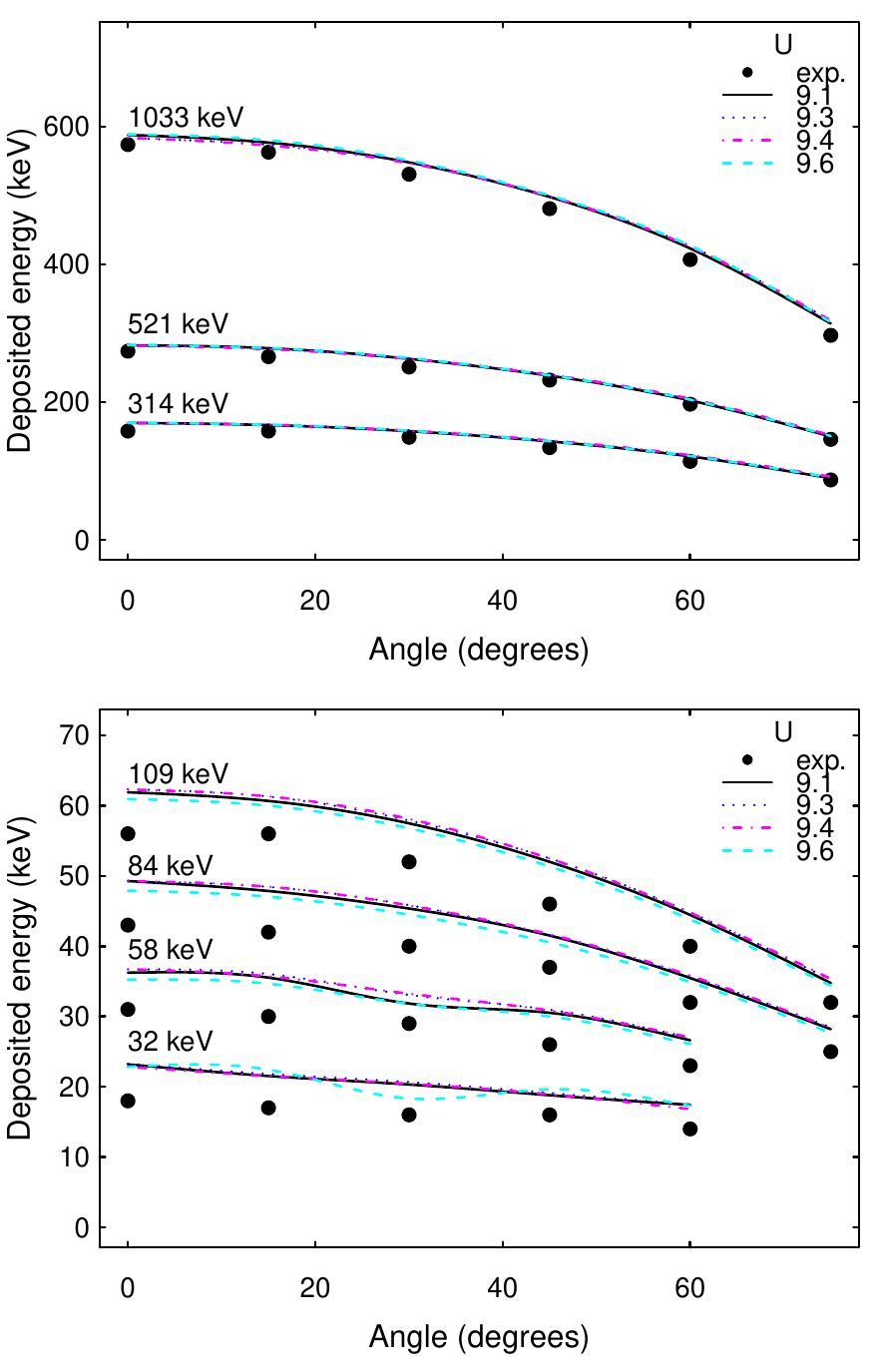}}
\caption{Total energy deposited in uranium: experimental data from
\cite{sandia80} (black dots) and simulations with Geant4 versions 9.1, 9.3, 9.4 and 9.6,
using models based on EEDL and EPDL evaluated data libraries.
The error bars of the experimental data points are not visible, when they are
smaller than the symbol size.
The results of simulations with Geant4 versions 9.2 and 9.5 are not shown for better clarity of the plots, since they 
are very close to those of the 9.6 version.}
\label{fig_U80}
\end{figure}

The total deposited energy produced by simulations with the electron-photon
models based on EEDL-EPDL evaluated data libraries and default multiple
scattering settings are shown in
Fig.~\ref{fig_Be80}-\ref{fig_U80} for all test cases and Geant4 versions.
The plots also report the experimental data of \cite{sandia80}.

The p-values resulting from the $\chi^2$ test over all experimental configurations
are listed in Tables \ref{tab_pvalue80_liv}, \ref{tab_pvalue80_pen} and
\ref{tab_pvalue80_std} for Geant4 electron-photon interaction
models based on EEDL-EPDL, originating from Penelope and implemented in the
``standard'' electromagnetic package respectively.
The number of test cases for which the hypothesis of equivalent simulated and 
experimental total energy deposition is rejected or not rejected by the 
$\chi^2$ test are summarized in Table \ref{tab_pass80}.

\begin{figure} [htbp]
\centerline{\includegraphics[angle=0,width=9cm]{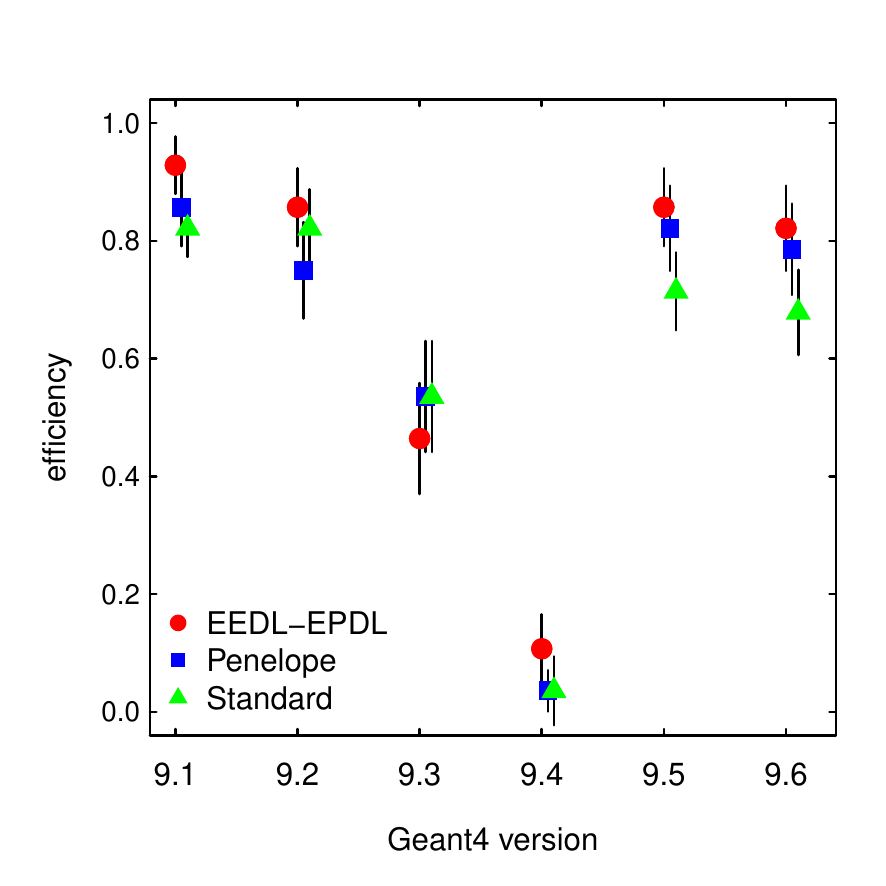}}
\caption{Efficiency of Geant4 simulation configurations for reproducing
experimental total deposited energy,
as a function of Geant4
version. The efficiency is shown for three sets of Geant4 electron-photon models: based on EEDL-EPDL
evaluated data libraries (red circles), originating from Penelope (blue squares)
and in the "standard" package (green triangles).
%The efficiency is defined as the fraction of test cases in which the $\chi^2$
%test does not reject the null hypothesis at 0.01 level of significance.
The symbols representing the data points are slightly shifted along the x-axis
to improve the clarity of the plot.}
\label{fig_eff80}
\end{figure}

The efficiency of different Geant4 configurations is plotted in Fig.~\ref{fig_eff80} as a
function of Geant4 version for the three  electron-photon model option
examined in this paper. 
%The number of test cases that pass the $\chi^2$ test, i.e. with p-value greater
%than the defined 0.01 significance level, is reported in Table~\ref{tab_pass80}.

\tabcolsep=2pt
% Table generated by Excel2LaTeX from sheet 'p-value'
\begin{table} [htbp]
  \centering
  \caption{P-value of the $\chi^{2}$ tests for total energy deposition: simulation with electron-photon models based on EEDL-EPDL}
    \begin{tabular}{lrc|cccccc}
    \hline
    \multicolumn{2}{c}{\textbf{Target}}  & \textbf{Angle} & \multicolumn{6}{c}{\textbf{Geant4 version}} \\
          &Z       &{\footnotesize (degrees)}  & \textbf{9.1} & \textbf{9.2} & \textbf{9.3} & \textbf{9.4} & \textbf{9.5} & \textbf{9.6} \\
\hline
    Be    & 4 & 1	&0.479	&0.436	&0.603	& $<$0.001	&0.386	&0.388 \\
    Be    & 4 & 16  & 0.991 & 0.982 & 0.988 & $<$0.001 & 0.944 & 0.944 \\
    Be    & 4 & 31 & 0.876 & 0.911 & 0.438 & $<$0.001 & 0.905 & 0.903 \\
    Be    & 4 & 46 & 0.776 & 0.893 & 0.012 & $<$0.001 & 0.884 & 0.873 \\
    Be    & 4 & 61  & 0.791 & 0.961 & $<$0.001 & $<$0.001 & 0.765 & 0.764 \\
    Be    & 4 & 76  & 0.008 & 0.040 & $<$0.001 & $<$0.001 & $<$0.001 & $<$0.001 \\
    Be    & 4 & 83.5   & $<$0.001 & $<$0.001 & $<$0.001 & $<$0.001 & $<$0.001 & $<$0.001 \\
    C     & 6 & 0  & 0.207 & 0.088 & 0.160 & $<$0.001 & 0.086 & 0.078 \\
    C     & 6 & 15  & 0.188 & 0.073 & 0.171 & $<$0.001 & 0.081 & 0.070 \\
    C     & 6 & 30  & 0.062 & 0.009 & 0.073 & $<$0.001 & 0.020 & 0.015 \\
    C     & 6 & 45  & 0.078 & 0.004 & 0.167 & $<$0.001 & 0.043 & 0.026 \\
    C     & 6 & 60  & 0.026 & $<$0.001 & 0.065 & $<$0.001 & 0.037 & 0.006 \\
    C     & 6 & 75  & 0.948 & 0.149 & 0.003 & $<$0.001 & 0.951 & 0.925 \\
    C     & 6 & 82.5   & 0.671 & 0.858 & $<$0.001 & $<$0.001 & 0.027 & 0.017 \\
    Al    & 13 & 0  & 0.670 & 0.849 & 0.010 & $<$0.001 & 0.831 & 0.759 \\
    Al    & 13 & 15  & 0.458 & 0.730 & 0.001 & $<$0.001 & 0.588 & 0.478 \\
    Al    & 13 & 30  & 0.455 & 0.829 & $<$0.001 & $<$0.001 & 0.519 & 0.480 \\
    Al    & 13 & 45  & 0.578 & 0.941 & $<$0.001 & $<$0.001 & 0.338 & 0.272 \\
    Al    & 13 & 60  & 0.725 & 0.781 & $<$0.001 & $<$0.001 & 0.138 & 0.024 \\
    Al    & 13 & 75   & 0.111 & 0.519 & $<$0.001 & $<$0.001 & $<$0.001 & $<$0.001 \\
    Ti    & 22 & 0  & 0.301 & 0.033 & 0.023 & $<$0.001 & 0.356 & 0.341 \\
    Ti    & 22 & 15  & 0.696 & 0.051 & 0.076 & $<$0.001 & 0.762 & 0.749 \\
    Ti    & 22 & 30  & 0.246 & 0.227 & 0.001 & $<$0.001 & 0.185 & 0.177 \\
    Ti    & 22 & 45  & 0.159 & 0.206 & $<$0.001 & $<$0.001 & 0.045 & 0.050 \\
    Ti    & 22 & 60  & 0.214 & 0.223 & $<$0.001 & $<$0.001 & 0.021 & 0.019 \\
    Ti    & 22 & 75  & 0.345 & 0.330 & $<$0.001 & $<$0.001 & 0.002 & 0.002 \\
%    Mo    & 42 & 0 & 0.006 & 0.002 & 0.022 & 0.002 & $<$0.001 & 0.001 & 0.001 \\
%    Mo    & 42 & 15 & 0.009 & 0.001 & 0.020 & 0.003 & $<$0.001 & 0.002 & 0.002 \\
%    Mo    & 42 & 30 & 0.003 & 0.001 & 0.011 & 0.001 & $<$0.001 & $<$0.001 & $<$0.001 \\
%    Mo    & 42 & 45 & 0.001 & 0.001 & 0.011 & 0.002 & $<$0.001 & 0.001 & $<$0.001 \\
%    Mo    & 42 & 60 & $<$0.001 & $<$0.001 & 0.008 & 0.001 & $<$0.001 & $<$0.001 & $<$0.001 \\
%    Mo    & 42 & 75 & $<$0.001 & 0.001 & 0.014 & 0.001 & $<$0.001 & $<$0.001 & $<$0.001 \\
   Ta   & 73 & 0 		& 0.066	& 0.176	& 0.055	& 0.039	& 0.129	& 0.111 \\
   Ta   & 73 & 30 		& 0.092	& 0.164	& 0.080	& 0.092	& 0.116	& 0.131 \\
   Ta   & 73 & 60	 	& 0.093	& 0.182	& 0.189	& 0.226	& 0.163	& 0.158 \\
%    U     & 92 & 0 & $<$0.001 & $<$0.001 & $<$0.001 & $<$0.001 & $<$0.001 & $<$0.001 & $<$0.001 \\
%    U     & 92 & 15 & $<$0.001 & $<$0.001 & $<$0.001 & $<$0.001 & $<$0.001 & $<$0.001 & $<$0.001 \\
%    U     & 92 & 30 & $<$0.001 & $<$0.001 & $<$0.001 & $<$0.001 & $<$0.001 & $<$0.001 & $<$0.001 \\
%    U     & 92 & 45 & $<$0.001 & $<$0.001 & $<$0.001 & $<$0.001 & $<$0.001 & $<$0.001 & $<$0.001 \\
%    U     & 92 & 60 & $<$0.001 & $<$0.001 & $<$0.001 & $<$0.001 & $<$0.001 & $<$0.001 & $<$0.001 \\
%    U     & 92 & 75 & 0.002 & $<$0.001 & $<$0.001 & $<$0.001 & $<$0.001 & $<$0.001 & $<$0.001 \\
    \hline
    \end{tabular}%
  \label{tab_pvalue80_liv}%
\end{table}%
\tabcolsep=6pt

\tabcolsep=2pt
% Table generated by Excel2LaTeX from sheet 'p-value R'
\begin{table}  [htbp]
  \centering
  \caption{P-value of the $\chi^{2}$ tests for total energy deposition: simulation with Penelope-like electron-photon models}
    \begin{tabular}{lrc|cccccc}
    \hline
   \multicolumn{2}{c}{\textbf{Target}}  & \textbf{Angle} & \multicolumn{6}{c}{\textbf{Geant4 version}} \\
          &Z       & (degrees)  & \textbf{9.1} & \textbf{9.2} & \textbf{9.3} & \textbf{9.4} & \textbf{9.5} & \textbf{9.6} \\
    \hline
    Be    & 4 & 1	&0.482	&0.448	&0.590	& $<$0.001	&0.393	&0.398 \\
    Be    & 4 & 16 & 0.993 & 0.986 & 0.981 & $<$0.001 & 0.949 & 0.948 \\
    Be    & 4 & 31 & 0.883 & 0.919 & 0.426 & $<$0.001 & 0.913 & 0.910 \\
    Be    & 4 & 46 & 0.797 & 0.893 & 0.010 & $<$0.001 & 0.895 & 0.883 \\
    Be    & 4 & 61 & 0.762 & 0.932 & $<$0.001 & $<$0.001 & 0.744 & 0.736 \\
    Be    & 4 & 76 & 0.005 & 0.036 & $<$0.001 & $<$0.001 & $<$0.001 & $<$0.001 \\
    Be    & 4 & 83.5 & $<$0.001 & $<$0.001 & $<$0.001 & $<$0.001 & $<$0.001 & $<$0.001 \\
    C     & 6 & 0 & 0.206 & 0.104 & 0.190 & $<$0.001 & 0.082 & 0.074 \\
    C     & 6 & 15 & 0.189 & 0.093 & 0.204 & $<$0.001 & 0.074 & 0.070 \\
    C     & 6 & 30 & 0.053 & 0.013 & 0.104 & $<$0.001 & 0.018 & 0.014 \\
    C     & 6 & 45 & 0.067 & 0.008 & 0.226 & $<$0.001 & 0.036 & 0.022 \\
    C     & 6 & 60 & 0.018 & $<$0.001 & 0.060 & $<$0.001 & 0.026 & 0.005 \\
    C     & 6 & 75 & 0.892 & 0.214 & 0.001 & $<$0.001 & 0.977 & 0.975 \\
    C     & 6 & 82.5 & 0.845 & 0.929 & $<$0.001 & $<$0.001 & 0.062 & 0.048 \\
    Al    & 13 & 0 & 0.795 & 0.868 & 0.024 & $<$0.001 & 0.890 & 0.828 \\
    Al    & 13 & 15 & 0.612 & 0.739 & 0.003 & $<$0.001 & 0.725 & 0.626 \\
    Al    & 13 & 30 & 0.669 & 0.851 & 0.001 & $<$0.001 & 0.736 & 0.686 \\
    Al    & 13 & 45 & 0.827 & 0.971 & $<$0.001 & $<$0.001 & 0.586 & 0.610 \\
    Al    & 13 & 60 & 0.797 & 0.613 & $<$0.001 & $<$0.001 & 0.317 & 0.115 \\
    Al    & 13 & 75 & 0.499 & 0.834 & $<$0.001 & $<$0.001 & $<$0.001 & $<$0.001 \\
    Ti    & 22 & 0 & 0.074 & 0.008 & 0.069 & $<$0.001 & 0.100 & 0.087 \\
    Ti    & 22 & 15 & 0.113 & 0.009 & 0.146 & $<$0.001 & 0.184 & 0.149 \\
    Ti    & 22 & 30 & 0.334 & 0.094 & 0.059 & $<$0.001 & 0.534 & 0.465 \\
    Ti    & 22 & 45 & 0.334 & 0.077 & 0.025 & $<$0.001 & 0.493 & 0.484 \\
    Ti    & 22 & 60 & 0.440 & 0.048 & 0.032 & $<$0.001 & 0.720 & 0.651 \\
    Ti    & 22 & 75 & 0.451 & 0.096 & 0.111 & $<$0.001 & 0.704 & 0.712 \\
%    Mo    & 42 & 0 & 0.389 & 0.182 & 0.295 & 0.042 & 0.589 & 0.502 \\
%    Mo    & 42 & 15 & 0.623 & 0.296 & 0.513 & 0.056 & 0.879 & 0.818 \\
%    Mo    & 42 & 30 & 0.380 & 0.130 & 0.277 & 0.024 & 0.577 & 0.479 \\
%    Mo    & 42 & 45 & 0.295 & 0.056 & 0.140 & 0.012 & 0.505 & 0.344 \\
%    Mo    & 42 & 60 & 0.706 & 0.268 & 0.471 & 0.123 & 0.793 & 0.708 \\
%    Mo    & 42 & 75 & 0.526 & 0.133 & 0.444 & 0.382 & 0.426 & 0.367 \\
    Ta    & 73 & 0 & 0.874 & 0.487 & 0.861 & 0.842 & 0.805 & 0.824 \\
    Ta    & 73 & 30 & 0.924 & 0.679 & 0.813 & 0.890 & 0.751 & 0.728 \\
    Ta    & 73 & 60 & 0.851 & 0.681 & 0.601 & 0.623 & 0.706 & 0.755 \\
%    U     & 92 & 0 & $<$0.001 & $<$0.001 & $<$0.001 & $<$0.001 & $<$0.001 & $<$0.001 \\
%    U     & 92 & 15 & $<$0.001 & $<$0.001 & $<$0.001 & $<$0.001 & $<$0.001 & $<$0.001 \\
%    U     & 92 & 30 & $<$0.001 & $<$0.001 & $<$0.001 & $<$0.001 & $<$0.001 & $<$0.001 \\
%    U     & 92 & 45 & $<$0.001 & $<$0.001 & $<$0.001 & $<$0.001 & $<$0.001 & $<$0.001 \\
%    U     & 92 & 60 & $<$0.001 & $<$0.001 & $<$0.001 & $<$0.001 & $<$0.001 & $<$0.001 \\
%    U     & 92 & 75 & $<$0.001 & $<$0.001 & $<$0.001 & $<$0.001 & $<$0.001 & $<$0.001 \\
    \hline
    \end{tabular}%
  \label{tab_pvalue80_pen}%
\end{table}%
\tabcolsep=6pt

\tabcolsep=2pt
% Table generated by Excel2LaTeX from sheet 'p-value R'
\begin{table} [htbp]
  \centering
  \caption{P-value of the $\chi^{2}$ tests for total energy deposition: simulation with standard electron-photon models}
    \begin{tabular}{lrc|cccccc}
    \hline
   \multicolumn{2}{c}{\textbf{Target}}  & \textbf{Angle} & \multicolumn{6}{c}{\textbf{Geant4 version}} \\
          &Z       & (degrees)  & \textbf{9.1} & \textbf{9.2} & \textbf{9.3} & \textbf{9.4} & \textbf{9.5} & \textbf{9.6} \\
    \hline
    Be    & 4 & 1	& 0.466	& 0.438	&0.580	& $<$0.001	& 0.390	& 0.390 \\
    Be    & 4 & 16 & 0.992 & 0.985 & 0.986 & $<$0.001 & 0.941 & 0.941 \\
    Be    & 4 & 31 & 0.909 & 0.923 & 0.505 & $<$0.001 & 0.910 & 0.913 \\
    Be    & 4 & 46 & 0.891 & 0.931 & 0.021 & $<$0.001 & 0.938 & 0.930 \\
    Be    & 4 & 61 & 0.954 & 0.996 & $<$0.001 & $<$0.001 & 0.898 & 0.903 \\
    Be    & 4 & 76 & 0.090 & 0.213 & $<$0.001 & $<$0.001 & 0.002 & 0.001 \\
    Be    & 4 & 83.5 & $<$0.001 & $<$0.001 & $<$0.001 & $<$0.001 & $<$0.001 & $<$0.001 \\
    C     & 6 & 0 & 0.165 & 0.115 & 0.180 & $<$0.001 & 0.062 & 0.059 \\
    C     & 6 & 15 & 0.149 & 0.096 & 0.191 & $<$0.001 & 0.054 & 0.052 \\
    C     & 6 & 30 & 0.032 & 0.016 & 0.092 & $<$0.001 & 0.009 & 0.007 \\
    C     & 6 & 45 & 0.031 & 0.011 & 0.210 & $<$0.001 & 0.013 & 0.008 \\
    C     & 6 & 60 & 0.003 & $<$0.001 & 0.072 & $<$0.001 & 0.005 & $<$0.001 \\
    C     & 6 & 75 & 0.582 & 0.167 & 0.005 & $<$0.001 & 0.955 & 0.946 \\
    C     & 6 & 82.5 & 0.950 & 0.808 & $<$0.001 & $<$0.001 & 0.206 & 0.224 \\
    Al    & 13 & 0 & 0.818 & 0.828 & 0.023 & $<$0.001 & 0.840 & 0.731 \\
    Al    & 13 & 15 & 0.718 & 0.793 & 0.004 & $<$0.001 & 0.653 & 0.500 \\
    Al    & 13 & 30 & 0.801 & 0.901 & 0.002 & $<$0.001 & 0.822 & 0.753 \\
    Al    & 13 & 45 & 0.925 & 0.979 & $<$0.001 & $<$0.001 & 0.745 & 0.756 \\
    Al    & 13 & 60 & 0.790 & 0.596 & $<$0.001 & $<$0.001 & 0.463 & 0.046 \\
    Al    & 13 & 75 & 0.702 & 0.899 & $<$0.001 & $<$0.001 & 0.003 & $<$0.001 \\
    Ti    & 22 & 0 & 0.046 & 0.017 & 0.054 & $<$0.001 & 0.070 & 0.060 \\
    Ti    & 22 & 15 & 0.064 & 0.016 & 0.120 & $<$0.001 & 0.122 & 0.117 \\
    Ti    & 22 & 30 & 0.284 & 0.143 & 0.055 & $<$0.001 & 0.446 & 0.444 \\
    Ti    & 22 & 45 & 0.232 & 0.141 & 0.042 & $<$0.001 & 0.502 & 0.488 \\
    Ti    & 22 & 60 & 0.240 & 0.097 & 0.056 & $<$0.001 & 0.729 & 0.655 \\
    Ti    & 22 & 75 & 0.316 & 0.068 & 0.253 & $<$0.001 & 0.805 & 0.799 \\
%    Mo    & 42 & 0 & 0.151 & 0.065 & 0.069 & 0.024 & 0.232 & 0.206 \\
%    Mo    & 42 & 15 & 0.369 & 0.167 & 0.158 & 0.048 & 0.508 & 0.453 \\
%    Mo    & 42 & 30 & 0.157 & 0.059 & 0.046 & 0.017 & 0.214 & 0.130 \\
%    Mo    & 42 & 45 & 0.073 & 0.013 & 0.016 & 0.006 & 0.154 & 0.083 \\
%    Mo    & 42 & 60 & 0.371 & 0.118 & 0.132 & 0.078 & 0.497 & 0.254 \\
%    Mo    & 42 & 75 & 0.232 & 0.080 & 0.183 & 0.413 & 0.446 & 0.283 \\
    Ta    & 73 & 0 & 0.686 & 0.411 & 0.685 & 0.760 & 0.654 & 0.676 \\
    Ta    & 73 & 30 & 0.589 & 0.514 & 0.586 & 0.723 & 0.574 & 0.654 \\
    Ta    & 73 & 60 & 0.525 & 0.310 & 0.247 & 0.344 & 0.612 & 0.580 \\
%    U     & 92 & 0 & $<$0.001 & $<$0.001 & $<$0.001 & $<$0.001 & $<$0.001 & $<$0.001 \\
%    U     & 92 & 15 & $<$0.001 & $<$0.001 & $<$0.001 & $<$0.001 & $<$0.001 & $<$0.001 \\
%    U     & 92 & 30 & $<$0.001 & $<$0.001 & $<$0.001 & $<$0.001 & $<$0.001 & $<$0.001 \\
%    U     & 92 & 45 & $<$0.001 & $<$0.001 & $<$0.001 & $<$0.001 & $<$0.001 & $<$0.001 \\
%    U     & 92 & 60 & $<$0.001 & $<$0.001 & $<$0.001 & $<$0.001 & $<$0.001 & $<$0.001 \\
%    U     & 92 & 75 & $<$0.001 & $<$0.001 & $<$0.001 & $<$0.001 & $<$0.001 & $<$0.001 \\
    \hline
    \end{tabular}%
  \label{tab_pvalue80_std}%
\end{table}%
\tabcolsep=6pt

% Table generated by Excel2LaTeX from sheet 'p-value R'
\begin{table}[htbp]
  \centering
  \caption{Number of test cases of total energy deposition that pass the $\chi^2$ test }
    \begin{tabular}{llcccccc}
    \hline
    &  & \multicolumn{6}{c}{\textbf{Geant4 version}} \\
    & \textbf{Geant4 models} & \textbf{9.1} & \textbf{9.2} & \textbf{9.3} & \textbf{9.4} & \textbf{9.5} & \textbf{9.6} \\
    \hline
    \multicolumn{1}{c}{\multirow{3}[0]{*}{Pass}} & EEDL-EPDL & 27    & 25    & 14    & 3     & 25    & 24 \\
    \multicolumn{1}{c}{} & Penelope & 27    & 24    & 18    & 3     & 26    & 25 \\
    \multicolumn{1}{c}{} & Standard & 27    & 27    & 19    & 3     & 24    & 23 \\
\hline
    \multicolumn{1}{c}{\multirow{3}[0]{*}{Fail}} & EEDL-EPDL & 2     & 4     & 15    & 26    & 4     & 5 \\
    \multicolumn{1}{c}{} & Penelope & 2     & 5     & 11    & 26    & 3     & 4 \\
    \multicolumn{1}{c}{} & Standard & 2     & 2     & 10    & 26    & 5     & 6 \\
    \hline
   \end{tabular}%
  \label{tab_pass80}%
\end{table}%

According to the results summarized in Fig.~\ref{fig_eff80}, the three options
of Geant4 electron-photon models exhibit a similar behaviour in all Geant4
versions regarding the total energy deposited in the calorimeter.
This results is confirmed by the statistical analysis comparing the 
outcome of the $\chi^2$ test with the same method as applied in 
section\ref{sec_sandia79_mod}: all statistical tests applied to contingency
tables comparing the three Geant4 electron-photon options produce
p-values greater than the significance level of 0.01.

The efficiency appears equivalent in all versions with the exception of Geant4
9.4, for which it is significantly lower, and Geant4 9.3, which produces an
intermediate result between the low efficiency of the 9.4 version and the higher
values of the other versions subject to test.
The results of the statistical analysis related to the evolution of the compatibility with 
experimental data over Geant4 versions are listed in Table \ref{tab_mcnemar80} 
for the physics configuration with EEDL-EPDL electron-photon settings.
Table \ref{tab_mcnemar80} reports the comparison of the outcome of the $\chi^2$
test with Geant4 9.1 with the outcome of subsequent versions; the data samples
are compared by means of McNemar's exact test only, since the use of the
$\chi^2$ asymptotic distribution for McNemar's test calculation would not be
justified due to the low number of entries in some cells of the 2x2 tables involved
in this evaluation.

% Table generated by Excel2LaTeX from sheet 'Conting R'
\begin{table}[htbp]
  \centering
  \caption{P-value of McNemar's test comparing the compatibility with experimental
total deposited energy simulated by Geant4 9.1 and later
versions}
    \begin{tabular}{lccccc}
    \hline
          & \multicolumn{5}{c}{\textbf{Geant4 version compared to 9.1}} \\
    \textbf{McNemar test } & \textbf{9.2} & \textbf{9.3} & \textbf{9.4} & \textbf{9.5} & \textbf{9.6} \\
    \hline
    Exact & 0.625 & $<$0.001 & $<$0.001 & 0.500 & 0.500 \\
%    Exact 1-side & 0.313 & $<$0.001 & $<$0.001 & 0.250 & 0.125 \\
    \hline
    \end{tabular}%
  \label{tab_mcnemar80}%
\end{table}%

These results suggest that differences due to the intrinsic capabilities of the
physics models and empirical algorithms implemented in Geant4
%which are visible in the longitudinal energy deposition profiles, 
are mitigated by
the characteristics of the geometrical configuration where the deposited energy
is scored: in the two validation scenarios evaluated in this paper, statistically significant 
differences in compatibility with experimental data 
associated with Geant4 electron-photon models and versions are
observed in the simulations corresponding to the experimental set-up
of \cite{sandia79}, which involves thin calorimeter layers,
while they are less visible in the coarser detector set-up of \cite{sandia80}.

\begin{figure} 
\centerline{\includegraphics[angle=0,width=9cm]{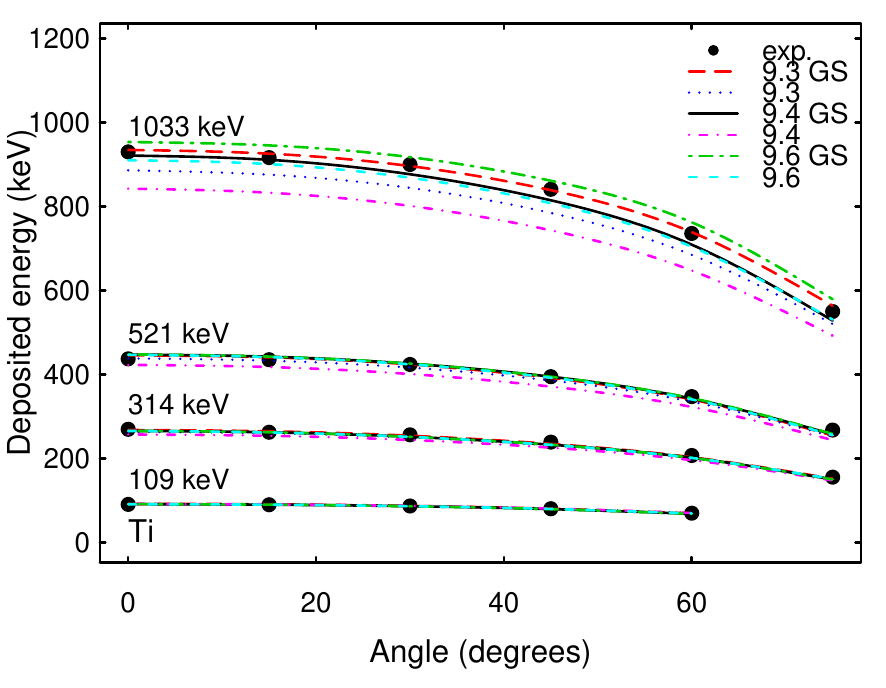}}
\caption{Total energy deposited in titanium produced with two multiple scattering models implemented in Geant4: Goudsmit-Saunderson and variants of the Urban model. The plot shows experimental data from
\cite{sandia80} (black dots) and simulations with Geant4 versions  9.3, 9.4 and 9.6,
using models based on EEDL and EPDL evaluated data libraries.
The error bars of the experimental data points are not visible, when they are
smaller than the symbol size.
The results of simulations with Geant4 version 9.5 are not shown for better clarity of the plot, since they 
are very close to those of the 9.6 version.}
\label{fig_Ti80GS}
\end{figure}

The replacement of the default ``Urban'' multiple scattering model with 
the Goudsmit-Saunderson model leads to different conclusions with respect
to what is observed in the thin layer scenario discussed in section 
\ref{sec_goudsmit}.
The p-values resulting from the $\chi^2$ test comparing the simulation results
with Goudsmit-Saunderson multiple scattering model to the experimental total
deposited energy of \cite{sandia80} are listed in Table \ref{tab_pvgoudsmit80},
and the associated ``efficiency'' is reported in Table \ref{tab_effGS80}.
One can observe that the compatibility with experimental data is approximately
constant over the examined Geant4 versions, when the Goudsmit-Saunderson
multiple scattering algorithm is used in the simulation, while large
discrepancies are observed with the variants of the Urban model corresponding to
different Geant4 versions.
According to Table \ref{tab_effGS80}, better compatibility with experimental
data is achieved with the Urban95 model used by default in Geant4 9.5 and
9.6, while the Goudsmit-Saunderson multiple scattering model provides more
accurate total deposited energy simulations with Geant4 9.3 and 9.4.
An example of the total deposited energy produced with the Goudsmit-Saunderson 
and variants of the Urban model is illustrated in Fig. \ref{fig_Ti80GS}.

The statistical significance of these qualitative observations is summarized in
Table \ref{tab_mcnemar80GS} for simulations using the EEDL-EPDL
electron-photon settings: the two models contribute to a
different compatibility with total deposited energy measurements in the Geant4
9.4 environment at a 0.01 significance level, while in the other environments the
associated compatibility with experimental data is statistically equivalent.

% Table generated by Excel2LaTeX from sheet 'p-value R'
\begin{table}[htbp]
  \centering
  \caption{P-values of the $\chi^2$ test for total deposited energy: simulations with electron-photon models based on EEDL-EPDL and Goudsmit-Saunderson multiple scattering model}
    \begin{tabular}{lrc|cccc}
    \hline
    \multicolumn{2}{c}{\textbf{Target}} & \textbf{angle} & \multicolumn{4}{c}{\textbf{Geant4 version}} \\
          & Z     & (degrees) & \textbf{9.3} & \textbf{9.4} & \textbf{9.5} & \textbf{9.6} \\
 \hline
    Be    & 4     & 1 & $<$0.001 & $<$0.001 & $<$0.001 & $<$0.001 \\
    Be    & 4     & 16 & 0.864 & 0.904 & 0.848 & 0.846 \\
    Be    & 4     & 31 & 0.302 & 0.583 & 0.323 & 0.324 \\
    Be    & 4     & 46 & 0.078 & 0.592 & 0.046 & 0.046 \\
    Be    & 4     & 61 & 0.012 & 0.908 & $<$0.001 & $<$0.001 \\
    Be    & 4     & 76 & 0.034 & 0.006 & $<$0.001 & $<$0.001 \\
    Be    & 4     & 83.5 & 0.699 & $<$0.001 & 0.005 & 0.005 \\
    C     & 6     & 0 & 0.026 & 0.017 & 0.010 & 0.010 \\
    C     & 6     & 15 & 0.038 & 0.028 & 0.011 & 0.011 \\
    C     & 6     & 30 & 0.006 & 0.008 & $<$0.001 & $<$0.001 \\
    C     & 6     & 45 & $<$0.001 & 0.005 & $<$0.001 & $<$0.001 \\
    C     & 6     & 60 & $<$0.001 & 0.006 & $<$0.001 & $<$0.001 \\
    C     & 6     & 75 & $<$0.001 & 0.314 & $<$0.001 & $<$0.001 \\
    C     & 6     & 82.5 & $<$0.001 & 0.006 & $<$0.001 & $<$0.001 \\
    Al    & 13    & 0 & 0.033 & 0.449 & 0.060 & 0.060 \\
    Al    & 13    & 15 & 0.096 & 0.550 & 0.170 & 0.170 \\
    Al    & 13    & 30 & 0.007 & 0.251 & 0.017 & 0.013 \\
    Al    & 13    & 45 & $<$0.001 & 0.013 & $<$0.001 & $<$0.001 \\
    Al    & 13    & 60 & $<$0.001 & $<$0.001 & $<$0.001 & $<$0.001 \\
    Al    & 13    & 75 & $<$0.001 & $<$0.001 & $<$0.001 & $<$0.001 \\
    Ti    & 22    & 0 & 0.868 &   0.783    & 0.547 & 0.570 \\
    Ti    & 22    & 15 & 0.916 &   0.956    & 0.589 & 0.600 \\
    Ti    & 22    & 30 & 0.998 &  0.722     & 0.842 & 0.830 \\
    Ti    & 22    & 45 & 0.931 &   0.486    & 0.607 & 0.621 \\
    Ti    & 22    & 60 & 0.807 &  0.245     & 0.231 & 0.261 \\
    Ti    & 22    & 75 & 0.054 & 0.027      & 0.010 & 0.010 \\
    Ta    & 73    & 0 & 0.022 & 0.547 & 0.461 & 0.461 \\
    Ta    & 73    & 30 & 0.053 & 0.005 & 0.005 & 0.005 \\
    Ta    & 73    & 60 & 0.214 & 0.729 & 0.744 & 0.729 \\
    \hline
    \end{tabular}%
  \label{tab_pvgoudsmit80}%
\end{table}%

\tabcolsep=4pt
\begin{table}
  \centering
  \caption{Efficiency for total energy deposition  with Goudsmit-Saunderson and Urban multiple scattering model}
    \begin{tabular}{rcccc}
    \hline
     & \multicolumn{4}{c}{\textbf{Geant4 version}} \\
    \textbf{Model} & \textbf{9.3} & \textbf{9.4} & \textbf{9.5} & \textbf{9.6} \\
 \hline
    Goudsmit & \multirow{2}[0]{*}{0.66 $\pm$ 0.09} & \multirow{2}[0]{*}{0.66 $\pm$ 0.09} & \multirow{2}[0]{*}{0.52 $\pm$ 0.09} & \multirow{2}[0]{*}{0.52 $\pm$ 0.09} \\
    Saunderson &       &       &       &  \\
    Urban & 0.48 $\pm$ 0.09  & 0.10 $\pm$ 0.06  & 0.86 $\pm$ 0.06 & 0.83 $\pm$ 0.07 \\
    \hline
    \end{tabular}%
  \label{tab_effGS80}%
\end{table}%
\tabcolsep=6pt

% Table generated by Excel2LaTeX from sheet 'ContingGS R'
\begin{table}[htbp]
  \centering
  \caption{P-values of McNemar test comparing the compatibility with experimental energy deposition profiles of simulations using Urban or Goudsmit-Saunderson multiple scattering models}
    \begin{tabular}{lcccc}
    \hline
          & \multicolumn{4}{c}{\textbf{Geant4 version}} \\
    \textbf{McNemar test } & \textbf{9.3} & \textbf{9.4} & \textbf{9.5} & \textbf{9.6} \\
   \hline
    Exact  & 0.049 & $<$0.001 & 0.057 & 0.092 \\
    \hline
    \end{tabular}%
  \label{tab_mcnemar80GS}%
\end{table}%

% Table generated by Excel2LaTeX from sheet 'p-value'
%\begin{table}[htbp]
%  \centering
%  \caption{Add caption}
%    \begin{tabular}{lccccccc}
%    \hline
%    \multirow{2}[0]{*}{\textbf{Model} & \multicolumn{7}{c}{\textbf{Geant4 version}} \\
%          & \textbf{8.1 & \textbf{9.1 & \textbf{9.2} & \textbf{9.3} & \textbf{9.4} & \textbf{9.5} & \textbf{9.6} \\
%\hline
%    \textbf{EEDL-EPDL} & 0.82  & 0.93  & 0.86  & 0.46  & 0.11  & 0.86  & 0.82 \\
%    \textbf{Penelope} & 0.75  & 0.86  & 0.75  & 0.54  & 0.04  & 0.82  & 0.79 \\
%    \textbf{Standard} & 0.68  & 0.82  & 0.82  & 0.54  & 0.04  & 0.71  & 0.68 \\
%    \hline
%    \end{tabular}%
%  \label{tab:addlabel}%
%\end{table}%

% -----------------------------------------------------------------------------------------

\section{Conclusion}

The extensive investigation of the capability of Geant4 to simulate the energy deposited
by electrons summarized in this paper updates the results published in \cite{tns_sandia} to reflect
the improvements to Geant4 electromagnetic physics mentioned in
\cite{em_mc2010,em_chep2010,em_radecs2011,em_chep2012}.
Simulations involving different Geant4 physics modeling options and Geant4
versions have been compared with high precision experimental measurements,
concerning electrons of energy up to approximately 1~MeV, various beam angles and
materials.
The validation of Geant4 simulation capabilities is quantified by means of the
statistical analysis of simulated and experimental distributions.

The investigation was concerned with two experimental observables: the longitudinal energy
deposition profile in thin layers (of approximately 5 to 90~$\mu$m thickness) as
a function of penetration depth, and the total energy deposited in larger
volumes (of approximately 0.5-5~mm thickness).

Largely different conclusions can be drawn regarding
the compatibility of the simulation with experimental measurements in the two scenarios.
Simulations involving a coarser geometry exhibit limited sensitivity to
different physics modeling options and the evolution of Geant4 electromagnetic
physics, while the capability of simulating the energy deposited in layers of a 
few tens of micrometers thickness appears to deteriorate in later versions with
respect to the results achieved by Geant4 9.1.
Regarding energy deposition profiles in thin layers, equivalent or better
accuracy of simulations based on later Geant4 versions with respect to 9.1 
is excluded with 0.01 significance.

The implementation of electron multiple scattering in Geant4 significantly affects the 
accuracy of energy deposition.
The Urban93 model appears responsible for degraded accuracy of the energy 
deposition simulation with respect to simulations using other variants of the Urban model.
The Goudsmit-Saunderson model, specialized for the simulation of electron
multiple scattering, is responsible for degraded accuracy with respect to the
Urban95 model in both experimental scenarios; with respect to the Urban92 and
Urban93 models applied by default in Geant4 9.3 and 9.4, it contributes to
comparable inaccuracy in the thin layer scenario, but to less inaccurate results
in the coarse grained one.

Statistically superior accuracy is achieved with Geant4 9.1 by using
electron-photon models based on EEDL-EPDL, rather than other modeling
alternatives, in the simulation of energy deposition profiles in thin layers.
In the same scenario, electron-photon models in Geant4 standard electromagnetic
package and models reengineered from Penelope exhibit statistically equivalent
behavior.
All electron-photon modeling alternatives produce equivalent results in 
simulations of coarse-grained detectors.
No significant difference is associated with implementations reengineered from
Penelope 2001 and 2008.

It is worthwhile to stress that these conclusions pertain to the experimental
scenarios studied in this paper and should not be taken as general reflections 
of the performance of different Geant4 versions and physics models for all experimental scenarios.

From the perspective of using Geant4 in experimental applications dealing with
the energy deposition by electrons in similar scenarios to those considered here, the simulation
requirements of coarse-grained detectors can be satisfied by most of the Geant4
models and versions evaluated in this paper.
Experiments concerned with accurate simulation of the energy deposition patterns
produced by low energy electrons may prefer Geant4 9.1, using EEDL-EPDL based
electron-photon models, with respect to later versions and other physics
options.

The risk of negative improvements as a result of evolutions in Geant4
electromagnetic physics could be mitigated by the adoption of sound software
engineering methods in support of physics modeling in Geant4 development
process.
Best practices in change management and software testing are embedded in
established software process frameworks such as the Unified Model \cite {up},
CMMI (Capability Maturity Model Integration) \cite{cmmi} and ISO~15504
\cite{iso15504},
or documented in specialized standards such as \cite{ieee_vv}.
%established frameworks, such as , support their introduction in existing software projects.
The adoption of a more agile software design of Geant4 electromagnetic physics
\cite{chep2012_refactoring}, characterized by classes with focused
responsibilities and minimal dependencies, facilitates both the validation of
Geant4 physics functionality and the transparency of the related change
management process.
Regular use of statistical methods is recommended to quantify the 
compatibility with experimental measurements in the course of Geant4
evolution.

% ------------------------------------------------------------------------------
\section*{Acknowledgment}
%The authors express their gratitude to CERN for support to the research
%described in this paper.
The Geant4-based simulation application used for this paper 
encompasses code implemented by Anton Lechner and Manju Sudhakar
for the study published in \cite{tns_sandia}.

The authors thank Nicholas Styles for proofreading the manuscript and valuable comments,
Robert Andritschke and Stefanie Granato for helpful discussions.

The CERN Library, in particular Tullio Basaglia, has provided helpful assistance
and essential reference material for this study.

% ------------------------------------------------------------------------------

% ------------------------------------------------------------------------


\begin{thebibliography}{199}

% Sandia

\bibitem{sandia79}
G. Lockwood et al.,
``Calorimetric Measurement of Electron Energy Deposition in Extended Media -
Theory vs Experiment'',
SAND79-0414 Report, Sandia National Laboratories, Albuquerque, 1980.

\bibitem{lockwood73}
G. J. Lockwood, G. H. Miller and J. A.  Halbleib, 
``Absolute Measurement of Low Energy Electron Deposition Profiles 
in Semi-Infinite Geometries''
\emph{IEEE Trans. Nucl. Sci.}, vol. 20, no. 6,  pp. 326-330, 1973.

\bibitem{sandia80}
G. J. Lockwood et al., 
``Electron Energy and Charge Albedos - Calorimetric Measurement vs Monte Carlo Theory”, 
SAND80-1968 Report, Sandia National Laboratories, Albuquerque, 1981.

\bibitem{lockwood75}
G. J. Lockwood, G. H. Miller and J. A.  Halbleib, 
``Simultaneous Integral Measurement of Electron Energy and Charge Albedos ''
\emph{IEEE Trans. Nucl. Sci.}, vol. 22, no. 6,  pp. 2537-2542, 1975.

\bibitem{lockwood76}
G. J. Lockwood, G. H. Miller and J. A.  Halbleib, 
``Electron Energy Deposition in Multilayer Geometries'',
\emph{IEEE Trans. Nucl. Sci.}, vol. 23, no. 6,  pp. 1862-1866 , 1976.

\bibitem{tiger}
J. A. Halbleib and T. A. Mehlhorn,
``ITS : The Integrated TIGER
Series of Coupled Electron/Photon Monte Carlo Transport Codes'',
Sandia National Laboratories Report No. SAND84-0573, Albuquerque,
November 1984.

\bibitem{kawrakow}
I. Kawrakow and D. W. O. Rogers,
``The EGSnrc System, a Status Report''
in \textit{Proc. of the Monte Carlo 2000 Conf.}, Springer, Berlin, 2001.

\bibitem{chibani}
O. Chibani and X. A. Li,
``Monte Carlo dose calculations in homogeneous media and at interfaces: 
A comparison between GEPTS, EGSnrc, MCNP, and measurements'',
\textit{Med. Phys.}, vol. 29, no. 5, pp. 835-847, 2002.

\bibitem{penelopebench} 
J. Sempau, J. M. Fernandez-Varea, E. Acosta and F. Salvat, 
``Experimental benchmarks of the Monte Carlo code Penelope'', 
\emph{Nucl. Instrum. Methods B}, vol. 207, pp. 107-123, 2003.

\bibitem{jun}
S. Jun,
``Benchmark Study for Energy Deposition by Energetic
Electrons in Thick Elemental Slabs: Monte Carlo Results and Experiments'',
\emph{IEEE Trans. Nucl. Sci.}, vol. 50, pp. 1732-1739, 2003.

\bibitem{carrier}
J. F.~Carrier, L.~Archambault and L. Beaulieu,
``Validation of GEANT4, an object-oriented Monte Carlo toolkit, for
simulations in medical physics'',
\textit{Med. Phys.}, vol. 31, no. 3, pp. 484-492, 2004.

\bibitem{ivanchenko}
V. N. Ivanchenko,
``Geant4: physics potential for instrumentation in space and medicine'',
\textit{Nucl. Instrum. Meth. A}, vol. 525, pp 402-405, 2004.

\bibitem{kim}
H. K. Kim and O. Kum,
``Development of a Parallel Electron and Photon Transport (PMCEPT)
Code II: Absorbed Dose Computation in Homogeneous and Heterogeneous
Media'', 
\textit{J. Korean Phys. Society}, 
vol. 49, no. 4, pp. 1640-1651, 2006.

\bibitem{kadri}
O. Kadri et al.,
``Geant4 simulation of electron energy deposition in extended media'',
\textit{Nucl. Instrum. Meth. B}, vol. 258, no. 2, pp. 381-387, 2007.

\bibitem{tns_sandia}
A. Lechner, M. G. Pia, and M. Sudhakar
``Validation of Geant4 low energy electromagnetic processes against precision
measurements of electron energy deposit'',
\textit{IEEE Trans. Nucl. Sci.}, vol. 56, no. 2, pp. 398-416, 2009.

% MC codes and 
% Basic Geant4 references:

\bibitem{egs4}
W. R. Nelson, H. Hirayama, and D. W. O. Rogers, 
``The EGS4 Code System'',
SLAC-265 Report, Stanford, CA, 1985. 

\bibitem{egsnrc}
I. Kawrakow, E. Mainegra-Hing, D. W. O. Rogers, F. Tessier and B. R. B. Walters, 
``The EGSnrc Code System: Monte Carlo
Simulation of Electron and Photon Transport
NRCC PIRS-701, 5th printing, 2010. 

\bibitem{g4nim} 
S.~Agostinelli et al., 
``Geant4 - a simulation toolkit''
\textit{Nucl. Instrum. Meth. A}, vol. 506, no. 3, pp. 250-303, 2003.

\bibitem{g4tns}
J.~Allison et al., 
``Geant4 Developments and Applications'' 
\textit{IEEE Trans. Nucl. Sci.}, vol. 53, no. 1, pp. 270-278, 2006.

\bibitem{mcnp}
X-5 Monte Carlo Team, 
``MCNP -- A General Monte Carlo N-Particle Transport Code'', 
Los Alamos National Laboratory Report LA-UR-03-1987,  2003.

\bibitem{mcnpx}
L. Waters et al., 
``The MCNPX Monte Carlo Radiation Transport Code'', 
in \textit{AIP Conf. Proc.}, vol. 896, pp. 81-90, 2006.

\bibitem{penelope}
J.~Baro, J.~Sempau, J. M.~Fernandez-Varea, and F.~Salvat, 
``PENELOPE, an algorithm for Monte Carlo simulation of the penetration and
energy loss of electrons and positrons in matter'',
\emph{Nucl. Instrum. Meth. B}, vol. 100, no. 1, pp. 31-46, 1995.

% EM improvements

\bibitem{em_mc2010}
V. N. Ivanchenko et al.,
``Recent Improvements in Geant4 Electromagnetic Physics Models and Interfaces'',
\textit{Progr. in Nucl. Sci Technol.}, vol. 2, pp.898-903, 2011.

\bibitem{em_chep2010}
A. Sch\"{a}licke et al., 
``Geant4 electromagnetic physics for the LHC and other HEP applications'',
\textit{J. Phys.: Conf. Ser.} vol. 331, pp. 032029, 2011.

\bibitem{em_radecs2011}
J. Allison et al.,
``New Geant4 model and interface developments for improved space electron transport simulations: First results'',
in \textit{Proc. 12th Eur. Conf. Radiation and Its Effects on Components and Systems (RADECS)}
pp. 115-118, 2011.

\bibitem{em_chep2012}
J. Allison et al.,
``Geant4 electromagnetic physics for high statistic simulation of LHC experiments'',
\textit{J. Phys.: Conf. Ser.}, vol. 396, pp. 022013, 2012.

\bibitem{kadri_goudsmit}
O. Kadri, V. Ivanchenkob, F. Gharbi, and A. Trabelsi,
``Incorporation of the Goudsmit-Saunderson electron transport theory in the Geant4 Monte Carlo code'',
\textit{Nucl. Instrum. Meth. B}, vol. 267, no. 23-24, pp. 3624-3632, 2009.

% Databases

\bibitem{eedl}
S.~T.~Perkins, D.~E.~Cullen, and S.~M.~Seltzer,
``Tables and Graphs of Electron-Interaction Cross
Sections from 10 eV to 100 GeV Derived from the LLNL Evaluated
Electron Data Library (EEDL)'', 
UCRL-50400 Vol. 31, 1997.

\bibitem{epdl97}	
D. Cullen et al., 
``EPDL97, the Evaluated Photon Data Library'', 
Lawrence Livermore National Laboratory Report UCRL-50400, Vol. 6, Rev. 5, 1997.


% Geant4 LowE

\bibitem{lowe_e} 
J. Apostolakis, S. Giani, M. Maire, P. Nieminen, M.G. Pia, L. Urban,
``Geant4 low energy electromagnetic models for electrons and photons''
\textit{INFN/AE-99/18}, Frascati, 1999. 

\bibitem{lowe_chep}
S. Chauvie, G. Depaola, V. Ivanchenko, F. Longo, P. Nieminen and M. G. Pia,
``Geant4 Low Energy Electromagnetic Physics'',
in \textit{Proc. Computing in High Energy and Nuclear Physics}, 
Beijing, China, pp. 337-340, 2001.

\bibitem{lowe_nss}
S. Chauvie et al., 
``Geant4 Low Energy Electromagnetic Physics'',
in \textit{2004 IEEE Nucl. Sci. Symp. Conf. Rec.}, pp. 1881-1885, 2004.

% Geant4 standard EM
\bibitem{standard}
V.~N.~Ivanchenko, M.~Maire, and L.~Urban, 
``Geant4 Standard electromagnetic package for HEP applications'', 
\emph{Conf. Rec. 2004 IEEE Nuclear Science Symposium}, N33-179.

% NIST
\bibitem{csda}
M. J. Berger, J. S. Coursey, M. A. Zucker and J. Chang,
``Significance of Calculated Quantities'', [Online].
Available: http://physics.nist.gov/PhysRefData/Star/Text/appendix.html.

\bibitem{estar}
M. J. Berger, J. S. Coursey, M. A. Zucker and J. Chang,
``ESTAR Stopping-Power and Range Tables for Electrons'', [Online]. 
Available: http://physics.nist.gov/PhysRefData/Star/Text/ESTAR.html.

% Multiple scattering

\bibitem{urban2002}
L. Urban, 
``Multiple scattering model in Geant4'',
CERN-OPEN-2002-070, Geneva, Switzerland, 2002.

\bibitem{urban}
L. Urban,
``A Model for multiple scattering in Geant4'',
in \emph{Proc. of The Monte Carlo Method: Versatility Unbounded in a Dynamic
Computing World}, 
on CD-ROM, American Nuclear Society, La Grange Park, IL, 2005.

\bibitem{urban2006}
L. Urban, 
``A model of multiple scattering in Geant4'', 
CERN-OPEN-2006-077, Geneva, Switzerland, 2006.

\bibitem{lewis}
H. W. Lewis, 
``Multiple Scattering in an Infinite Medium",
\textit{Phys. Rev.}, vol. 78, pp. 526-529, 1950.

\bibitem{elles}
S. Elles, V. N. Ivanchenko, M. Maire and L. Urban,
``Geant4 and Fano cavity: where are we?'',
\textit{J. Phys.: Conf. Series}, vol. 102, pp. 1-8, 2009.


% Penelope 2001-2008

\bibitem{penelope2001}
F. Salvat, J. M. Fernandez-Varea, E. Acosta, and J. Sempau,
``Penelope - A code system for Monte Carlo simulation of electron and
photon transport'', 
Proc. NEA Workshop 19, 2001.

\bibitem{penelope2008}
F. Salvat, J. M. Fernandez-Varea, and J. Sempau,
``Penelope - A code system for Monte Carlo simulation of electron and
photon transport'', 
Proc. NEA Workshop 6416, 2008.

% Rayleigh
\bibitem{tns_rayleigh}
M. Batic, G. Hoff, M. G. Pia, and P. Saracco,
``Photon elastic scattering simulation: validation and improvements to Geant4'',
\textit{IEEE Trans. Nucl. Sci.}, vol. 59, no. 4, pp. 1636-1664, 2012.

\bibitem{tns_relax}
S. Guatelli, A. Mantero, B. Mascialino, P. Nieminen, and M. G. Pia, 
``Geant4 Atomic Relaxation'', 
\emph{IEEE Trans. Nucl. Sci.}, vol. 54, no. 3, pp. 585-593, 2007.

% ---- Statistics

\bibitem{gof1}
G. A. P. Cirrone et al., 
``A Goodness-of-Fit Statistical Toolkit'', 
\emph{IEEE Trans. Nucl. Sci.}, vol. 51, no. 5, pp. 2056-2063, 2004.

\bibitem{gof2}
B. Mascialino, A. Pfeiffer, M. G. Pia, A. Ribon, and P. Viarengo, 
``New developments of the Goodness-of-Fit Statistical Toolkit'', 
\emph{IEEE Trans. Nucl. Sci.}, vol. 53, no. 6, pp. 3834-3841,  2006.

\bibitem{R}
R Core Team,
``R: A language and environment for statistical computing'' R Foundation for
Statistical Computing, Vienna, Austria, ISBN 3-900051-07-0, 2012. 
[Online]. Available: http://www.R-project.org/.

\bibitem{bock}
R. K. Bock and W. Krischer,
``The Data Analysis BriefBook '',
Ed. Springer, Berlin, 1998. 

\bibitem{pearson}
K. Pearson,
``On the $\chi^2$ test of Goodness of Fit'',
\textit{Biometrika}, vol. 14, no. 1-2, pp. 186-191, 1922.

\bibitem{fisher}
R. A. Fisher,
``On the interpretation of  $\chi^2$ from contingency tables, and the calculation of P'',
\textit{J. Royal Stat. Soc.}, vol. 85, no. 1, pp. 87-94, 1922. 

\bibitem{barnard}
 G. A. Barnard,
`` Significance tests for 2 × 2 tables'',
\textit{Biometrika}, vol. 34, pp. 123-138, 1947.

\bibitem{boschloo}
R. D. Boschloo, 
``Raised Conditional Level of Significance for the 2$\times$2-table when Testing the Equality of Two Probabilities'', \textit{Stat. Neerlandica}, vol. 24, pp. 1-35, 1970.

\bibitem{suissa}
S. Suissa,  and  J. J. Shuster,
``Exact Unconditional Sample Sizes for the 2$\times$2 Binomial Trial'',
\textit{J. Royal Stati. Soc., Ser. A}, vol. 148, pp. 317-327, 1985.

\bibitem{andres_1994}
A. Martin Andres and  A. Silva Mato,   
``Choosing the optimal unconditioned test for comparing two independent proportions'',
\textit{Comp. Stat. Data Anal.},  vol. 17, pp. 555-574, 1994.

\bibitem{andres_2004}
A. Martin Andres, A. Silva Mato, J. M . Tapia Garcia, and M. J. Sanchez Quevedo,  
``Comparing the asymptotic power of exact tests in 2 × 2 tables'',
\textit{Comp. Stat. Data Anal.},  vol.  47, pp. 745-756, 2004.

\bibitem{agresti}
A. Agresti, 
``A Survey of Exact Inference for Contingency Tables'', 
\textit{Stat. Sci.}, vol. 7, pp. 131-153, 1992.

\bibitem{mcnemar}
Q. McNemar, 
`` Note on the sampling error of the difference between correlated proportions or percentages'',
\textit{Psychometrika}, vol. 12, pp. 153-157, 1947. 

\bibitem{bennett}
B. M. Bennett and R. E. Underwood,
``Note: On McNemar's Test for the 2 $\times$ 2 Table and Its Power Function'',
\textit{Biometrics}, vol. 26, no. 2, pp. 339-343, 1970.  

\bibitem{yates}
F. Yates,
``Contingency table involving small numbers and the $\chi^2$ test'',
\emph{J. Royal Stat. Soc. Suppl.}, vol. 1, pp. 217-235, 1934.

\bibitem{lui_2001}
K. J. Lui,
``Notes on Testing Equality in Dichotomous Data with Matched Pairs'',
\textit{Biometr. J.}, vol. 43, no. 3, pp. 313–321, 2001.

% Goudsmit-Saunderson

\bibitem{goudsmit1}
S. Goudsmit and J. L. Saunderson,
``Multiple Scattering of Electrons'',
\textit{Phys. Rev.}, vol. 58, pp. 24-29,  1940.

\bibitem{goudsmit2}
S. Goudsmit and J. L. Saunderson,
``Multiple Scattering of Electrons. II'',
\textit{Phys. Rev.}, vol. 58, pp. 36-42,  1940.

% Sw process
\bibitem{up}
I. Jacobson, J. Booch, and J. Rumbaugh, 
``The Unified Software Development Process'', 
1st ed., Ed.: Addison-Wesley, 1999.

\bibitem{cmmi}
M. B. Chrissis, M. Konrad and S. Shrum,
``CMMI\textregistered ~for Development'', 
Ed.: Addison-Wesley, 2011.

\bibitem{iso15504}
ISO/IEC Joint Technical Committee 1, 
``ISO/IEC DTR 15504 Part 5: An Assessment Model and Indicator Guidance'', 
Ed: J. M. Simon, 1999.

\bibitem{ieee_vv}
IEEE Computer Society,
``IEEE Standard for Software Verification and Validation'', 
IEEE Std 1012-2004, 2005.

\bibitem{chep2012_refactoring}
M. Batic et al., 
``Refactoring, reengineering and evolution: paths to Geant4 uncertainty quantification and performance improvement'',
\textit{J. Phys.: Conf. Ser.}, vol. 396, pp. 022038, 2012.

\end{thebibliography}
\end{document}